\documentclass[%
 reprint,
 amsmath,amssymb,
 aps,
amsfonts,
prb,
prstab,
prstper,
floatfix,
]{revtex4-2}

\usepackage{graphicx}
\usepackage{dcolumn}
\usepackage{bm}
\usepackage{hyperref}
\usepackage[mathlines]{lineno}

\usepackage[caption=false]{subfig}
\usepackage{comment}

\usepackage[
scale=0.7, marginratio={1:1, 2:3}, ignoreall,
text={7in,10in},centering,
margin=0.24in, 
total={6.5in,8.75in}, top=1.2in, left=0.6in,right = 0.6in, includefoot, 
]{geometry}

\usepackage{setspace}
\hypersetup{
	colorlinks=true,
	linkcolor=blue, 
	urlcolor=cyan,
	citecolor=green 
}

\usepackage{tikz}
\usetikzlibrary{shapes.geometric, arrows}

\tikzstyle{startstop} = [rectangle, rounded corners, minimum width=3cm, minimum height=1cm,text centered, draw=black, fill=red!30]
\tikzstyle{io} = [trapezium, trapezium left angle=70, trapezium right angle=110, minimum width=3cm, minimum height=1cm, text centered, draw=black, fill=blue!30]
\tikzstyle{process} = [rectangle, minimum width=3cm, minimum height=1cm, text centered, text width=3cm, draw=black, fill=orange!30]
\tikzstyle{decision} = [diamond, minimum width=3cm, minimum height=1cm, text centered, draw=black, fill=green!30]
\tikzstyle{arrow} = [thick,->,>=stealth]

\begin{document}

\preprint{APS/123-QED}

\title{Universal Relations for rapidly rotating neutron stars\\ using supervised machine-learning techniques}

\author{\large Grigorios Papigkiotis}
\email{gpapigki@auth.gr}

\author{\large George Pappas}%

 \email{gpappas@auth.gr}
\affiliation{\large Department of Physics, Aristotle University of Thessaloniki, Thessaloniki 54124, Greece\\%
}%






\begin{abstract}
As some of the most compact stellar objects in the universe, neutron stars are unique cosmic laboratories. The study of neutron stars provides an ideal theoretical testbed for investigating both physics at supra-nuclear densities as well as fundamental physics. Their global astrophysical properties however depend strongly on the star's internal structure, which is currently unknown due to uncertainties in the equation of state. In recent years, a lot of work has revealed the existence of universal relations between stellar quantities that are insensitive to the equation of state. At the same time, the fields of multimessenger astronomy and machine learning have both advanced significantly. As such, there has been a confluence of research into their combination and the field is growing. In this paper, we develop universal relations for rapidly rotating neutron stars, by using supervised machine learning methods, thus proposing a new way of discovering and validating such relations. The analysis is performed for tabulated hadronic, hyperonic, and hybrid EoS-ensembles that obey the multimessenger constraints and cover a wide range of stiffnesses. The relations discussed could provide an accurate tool to constrain the equation of state of nuclear matter when measurements of the relevant observables become available.

\end{abstract}

\maketitle


\section{\label{sec:introduction}Introduction}

Neutron stars (NSs) are the densest and most exotic, often rapidly rotating, compact astrophysical objects in the universe. Their study provides a meeting place of research between relativistic astrophysics and nuclear physics. They present an interesting astrophysical system described by the General Theory of Relativity (GR), and constitute a laboratory impossible to recreate in terrestrial environments. 

One of the open problems in nuclear astrophysics is the determination of fundamental interactions at extremely high densities reached in NS cores. These regimes of ultra-dense and cold nuclear matter are still poorly known. In this context, the relation between the structure of a NS, its global properties (mass, radius, etc.), and the microphysics, namely the equation of state (EoS), is crucial for understanding and testing various astrophysical scenarios. Assuming that a perfect fluid describes nuclear matter at the star's interior, the fundamental microphysics properties are encapsulated by a barotropic EoS, which corresponds to a relation between the pressure and the energy density of matter \cite{camenzind_compact_2007,breu_maximum_2016, poisson_gravity_2014,rezzolla_physics_2018,steiner2010equation,lattimer2011neutron,zdunik2013maximum, ozel_masses_2016}. Such an EoS remains an open question in nuclear physics.

Over the years, a wide variety of different models for the EoS have been proposed by the nuclear physics community. Approaches and models differ by the assumed NS interior composition, the nucleon interaction properties, and the methodologies involved in solving the related many-body problem. In each case, modeling NSs and deriving their properties directly depends on the particular EoS. Once the EoS is specified, one can compute stellar models using the framework of GR. Each EoS has a specific prediction for the relation between the mass and the equatorial radius of a NS ($M=M(R_{eq})$), and this $1-1$ correspondence could be used to identify the most promising EoS candidates by using astrophysical observations of NSs. Electromagnetic sources, and the most prominent gravitational-wave (GW) observations, will provide information to constrain these models and eventually help us understand the behavior of matter inside the neutron  star's inner core \cite{ferrari_general_2020,steiner2010equation,lattimer2011neutron, haensel_neutron_2007,zdunik2013maximum,ozel_masses_2016}.

Although NS properties are related to the EoS \cite{lattimer2011neutron, zdunik2013maximum,steiner2010equation,ozel_masses_2016}, it could be possible to find ways to describe stellar objects and their properties in manners that are not very sensitive to the specific choice of the EoS. In recent years, there has been a lot of work in that direction, i.e., looking for EoS-insensitive relations between stellar parameters. The motivation behind this is twofold. On the one hand, having observables that are insensitive to the specific characteristics of the EoS can be very useful in our efforts to measure astrophysically the various properties of NSs, such as the NS radius, the moment of inertia, the tidal deformability, or the multipole moments, while at the same time reducing the errors and model uncertainties significantly. 
Having then measured some of the difficult-to-measure parameters, one could try to use some of them to solve the inverse problem of constraining the EoS.
On the other hand, appropriate observable quantities and the relevant EoS independent relations could help circumvent the unknowns of nuclear physics at supranuclear densities, thus providing a framework for NSs to be used for detecting possible effects due to modifications in GR \cite{breu_maximum_2016,rezzolla_physics_2018}. 
Such relations have been found for stationary isolated NSs, as well as fully dynamical and merging binary systems \cite{rezzolla_physics_2018,breu_maximum_2016,riahi_universal_2019}.

In what follows, we give a brief incomplete outline of the development of these relations. 
One of the first instances that such a relation was introduced, was when Ravenhall \& Pethick \cite{ravenhall1994neutron} accentuated an apparently EoS insensitive relation which connects the normalized moment of inertia $I/MR_{eq}^2$ with the stellar compactness expressed in geometric units $\mathcal{C}=M/R_{eq}$. This relation was produced considering EoSs without an extreme softening at supranuclear densities. It was later modified by Lattimer \& Prakash \cite{lattimer2001neutron}, and Bejger \& Haensel \cite{bejger2002moments}, and then employed by Lattimer \& Schutz \cite{lattimer2005constraining} in order to estimate the NS's radius using combined data measurements from mass and moment of inertia of a pulsar in a binary system. Similar relations were also suggested by Breu \& Rezzola \cite{breu_maximum_2016} in both the case of slowly as well as rapidly rotating equilibrium configurations. In the field of asteroseismology, it was recognized by Andersson \& Kokkotas \cite{andersson1996gravitational, andersson1998towards}, and reexamined by Benhar, Ferrari \& Gualtieri \cite{benhar2004gravitational}, that a tight correlation between the $f,w$ fundamental oscillation modes and stellar average density, manifests some EoS independence. Furthermore, Bejger, in \cite{bejger2013parameters}, derived a pair of relations that connect the maximum and minimum masses ($M_{max},M_{min}$) of a rotating star to the maximum mass of a non-rotating stellar configuration, $M_{\text{tov}}$. Additional EoS-independent relations between the redshift (polar, forward, and backward) \cite{lattimer2001neutron,bejger2002moments} and the minimum and maximum compactness of the star were also presented \cite{bejger2013parameters}. Laarakkers \& Poisson \cite{laarakkers1999quadrupole} showed that for a fixed mass and EoS, the dependence between the quadrupole moment $Q$ and the star's angular momentum $J$ can be fitted by a quadratic formula, while Pappas \& Apostolatos \cite{Pappas:2012ns,Pappas:2012qg} demonstrated that in addition the spin octupole $S_3$ exhibits a cubic dependence in $J$, hinting to a Kerr-like behavior for the moments. Urbanec, Miller \& Stuchlik \cite{urbanec2013quadrupole} using the Hartle-Thorne, slow-rotation approximation found that there is a universal relation between the reduced quadrupole moment $\bar{Q}=-MM_{2}/J^2$ and the inverse compactness over two 
$R/2M$ for slowly rotating NSs. Considering slowly rotating NSs as well as quark stars, Yagi \& Yunes highlighted a new universal relation between the normalized quadrupole moment $\bar{Q}$ and the stellar compactness $\mathcal{C}\equiv M/R$ \cite{yagi2017approximate}.

In more recent years, Yagi \& Yunes \cite{yagi2013love} discovered some EoS insensitive relations involving the appropriately reduced moment of inertia $I$, quadrupole moment $Q$, and tidal love number $\lambda$, that had a better than $1\%$ accuracy in the slow-rotation limit, and assuming small tidal deformations. These expressions were called $I-\textrm{Love}-Q$ relations. Related work using Hartle-Thorne approximation has also been done by 
Baub{\"o}ck et al. \cite{baubock2013relations}. These relations can be a useful tool in order to remove degeneracies appearing in modeling GW signals coming from inspiraling binaries \cite{yagi_i-love-q_2013a,maselli2013equation}. 

However, in the case of rapid rotation, Doneva, Yazadjiev, Stergioulas \& Kokkotas \cite{doneva2013breakdown} showed that the EoS insensitive behavior between $I$ and $Q$ weakens, assuming sequences of models with constant spin frequency. The universality of the $I-Q$ relation for rapidly rotating NSs was revisited by Pappas \& Apostolatos \cite{pappas2014effectively} who noted that universality is recovered when rotation is parameterized by the normalized spin $\chi=J/M^2$, where $J$ and $M$ are the angular momentum and mass of the star expressed in geometric units, instead of the spin frequency. This result was immediately reproduced and extended by Chakrabarti et al. \cite{chakrabarti2014q} who also presented extensions of the $I-Q$ relations considering dimensionless quantities such as $M\times f \ \textrm{and} \ R_{eq}\times f$. These works highlighted the importance of using appropriate dimensionless quantities for producing universal relations, which hold for even extremely rapidly rotating NSs. However some confusion on this topic still remains in the literature.
 
Cipolletta et al. \cite{cipolletta2017last} derived useful relations for the binding energy of non-rotating and rotating configurations as a function of the gravitational mass and the spin parameter $\chi$. In addition, they presented an expression between the maximum mass along the secular instability limit and $\chi$. Riahi, Kalantari \& Rueda \cite{riahi_universal_2019} also presented universal relations connecting the mass and frequency along the mass-shedding sequence to the mass and radius of the non-rotating NS configuration with the same energy density. Luk et al. \cite{luk2018universal} investigated the last stable circular orbit of a test particle around rapidly rotating NSs. They highlighted a pair of fitting formulas that relate the radius and orbital frequency of this orbit to the rotation frequency and mass of the rapidly rotating NS. Furthermore, Haskell et al.\cite{haskell2013universality} have shown that the EoS independent relation between $I$ and $Q$ is broken for stars with long spin periods ($P\gtrsim 10 \ \textrm{s}$) and strong magnetic fields ($B\gtrsim 10^{12} \ \textrm{Gauss}$).

On a slightly different direction, Pappas \& Apostolatos \cite{pappas2014effectively}, and Stein, Yagi \& Yunes \cite{stein2014three}, investigated EoS insensitive three hair relations ($M,\chi,\bar{Q}$) for NSs, that where later extended to quark stars as well (Yagi et al. \cite{yagi2014effective}, Chatziioannou, Yagi \& Yunes \cite{chatziioannou2014toward}), which involved higher order multipole moments. Furthermore, Doneva, Yazadjiev, Staykov \& Kokkotas \cite{doneva2014universal}, Kleihaus, Kunz \& Mojica \cite{kleihaus2014quadrupole}, Pani \& Berti \cite{pani2014slowly}, Pappas et al. \cite{Pappas:2018csu}, and Yagi \& Stepniczka \cite{Yagi:2021loe}, to name a few, have investigated universality in modified theories of gravity as well. 

Many efforts have also been made to provide some justification for the observed universality. Yagi \& Yunes initially suggested that $I,\lambda, \textrm{and}, Q$ are mainly determined by the properties of nuclear matter at low-mass densities. The idea was that the effect is due to the fact that nuclear-physics experiments better constrain realistic EoSs at these regimes and therefore the various EoSs do not differ significantly \cite{yagi_i-love-q_2013}. Subsequent investigations \cite{stein2014three,yagi2014love} demonstrated that the universality is strongly connected to having a structure of homologous isodensity profiles inside the stars, a property that holds exactly for Newtonian constant density configurations but is also approximately true for realistic NSs in GR. Martinon et al. \cite{martinon2014rotating}, and Sham et al. \cite{sham2015unveiling} proposed that the ultradense EoSs are stiff enough, that can be considered as an expansion around the incompressible limit. Therefore, low-mass stellar objects, composed chiefly of soft matter at low densities, would depend more sensitively on the EoS, while the EoS-independence appears towards higher compactnesses, where nuclear matter approaches the limit of an incompressible fluid. The two pictures are compatible and complementary to each other, since a homologous, almost constant-ellipticity profile for the isodensity surfaces is what one finds for stars in the incompressible limit.  

Lastly, concerning the rotating star's surface, Morsink et al., \cite{Morsink:2007tv} as well as AlGendy \& Morsink \cite{AlGendy:2014eua} 
proposed EoS-insensitive formulas related to the equatorial radius $R_{eq}$, assuming that the fitting coefficients depend on both the stellar compactness $\mathcal{C}$ and the dimensionless spin $\sigma=\Omega^2R^3_{eq}/GM$. Silva, Pappas, Yunes \& Yagi \cite{Silva:2020oww} 
reviewed these coefficients for a wider range of NS models, and suggested a new fitting formula based on an elliptical isodensity approximation \cite{1993ApJS...88..205L}. 

In this work, we revisit some established universal relations and produce some new ones. Along the way, we develop a new approach to generating such relations using machine-learning techniques. The complete analysis is performed using statistical evaluation methods to find the best universality patterns and supervised machine-learning techniques, such as Cross-Validation and Linear-Regression, to determine the best functional form that verifies the correlated data. The primary investigation is carried out for a wide range of astrophysically relevant slowly rotating as well as rapidly rotating models with frequencies ranging from a few hundred $Hz$ up to $kHz$, close to the mass-shedding limit. 
More specifically, we present here universal relations related to the star's reduced quadrupole deformation $\bar{Q}$, the normalized moment of inertia $\bar{I}$ as well as the star's reduced octupole deformation $\bar{S_3}$ for a wide range of parameters including stellar compactness and rotation. We also look into some less standard relations, such as relations about the inverse compactness $\mathcal{K}=\mathcal{C}^{-1}$, the fraction of kinetic to gravitational energy, and the rotational frequency $f$ of the star.

The plan of the paper is as follows. First, in section \ref{sec:level_eos}, we briefly review the mathematical setup and EoSs used to calculate our equilibrium stellar models ensemble. Then, section \ref{sec:machine_learning} is dedicated to presenting the statistical evaluation test and the machine-learning framework used to extract our results. Then in section \ref{sec:results}, we present our main EoS-insensitive fitting formulas, while in section \ref{sec:conclusions}, we summarize our results and present our concluding remarks. Finally, in Appendix \ref{app:rest_universal}, we present some additional universal relations beyond those discussed in Section \ref{sec:results}. Unless stated otherwise, we work in geometric units with $G=c=1$.

\section{\label{sec:level_eos}Mathematical setup and Equation of state models}

Most significant uncertainties about NS bulk properties, are related to the unknown interactions present at high-density regions. As soon as the energy density significantly exceeds the nuclear saturation density $\rho_ {0} = 2.8\times 10^{14} gr / cm^{3}$ of the ordinary standard symmetric nuclear matter found in heavy atomic nuclei, the structure and composition of a NS become more uncertain \cite{haensel_neutron_2007,steiner2010equation,lattimer2011neutron}. The construction of sequences of static and rotating NSs directly depends on the resulting EoS. 
Different proposed EoSs have strikingly different bulk properties. When the EoS is provided, it can be used as an input for integrating Einstein's Field equations. Therefore, the EoS is essential for describing the macroscopic properties of NS physics~\cite{haensel_neutron_2007,friedman_rotating_2013,shapiro_black_2008}. In this work, we have constructed a large number of equilibrium NS models for a variety of EoSs, that are either non-rotating or uniformly rotating with frequencies from a few hundred $Hz$ up to the range of $\sim kHz$ close to the mass-shedding limit. For non-rotating configurations, solutions can be obtained after the integration of hydrostatic equilibrium equations in spherical symmetry~\cite{haensel_neutron_2007,oppenheimer_massive_1939}, while for rotating ones, we have used a two-dimensional numerical code for integrating the non-linear elliptic type field equations, together with the hydrostationary equilibrium equation \cite{cook_spin-up_1992,friedman_rotating_2013}.

More precisely, we consider the stellar matter as perfect fluid with local isotropy and energy-momentum tensor \cite{friedman_rotating_2013, rezzolla2013relativistic} 
\begin{equation}
\label{Tmn}
T^{a\beta}=\left(\epsilon+P\right)u^au^{\beta}+Pg^{a\beta},
\end{equation}
where $u^a$ is the fluid 4-velocity, $g^{a\beta}$ is the metric tensor, while $\epsilon$ and $P$ are scalar quantities that correspond to the fluid's total energy density and pressure, respectively. Considering non-rotating NSs, we take a spherically symmetric metric
\cite{friedman_rotating_2013} 
\begin{equation}
\label{ds2_stat}
ds^2=- e^{2\nu(r)}dt^2+e^{2\lambda(r)}dr^2+r^2\left(d\theta^2+\sin^2\theta d\phi^2\right), 
\end{equation}
where $\nu(r)\ \text{and} \  \lambda(r)$ are time-independent metric functions of the radial coordinate $r$ as Birchoff's theorem suggests.
Since the metric tensor is time-independent, the matter inside the NS is in hydrostatic equilibrium ~\cite{haensel_neutron_2007,hobson_general_2006,poisson_gravity_2014}. Equilibrium models are then found as solutions of the Tolman-Oppenheimer-Volkoff (TOV) equations \cite{haensel_neutron_2007,oppenheimer_massive_1939}
\begin{subequations}
	\label{eq:tov_system}
	\begin{align}
	\frac{dm}{dr} & = 4\pi r^2 \epsilon(r),\\
	\frac{d\nu}{dr}& =\frac{m(r)+4\pi r^3 P(r)}{r(r-2m(r))},\\
	\frac{d P}{d r}& =-\left(\epsilon(r) +P(r)\right)\frac{d\nu}{dr},
	\end{align}
\end{subequations}
where the function $m(r)$ is identified as $m(r) = \frac{r}{2}\left(1-e^{-2\lambda (r)}\right)$. 
The TOV equations are supplemented by a cold dense nuclear matter EoS that provides a relation between the energy density and pressure, $\epsilon=\epsilon(P)$ ~\cite{haensel_neutron_2007,hobson_general_2006}.

As we have mentioned, there is still uncertainty about the EoS at high densities, and a large number of EoS models are presented in the literature. EoS models are based on different many-body nonrelativistic \cite{dutra2012skyrme} and relativistic \cite{dutra2014relativistic} theories employed to describe nuclear matter at ultra-high densities in $\beta$-equilibrium. More specifically, some nonrelativistic nuclear models that are taken into account are those using nuclear effective interaction forces (EI), cluster energy functionals (CEF), and unified Scyrme-Hartree-Fock nuclear forces (SHF) \cite{haensel_neutron_2007}. The relativistic methods encountered are based on the relativistic mean-field theory (RMF), the chiral perturbation theory (chPT), the perturbative Brueckner-Bethe-Goldstone quantum theory (BBG), the Brueckner-Hartree-Fock approximation with the continuous choice for the auxiliary single-particle potential (BHF), the chiral mean-field theory models (CMF), the non-perturbative functional renormalization group approach (NP-FRG), the Nambu-Jona-Lasinio (NJL) model \cite{yu_self-consistent_2020} within the mean-field approximation (NJL-MF) and, finally, the non-perturbative functional renormalization group approach (NP-FRG) within a two flavor quark-meson truncation in the local potential approximation (LPA) including vector interactions \cite{haensel_neutron_2007,malik_gw170817_2018}. 

Considering these theoretical features, we have used EoS models based on the \href{https://compose.obspm.fr/home}{comPOSE} database. In our EoS ensemble, hadronic, hyperonic, and hybrid models were used in tabulated form to describe the NS's interior, including the crust and the main core. The complete list (38 in total) of cold EoSs used in this work for each possible NS category examined is presented in the tables (\ref{tab:hadronic},\ref{tab:hyperonic},\ref{tab:hybrid}) given in Appendix ~\ref{app:eos_tables}. For all of them, $\beta$-equilibrium and zero temperature were assumed. Therefore,  the EoS reduces to a relation between the pressure and the energy density. In addition, many individual families are divided even further based on the physical theory used to describe the EoS data. Furthermore, for each EoS, we provide the matter composition at the star's core and also the NS properties, such as the non-rotating maximum mass, the corresponding radius, and the radius of $1.4 M_{\odot}$ NS.

\begin{figure}[!ht]
	\includegraphics[width=0.4\textwidth]{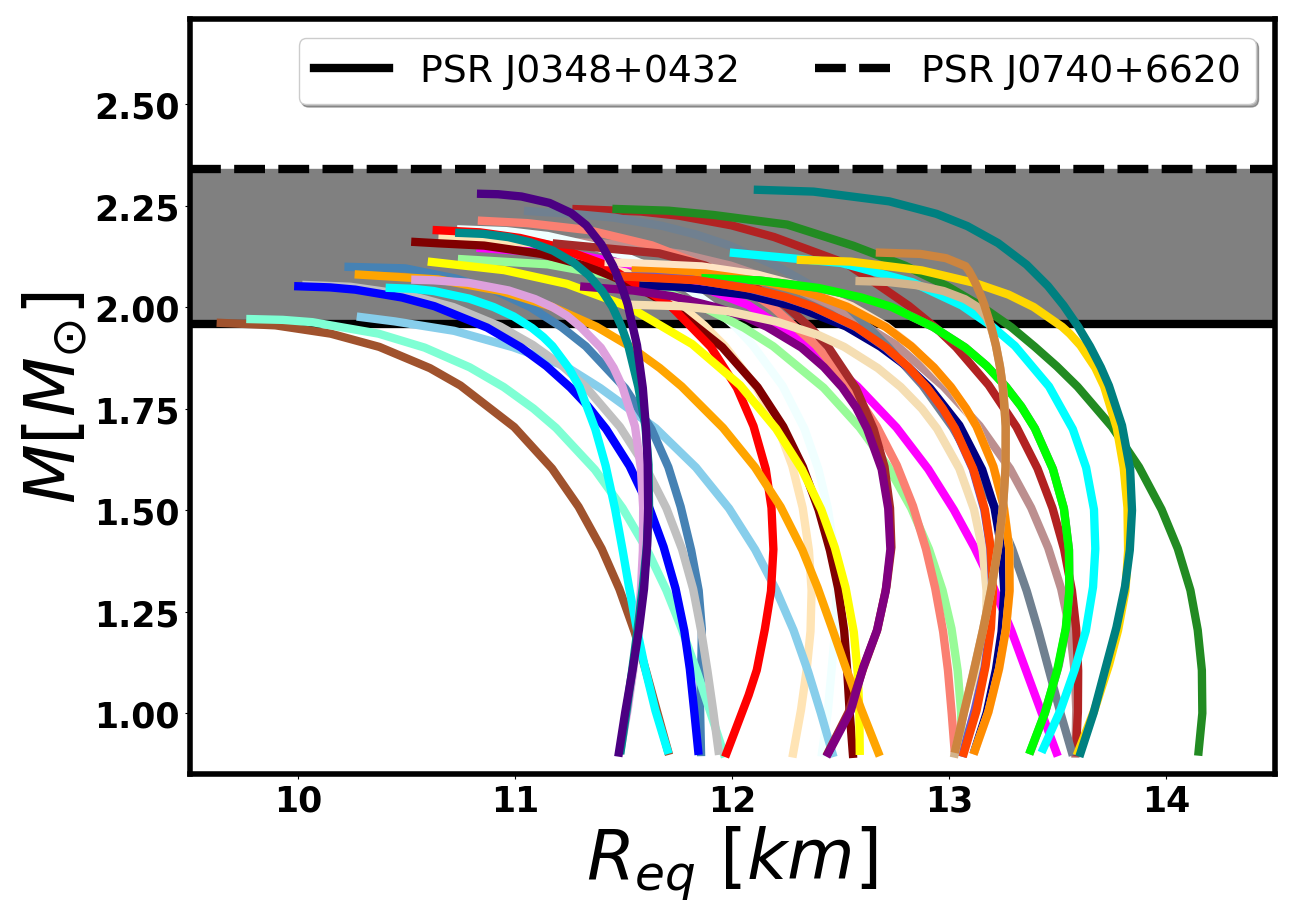}
	\caption{\label{fig:M_R} {$M=M(R_{eq})$} diagram for sequences of non-rotating NSs. 
    The different colors correspond to different EoSs according to the legend given in Fig.\ref{fig:color_band}. The same color-to-Eos map is used for all subsequent figures.}
\end{figure}

We have to note that each EoS selected in the above tables satisfies the lower limit for the maximum non-rotating mass for the PSR J0348+0432 ($M=2.01^{\ +0.04}_{\ -0.04} \ M_{\odot}$)~\cite{demorest2010two,antoniadis_massive_2013} and the PSR J0740+6620 ($M=2.14^{\ +0.20}_{\ -0.18} \ M_{\odot}$) ($2\sigma$ credible interval) \cite{cromartie_relativistic_2020} and produces a non-rotating maximum mass NS with radius $R_{{M_{\textrm{max}}}} \geq \ 9.60^{+0.14}_{-0.03}\ \textrm{km}$ as suggested by the GW170817 NS-NS merger analysis ~\cite{bauswein_neutron-star_2017, friedman_astrophysical_2020}. Furthermore, none of them has a maximum mass greater than $2.32 \  M_{\odot}$ according to the $2\sigma$ bound, assuming that the final GW170817 remnant was a black hole ~\cite{dietrich2020new,rezzolla_using_2018}. For all the EoS models considered, we verified that they satisfy the acceptability conditions, which ensure equilibrium. The imposed conditions are ~\cite{haensel_neutron_2007,shapiro_black_2008}:
\begin{itemize}
	\setlength\itemsep{0.09em}
	\item first law of thermodynamics $\ d\epsilon/d\rho=(\epsilon+P)/\rho$, where $\rho$ is the baryon mass density,  
	\item dominant energy condition $\ \epsilon  c^2> P$,
	\item microscopic stability $c_s^2= dP / d \epsilon \geq 0$ and causality $ c_s^2= dP / d \epsilon \leq c^2 $ which ensures that the speed of sound $c_s$ in the dense matter should not exceed the speed of light,
	\item and, finally, the Harrison-Zeldovich-Novikov stability condition $dM/d\epsilon_c \geq 0\ $, i.e., considering the $M-\epsilon_c$ curve, stars with $\epsilon_{c}>\epsilon_{c}(M_{max})$ have $dM/d\epsilon_c < 0$ and are unstable, thus not astrophysically relevant. Therefore, a NS with the maximum possible mass should have the maximum possible central energy density $\epsilon_c$.
\end{itemize}
In Fig.\ref{fig:M_R}, we present the Mass-Radius diagram for the cold EoS ensemble used in this work. 

Most models assume $n p e\mu$ composition in the stellar core (table \ref{tab:hadronic}). However, a significant number of EoSs also include other components of exotic matter, such as hyperons (table \ref{tab:hyperonic}) or quarks (table \ref{tab:hybrid}). It is worth mentioning that the $M - R_{eq}$ relation for an EoS is displayed only up to its maximum mass. As we see, the radius of non-rotating NSs mainly decreases with increasing mass, and there is a significant scattering in the predicted maximum masses. The horizontal lines indicate the $2\sigma$ lower and upper range for the masses of the two most massive radio pulsars known to date, PSR J0348+0432 ($M=2.01^{\ +0.04}_{\ -0.04} \ M_{\odot}$) \cite{antoniadis_massive_2013} (solid line, lower limit) and PSR J0740+6620 ($M=2.14^{\ +0.20}_{\ -0.18} \ M_{\odot}$) \cite{cromartie_relativistic_2020} (dashed line, upper limit).

In general though, most astrophysical objects are often rotating and some times they even rotate rapidly. A rotating compact object is described by its mass $M$ and its angular momentum $J$ \cite{friedman_rotating_2013}. In this case, the spacetime is assumed to be stationary, axisymmetric, and asymptotically flat. These assumptions can be mathematically formulated by introducing two Killing vectors $t^a$ and $\phi^a$. 
In isotropic gauge, a stationary metric that satisfies the assumptions mentioned can be described by the line element \cite{butterworth_structure_1976,friedman_rotating_2013}

\begin{align}
\label{eq:ds_rot_2}
\nonumber ds^2=&-e^{(\gamma+\rho)}dt^2+e^{(\gamma-\rho)}r^2\sin^2\theta(d\phi-\omega dt)^2\\
&+e^{2a}(dr^2+r^2d\theta^2),
\end{align}
where the metric potentials $\gamma, \rho, \omega, \alpha$ are functions of the quasi-isotropic coordinates ($r,\theta$). In the case of uniformly rotating stellar models, the angular velocity $\Omega$ of the star is constant. The equation of hydrostationary equilibrium for a stationary, axisymmetric, and uniformly rotating perfect fluid star is given by \cite{friedman_rotating_2013}
\begin{equation}
\label{hydro_uniform}
\small
\frac{\nabla_a P}{\epsilon + P}=\nabla_{a} \ln u^t,
\end{equation}
where $u^a=u^t(t^a+\Omega\phi^a)$ is the 4-velocity of a fluid element expressed in terms of the timelike and spacelike Killing vectors $t^a$ and $\phi^a$ while,  
\begin{equation}
\small
u^t=\frac{e^{-(\rho+\gamma)/2}}{\sqrt{1-(\Omega-\omega)^2r^2\sin^2(\theta) \ e^{-2\rho}}}
\end{equation}
follows from the normalization condition $u^au_a = -1$ and the line element in Eq.\ref{eq:ds_rot_2}.

To integrate the non-linear Einstein field equations with the equation of the hydrostationary equilibrium \cite{cook_spin-up_1992,friedman_rotating_2013,paschalidis2017rotating}, many numerical methods have been developed since the 1970s \cite{wilson1972models,bonazzola1974exact,friedman1989implications,komatsu_rapidly_1989,komatsu_rapidly_1989_II,stergioulas1994comparing}. These equations can be solved numerically on a discrete grid using a combination of integral and finite differences techniques \cite{komatsu_rapidly_1989}. Komatsu Eriguchi and Hachisu (KEH) \cite{komatsu_rapidly_1989,komatsu_rapidly_1989_II} and Cook, Shapiro, and Teukolsky
(CST) \cite{cook_spin-up_1992,cook_rapidly_1994} developed an iterative numerical method using the integral representation with Green's functions. We use, for the numerical integration of the equations of structure and the field equations, the public \href{https://github.com/cgca/rns}{\textit{RNS}} 
library developed by Stergioulas and Friedman \cite{stergioulas1994comparing}, which is based on the aforementioned methods. 

More specifically, assuming a perfect fluid, the \textit{RNS} code solves for the NS's interior (matter and spacetime) and exterior (spacetime) configuration on a discrete grid with the radial coordinate $r$ compactified and equally spaced in the interval $s \in [0,1]$, using the definition $s \equiv r/(r + r_{eq} )$, and the angular coordinate $\mu = \cos(\theta)$ also equally spaced in the interval $\mu \in [0, 1]$. In the former definition for the compactified radial coordinate $s$, $r_{eq}$ corresponds to the coordinate radius of the star's surface at the equator. The star's center is at $s=0$, the surface is at $s=1/2$, whereas infinity is at $s=1$. The equatorial plane is located at $\mu=0$ ($\theta= \pi/2$) while the pole at $\mu=1$ ($\theta= 0$) \cite{friedman_rotating_2013,cook_spin-up_1992,butterworth_structure_1976,komatsu_rapidly_1989}. Therefore, around half of the computational grid is assigned to the star's interior, while the other half is assigned to the vacuum exterior. The usual choice for the angular grid is to be half of the radial one. In our calculations, we use a grid size of $\textrm{MDIV} \times \textrm{SDIV}=151 \times 301$, where $\textrm{MDIV}$ is the number of points in the $\mu$-direction (number of spokes in the angular direction) and $\textrm{SDIV}$ the number of points in the $s$-direction.

For a given EoS, assuming uniform rotation, the \textit{RNS} code can obtain unique equilibrium solutions once a central energy density $\epsilon_c$ and the axial ratio $r_{pol}/r_{eq}$ between the polar and the equatorial coordinate radii have been specified. Therefore, stellar models can then be computed along sequences in which the central energy density and the axial ratio are varied \cite{friedman_rotating_2013, stergioulas1994comparing}. Together with the star's metric functions that are calculated numerically in the interior and exterior region, the \textit{RNS} source code also computes the fluid configuration as well as various equilibrium quantities, such as the star's gravitational mass $M$, the proper mass $M_p$, the baryonic mass $M_b$, the angular momentum $J$, the equatorial radius $R_{eq}$, the moment of inertia $I$, the fraction of rotational to gravitational energy $T/|W|$, the spacetime's Geroch-Hansen multipole moments ($M_0\equiv M, S_1\equiv J,M_2,S_3,..$) \cite{geroch_multipole_1970_I,geroch_multipole_1970_II,hansen_multipole_1974,fodor_multipole_1989,rezzolla_physics_2018,Pappas:2012ns,Pappas:2012qg,Doneva2018book}, 
etc.
For each EoS included in our ensemble, we have calculated an extended sample of relatively slowly and rapidly rotating NSs for a wide range of astrophysically relevant models with central densities [$\epsilon_c\sim (3\times 10^{14}-3.1\times 10^{15}) \ gr/cm^3$] and masses starting from $\sim 0.9 M_{\odot}$ and up to the star's maximum mass $M_ {max}$. In total, our entire sample includes 11983 models of rotating NSs from a few hundred $Hz$ ($\simeq268 \ Hz$) up to $\sim 2 \ kHz$ close to the mass-shedding (keplerian) limit and 704 non-rotating and keplerian models in equilibrium.

\section{\label{sec:machine_learning}Correlation Matrix and Supervised Machine learning Techniques}

In this section, we will present the tools and processes that we have used in order to infer from the NS data the shot for universal relations.
  
To thoroughly investigate and identify the possible connections between the quantities that will be examined in the present work, we first use a statistical evaluation test method (correlation analysis) that brings to the surface the underlying connections between the quantities. The method is presented in the following section (\ref{sec:cor_matrix}). 

We then proceed to find the specific relations between the correlated quantities. Firstly, in order to indicate the best functional form that describes the data, we split the dataset into training and validation sets. Afterwards, we fit the model on the training set and evaluate its performance, using statistical tools (statistical metric functions), on the validation set. It should be noted that we appropriately use the whole dataset as a validation set. Then, the process is repeated and evaluated for various functional forms that could describe the data. 

Finally, for the function with satisfactory statistical scores (best functional form), we re-fit the model on the entire dataset to determine its coefficients (best coefficients). The complete analysis was performed using supervised machine learning methods such as the Cross-Validation evaluation test and the Least-Squares Regression formula, as elaborated in the following subsections (\ref{sec:cross_validation}, \ref{sec:regression}).

\subsection{\label{sec:cor_matrix} Correlation Analysis-Universality Patterns}

Motivated by the question of whether there may be useful universal relations between NS parameters that are yet undiscovered, we ventured to find a more systematic way of identifying the universality patterns that may exist between the various physical observable quantities. For this purpose, we decided to use tools from statistical analysis.

Correlation analysis is a method of statistical evaluation used to study the strength of a connection between two numerically measured, continuous variables $x,y$. The analysis shows the kind of relation (in terms of direction) and how strong the relationship between the two continuous variables is. By direction, we mean whether the variables are directly proportional or inversely proportional to each other, i.e., increasing the value of one variable might positively or negatively impact the value of the other variable. Pearson's correlation coefficient gives the main result of a correlation \cite{cowan_statistical_1998,james_statistical_2006}. Given the sample data $\{(x_i,y_i)\}_{i=1}^{N}$ consisting of $N$ pairs, the correlation coefficient is a dimensionless statistical metric function given by

\begin{equation}
\rho_{x,y}=\frac{\text{Cov}[x,y]}{\sigma_x\sigma_y},
\end{equation}
where $\text{Cov}[x,y]$ is the statistical covariance
\begin{equation}
\text{Cov}[x,y]=\frac{1}{N}\sum_{i=1}^N(x_i-\bar{x})(y_i-\bar{y})=E[x,y]-\bar{x}\bar{y}=\sigma_{xy},
\end{equation}
and $\sigma_x, \ \sigma_y$ represent the standard deviations defined as: 
$$\small \sigma_x= \left(\frac{1}{N} \sum_{i=1}^N (x_i-\bar{x})^2\right)^{1/2}, \ \sigma_y= \left(\frac{1}{N} \sum_{i=1}^N (y_i-\bar{y})^2\right)^{1/2}.$$ 

In the above definitions, $x_i,y_i$ are the individual sample points indexed with $i$ while $\bar{x},\bar{y}$, and $E[x,y]$ are the corresponding expectation values. First, we must note that Pearson's coefficient $\rho_{x,y}$ ranges from $-1$ to $+1$. The closer its value is to $+1$ or $-1$, the more closely the two examined variables ($x,y$) are related. When the correlation coefficient is positive, an increase in one variable also increases the other, while when the correlation coefficient is negative, the changes in the two variables are in opposite directions. If variables $x$ and $y$ are independent, then by definition, the correlation coefficient must be zero. However, if it is zero and we do not have specific information about the variables, we can only say there is no linear dependence. Other functional relations could exist between the variables \cite{cowan_statistical_1998,james_statistical_2006}.

For a random vector $\mathcal{X}=(X_1,...,X_p)$ the table containing all possible correlations between the pairs of the examined variables, considering $\text{Cor}(X_i,X_i)=1$, is defined as:
\begin{equation}
\text{Cor}(\mathcal{X})=
\begin{scriptsize}
\left[\begin{array}{cccc}
1 & \text{Cor}\left(X_{1}, X_{2}\right) & \ldots & \text{Cor}\left(X_{1}, X_{p}\right) \\
\text{Cor}\left(X_{2}, X_{1}\right) & 1 & & \text{Cor}\left(X_{2}, X_{p}\right) \\
\vdots & & \ddots & \vdots \\
\text{Cor}\left(X_{p}, X_{1}\right) & \text{Cor}\left(X_{p}, X_{2}\right) & \ldots & 1
\end{array}\right].
\end{scriptsize}
\end{equation}	
 
\begin{figure}[!h]
	\includegraphics[width=0.5\textwidth]{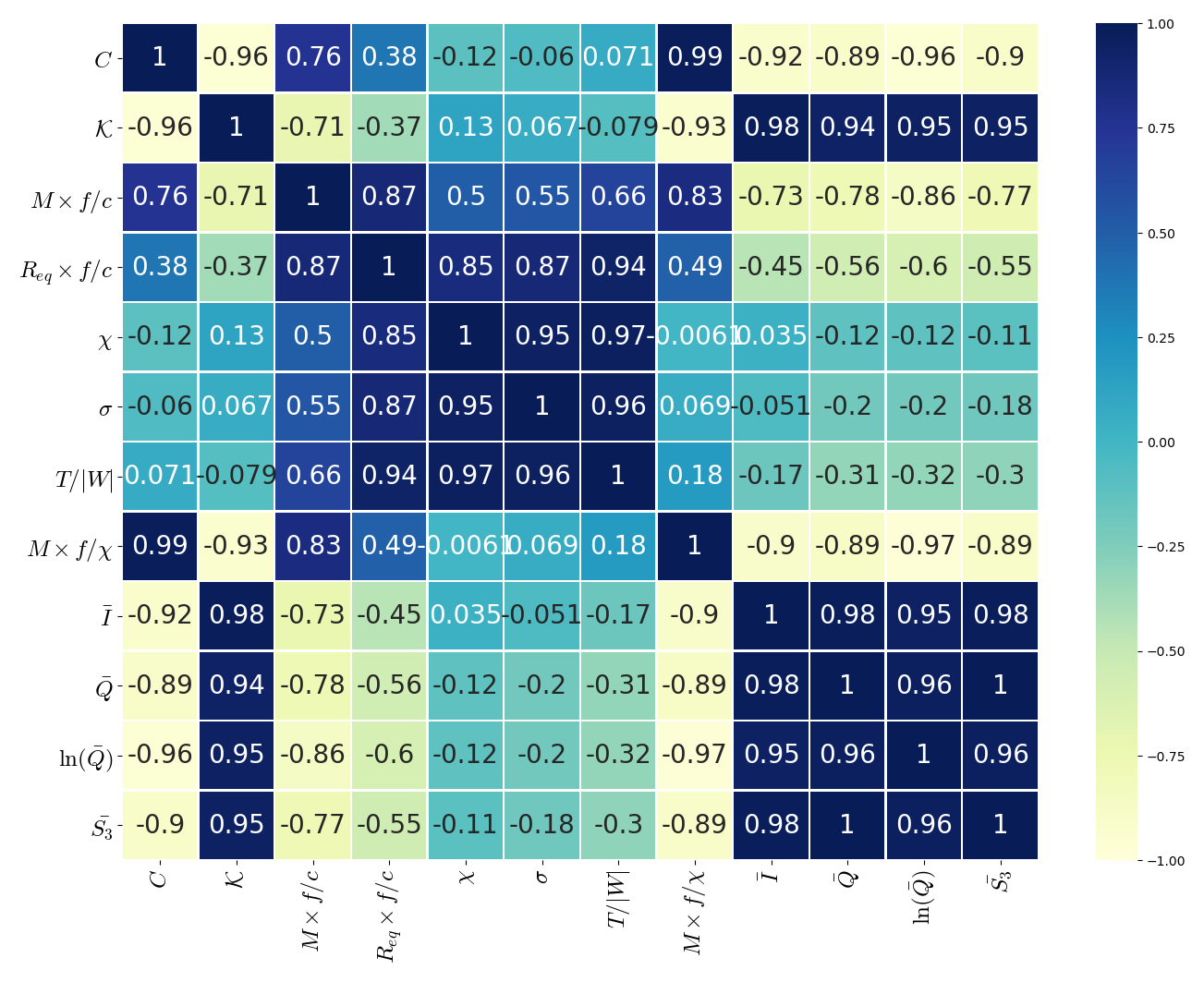}
	\caption{\label{fig:corr_matrix} Correlation Matrix for the physical quantities considered in section \ref{sec:results} for rapidly rotating stellar models.}
\end{figure}

For the various NS parameters that we have investigated for correlations, the resulting correlation matrix is given in Fig.\ref{fig:corr_matrix}.
The various observables presented in Fig.\ref{fig:corr_matrix} correspond to rapidly rotating stellar models and will be defined in the section \ref{sec:results}, where we highlight our EoS-insensitive relations.

\subsection{\label{sec:cross_validation} Cross-Validation}

We now proceed to briefly describe the evaluation process followed in order to find the best functional form that describes the data. 
In general, training a model to "learn" its parameters (by optimizing the prediction model function called the Loss function) and evaluating the model's statistical performance on the same dataset, is a methodological mistake. This is because, during training, the model associates patterns or features with duplicate labels in the training data, resulting in repeated labeling of new data points. This situation is designated as overfitting. Thus, it is possible to fail to predict anything useful on yet-unseen data. Consequently, the model would not have the appropriate generalization ability. Generalization is the ability of the model to correctly categorize new examples that differ from those used for training. More specifically, the variability of the input vectors may be such that the training data would comprise only a tiny fraction of all possible input vectors. Therefore, generalization is the central goal in machine learning \cite{bishop_pattern_2006} in order to be able to have a wider application. 

In order to compute the model that best describes the data, it is common practice to hold out a part of the available dataset, hidden from the training process, known as test set ($X_{\textrm{test}}, Z_{\textrm{test}}$). This is only possible when conducting a supervised (predictive) machine-learning experiment. By supervised machine learning task, we define the predictive approach that its goal is to learn a mapping from inputs $x$ to outputs $z$ given a labeled set of input-output pairs $\mathcal{D}=\{(x_i,z_i)\}_{i=1}^{N}$. 

\begin{figure}[!ht]
	\includegraphics[width=0.4\textwidth]{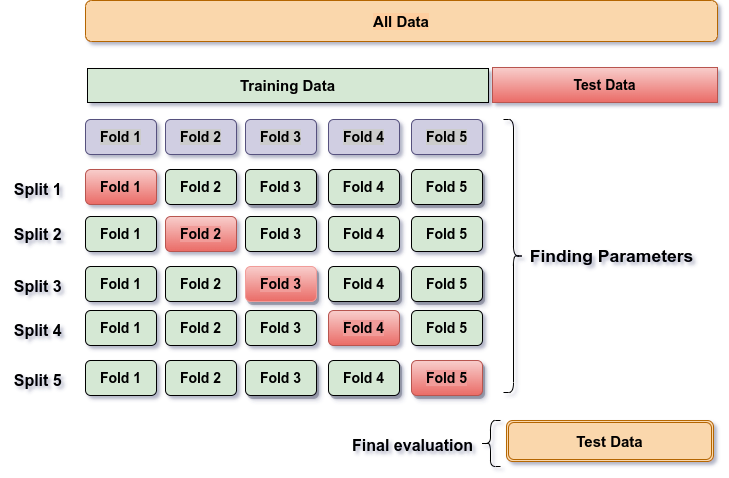}
	\caption{5-fold Cross-Validation splitting procedure. In this example, 20$\%$ of the data has been left out for each iteration and is shown as a test set. 
    The remaining 4 samples (folds) form the training set \cite{pedregosa_scikit-learn_2011,buitinck2013api}.}
	\label{fig:val2_flowchart}
\end{figure}

We define $\mathcal{D}$ as the training set and $N$ as the number of training examples \cite{murphy_machine_2012}. This evaluation process must be performed to select the best-fitting function that universally describes the data (generalization ability). The following figure shows a flowchart of a typical Cross-Validation workflow in model training. Grid search techniques can determine the best parameters \cite{pedregosa_scikit-learn_2011}.
\begin{figure}[!ht]
	\centering
	\includegraphics[width=0.4\textwidth]{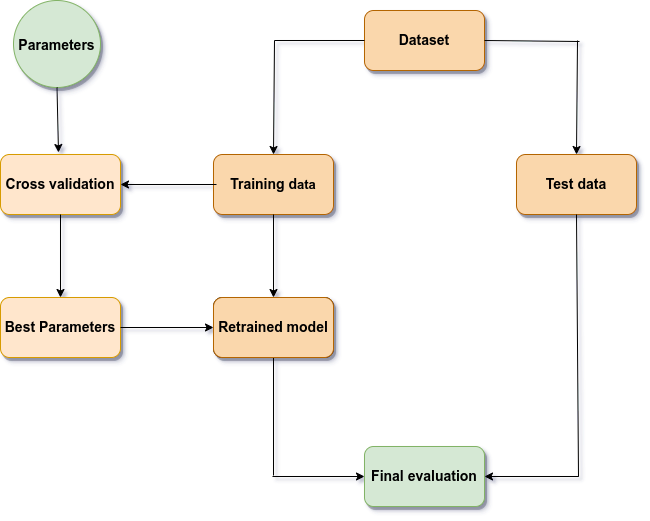}
	\caption{Best fit parameters: Cross-Validation flowchart \cite{pedregosa_scikit-learn_2011,buitinck2013api}.}
	\label{fig:val1_flowchart}
\end{figure}
The procedure illustrated in the flowchart (\ref{fig:val1_flowchart}) is repeated for different trial function models in order to find the functional form that best describes the data (best fitting function). More specifically, when evaluating different estimator parameters, i.e., different model functions in order to describe the data, there is a risk of overfitting on the test set used. This is because the fitting parameters can be tweaked until the estimator operates optimally on the test set. This way, information about the test set can leak into the model, and evaluation metrics no longer report on generalization performance. In order to solve this problem, an additional part of the dataset sample can be held out as a validation set. In the validation set, the training proceeds on the training set, after which evaluation is done on the validation set. When the experiment seems to be successful, a final evaluation can be done on the test set. In the most popular approach, defined as k-fold Cross-Validation, the training dataset is split into $k$ disjoint subsets (folds) \cite{james2013introduction,pedregosa_scikit-learn_2011}. Then, the following procedure is used for each one of the k-folds:
\begin{enumerate}
	\item A model function is trained using $k-1$ of the folds as training data.
	\item The resulting model function is validated on the $k$-th fold of the data as a test set.
\end{enumerate}
The performance estimation reported is then the average of the values computed during the k-fold Cross-Validation. This procedure can be computationally expensive but does not waste too much data (as is the case when producing an arbitrary validation dataset), which is a significant advantage in problems such as inverse inference where the number of samples is small. More clearly the process is given in Fig.(\ref{fig:val2_flowchart}) \cite{james2013introduction, pedregosa_scikit-learn_2011}.

There are a lot of different Cross-Validation iteration strategies that can be used to generate dataset splits. In this work, we use the 'Leave-One-Out'  Cross-Validation (LOOCV) method. LOOCV is a simple Cross-Validation method that provides train/test indices to split data into train/test sets. Each sample is used once as a test set (singleton), while the remaining samples form the training set. Therefore, each learning set is created by taking all the samples except one, the test set being the sample left out. Thus, for $n$ samples, we have $n$ different training sets and $n$ different test sets, i.e., the number of test sets is the same as the number of samples. LOOCV is a more computationally expensive but accurate validation method found in the literature. In addition, models constructed from LOOCV are virtually identical to each other, and the model built from the entire training set. \cite{james2013introduction,pedregosa_scikit-learn_2011}.

For estimating and evaluating the model's performance in the LOOCV test, we use the following statistical metric functions \cite{james2013introduction,pedregosa_scikit-learn_2011}.

\begin{itemize}
	\setlength\itemsep{0.09em}
	\item {\bf Max Error}: The max error function computes the maximum residual error (difference between the observed value and the estimated one), which captures the worst-case error between the predicted and actual values. Considering a perfectly fitted regression model, the training set's max error would be zero. However, this would be highly unlikely in the real world. This metric function shows the extent of error the model had when fitted.
	If $\hat{y}_i$ is the predicted value of the $i$-th sample and $y_i$ is the corresponding actual value, then the max error is defined as
	$$ \textrm{Max\_Error}(y,\hat{y})=\textrm{max}\left(|y_i-\hat{y}_i|\right),$$
	while the {\bf maximum deviation} is given as $$d_{\text{max}}(y,\hat{y}) =  \text{max} \left(\frac{|y_i-\hat{y}_i|}{\text{max}(\epsilon,|y_i|)} \right),$$
	where $\epsilon$ is an arbitrarily small yet strictly positive number to avoid undefined results when $y=0$. The result for $d_{\text{max}}(y,\hat{y})$ lies in the range $[0, 1]$. 

    In addition, other common statistical evaluation functions used are the {\bf Mean Absolute Error} (MAE)
    $$\textrm{MAE}(y,\hat{y}) = \frac{1}{n}\sum_{i=0}^{n-1}|y_i-\hat{y_i}|$$, and the {\bf Mean Squared Error} (MSE)
    $$\textrm{MSE}(y,\hat{y}) = \frac{1}{n}\sum_{i=0}^{n-1}(y_i-\hat{y_i})^2$$ computed as the average values over $n$ different test sets. At this point, we have to note that the max error, the maximum deviation, the MAE, and the MSE are slightly larger at the validation set compared to the corresponding quantities in the training set.
	\item {\bf Mean absolute percentage error metric}: The mean absolute percentage error (MAPE) is a metric evaluation function for regression problems. The idea of this metric function is to be sensitive to relative errors. Therefore, it is not expected to be changed by a global scaling of the target variable. For example, if $\hat{y}_i$ is the predicted value of the $i$-th sample and $y_i$ is the corresponding actual value, then the mean absolute percentage error (MAPE) estimated over the number of samples $n$ is defined as
	$$\textrm{MAPE}(y,\hat{y})=\frac{1}{n}\sum_{i=0}^{n-1}\frac{   |y_i-\hat{y}_i|}{\text{max}(\epsilon,|y_i|)}.$$
	Again, $\epsilon$ is an arbitrarily small but positive number used to avoid undetermined results when $y=0$. 
	Again, the result is in the range $[0, 1]$.
	\item {\bf Explained Variance}: Explained Variance is a regression score metric function. For example, if $\hat{y}$ is the estimated target output, $y$ the corresponding actual target output, and $\textrm{Var}$ is Variance, the square of the standard deviation, then the explained Variance is estimated as
	$$\text{Explained\_Variance}(y,\hat{y})=1-\frac{\textrm{Var}[y-\hat{y}]}{\textrm{Var}[y]}$$
	The best possible score at this metric function is 1.0, while lower values are worse.
	
\end{itemize}
Based on the Cross-Validation evaluation process, the reference criteria for selecting the fitting function are the statistical evaluation functions with the optimum results. To be precise, from all the possible functional forms tested to verify the data, the one with the optimal statistical metric evaluation functions is selected. 

\subsection{\label{sec:regression} Least-Squares Regression}

We now define the mathematical model used to adapt the best-fit function that describes the data. In this work, we use the linear Least-Squares regression approach to determine the best-approximating data fit. The Loss function that is optimized by the training procedure is the sum of the squares of differences between the given $z$-values (observed values) and the $\hat{z}$-values provided by the regression model. This procedure aims for the regression model to approximate the best-fitting function for the training data. For example, given a dataset that consists of $(x_i, y_i, z_i)$, $i=1,...n$ data where ($x_i,y_i$) are the independent variables and $z_i$ are the dependent ones, the best-fit optimal parameters $\hat{a}$, also known as optimizers, can be found by optimizing the Loss function \cite{burden_numerical_2015, ramachandran2009mathematical}:

\begin{equation}
\mathcal{L}=\sum_{i=1}^{n} ||z_i - \mathcal{F}(x_i,y_i;\hat{a})||^2=\sum_{i=1}^{n}r_i^2,
\end{equation}
with $r_i=z_i - \mathcal{F}(x_i,y_i;\hat{a})$. Depending on the case, the mathematical model function used has the form $\mathcal{F}(x,y;\hat{a})=\sum_{j=0}^{m}\hat{a}_j \ \mathcal{H}_j(x,y)$, where $m$ adjustable and uncorrelated optimizer parameters are held in the vector $\hat{a}$, i.e., $\hat{a}=[\hat{a}_0,\hat{a}_1,\hat{a}_2.....\hat{a}_{m-1},\hat{a}_m]$, while $\mathcal{H}_j(x,y)$ is a function of polynomial combinations of $x \ \textrm{and}\ y$. The regression coefficients $\hat{a}$ for the model that best fits the data should be given from the minimum of the Loss function by setting the corresponding partial derivative to zero. Since the model contains $m$ uncorrelated optimizer parameters $\hat{a}$, there are $m$ equations given by
\begin{equation}
\frac{\partial\mathcal{L}}{\partial \hat{a}_j}=-2\sum_{i=1}^n r_i\left(\frac{\partial r_i}{\partial \hat{a}_j}\right)=-2\sum_{i=1}^n r_i\left(\frac{\partial \mathcal{F}}{\partial \hat{a}_j}\right)=0,
\end{equation}
where $j=0,1,...,m$. Therefore, to compute the model optimizers $\hat{a}$, we need to set each partial derivative to zero and simultaneously solve the resulting equations system (normal equations) \cite{burden_numerical_2015}. Thus, for each particular universal correlation, it is necessary to derive a specific expression for the best possible fit regression model function $\mathcal{F}(x,y;\hat{a})$, as well as its partial derivatives in order to find the best-fit optimizers $\hat{a}$.

\section{\label{sec:results} Results-E\lowercase{o}S insensitive (Universal) Relations}

The correlation analysis described in the previous section, gave some candidates that are well known in the literature, such as the $\bar{I}-\bar{Q}$ pair or the $\bar{S}_3-\bar{Q}$ pair, as well as some new. It also gave some hints on possible improvements of the fit by the inclusion of an additional parameter in some cases. In this section, we will present our findings, which are either improvements on known relations or new relations altogether. The selection given in this section will also serve to demonstrate the algorithm for producing such relations.%

The presentation of the results will be organized in the following way, section \ref{sec:Universal Relations with Quadrupole Moment} is dedicated to proposing a universal relation for the reduced quadrupole moment $\bar{Q}$, section \ref{sec:Universal inverse stellar compactness} proposes a relation for the inverse stellar compactness $\mathcal{K}$, section \ref{sec:Universal Relation for the fraction of kinetic to gravitational energy} for the fraction of rotational to gravitational energy $\mathcal{E}=T/|W|$, and section \ref{sec:Universal Relations for the frequency} presents a relation for the reduced stellar rotational frequency $M\times \tilde{f}$, where $\tilde{f}=f/c$. Then, in section \ref{sec:inertia}, we suggest several EoS-independent relations for the reduced moment of inertia $\bar{I}=I/M^3$. Finally, in section \ref{sec:Universal Relations with Octupole Moment}, we present some EoS-insensitive relations for the reduced spin octupole moment $\bar{S_3}=-S_3/(\chi^3 M^4)$. 

We define a relation between some chosen parameters to be universal, when the relative errors in the validation set are $\lesssim \mathcal{O}(10\%)$.

\subsection{\label{sec:Universal Relations with Quadrupole Moment} 
A universal relation for the reduced quadrupole moment $\bar{Q}$}

The relativistic multipole moments characterize a spacetime's structure and physical properties (i.e., the geometry, the geodesics, and so on and so forth). However, in the case of the spacetime surrounding NSs, these moments depend on the star's internal structure, which is determined by the unknown EoS \cite{Laarakkers:1997hb, Morsink:1998mg, Pappas:2012ns, Pappas:2012qg, pappas2014effectively, Pappas:2015mba, Doneva2018book}.

Therefore, exploring EoS-insensitive relations concerning the multipole moments is quite important. In this section, we present a universal relation for the Geroch-Hansen mass quadrupole moment $M_2$ expressed in geometric units ($[M_2]=[km^2]$). Specifically, the relations are with respect to the dimensionless reduced quadrupole moment $\bar{Q}$ defined as $\bar{Q}\equiv -M_2 M/ J^2=-M_2/(M^3 \chi^2)$, where $\chi$ is the dimensionless angular momentum, and $M$ is the mass of the NS given in geometric units ($[M]=[\textrm{km}]$). 

To identify successful universal relations for the reduced quadrupole moment $\bar{Q}$, we have considered the following parameters, the dimensionless stellar compactness $\mathcal{C}$, and either the dimensionless angular momentum $\chi$ or the dimensionless spin $\sigma=\Omega^2 R_{eq}^3/ G M$, both taken as parameters referring to the star's rotation.

\begin{figure}[!h]
	\includegraphics[width=0.25\textwidth]{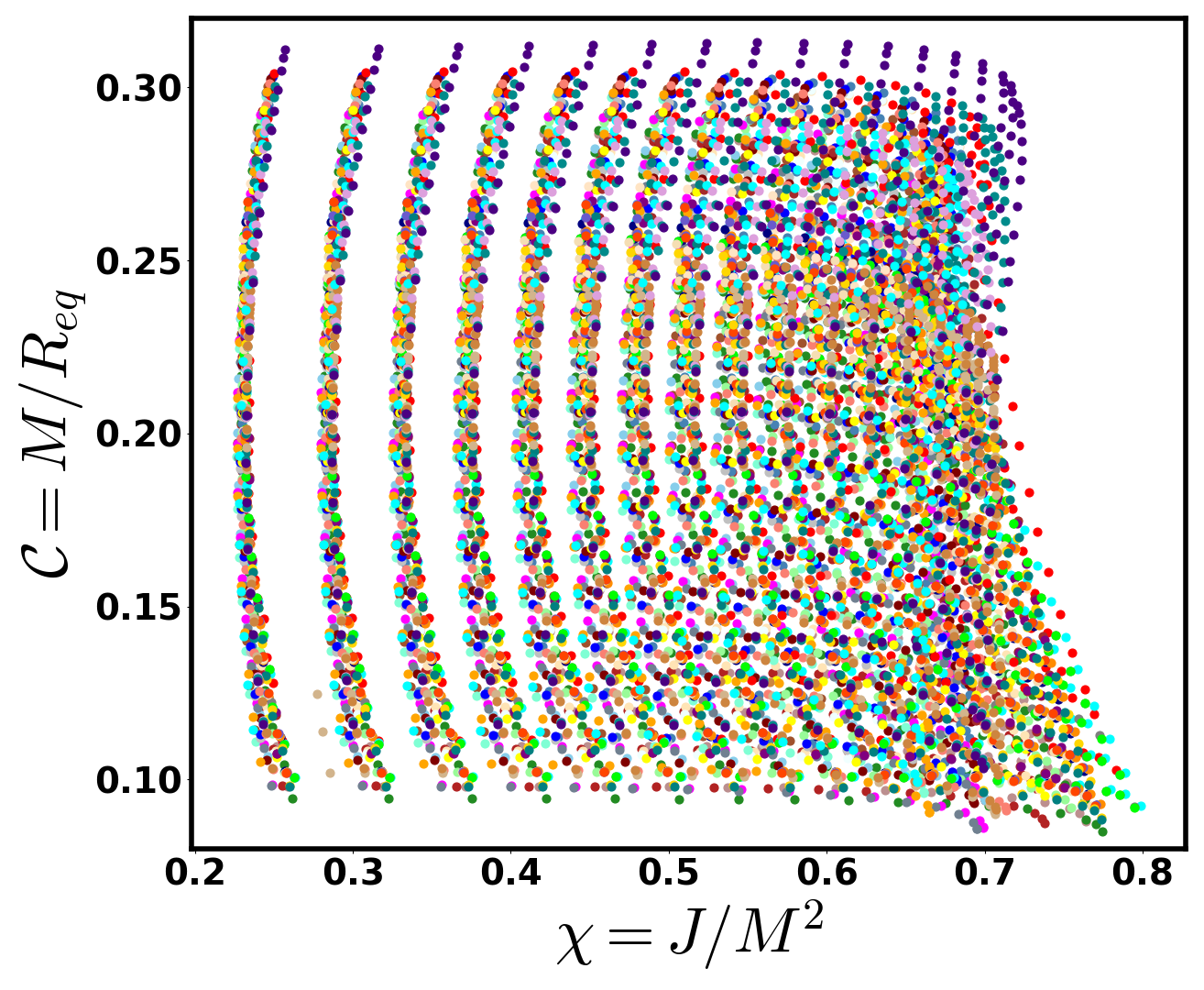}
	\includegraphics[width=0.25\textwidth]{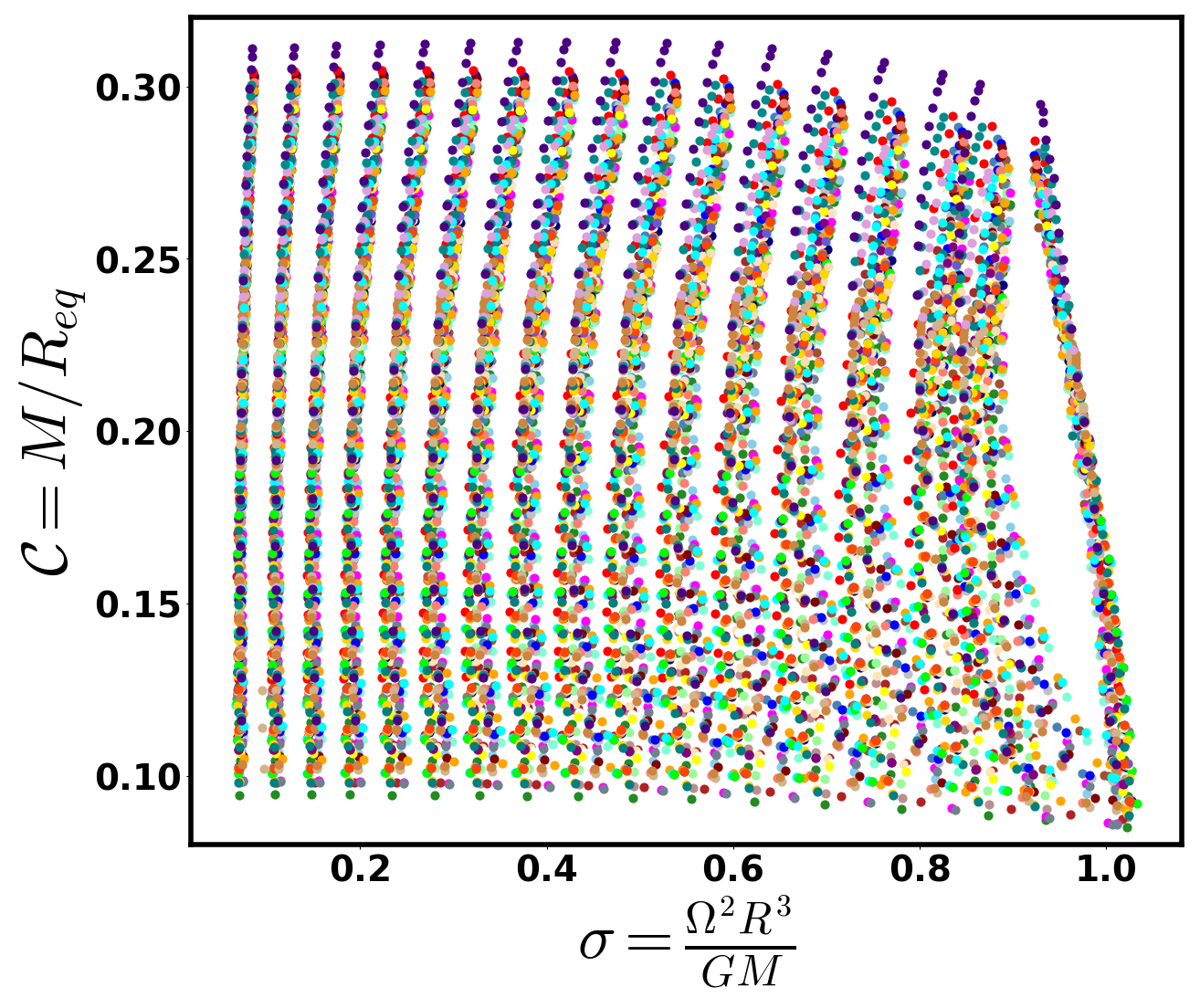}
	\caption{$\mathcal{C}-\chi$ and $\mathcal{C}-\sigma$ representations in the phase space that covers a wide range of rotation rates and stiffness. The comparison between the two plots indicates that $\sigma$ may be a better parameter than $\chi$.}
	\label{fig:c_x_sigma}
\end{figure}

For our sample of rapidly rotating models, these parameters are in the respective ranges, $0.085\leq \mathcal{C} \leq 0.313, \ 0.227\leq \chi \leq 0.799 , \ \textrm{and}\ 0.067\leq \sigma \leq 1.033$. In Fig.\ref{fig:c_x_sigma}, we present, for completeness, the $\mathcal{C}-\chi$ and $\mathcal{C}-\sigma$ distributions in the phase space for each EoS selected from our sample.

It has become clear in the literature that the EoS-independent behavior between observable quantities is sensitive to the choice one makes in selecting parameters \cite{rezzolla_physics_2018}. Therefore, having our choice of parameters being informed by our correlation analysis, and the results shown in Fig. \ref{fig:c_x_sigma}, where we can see that the $\chi$ parameterization does not give a 1-1 correspondence to $\mathcal{C}$, we select to further investigate only a relation between $\bar{Q}$ and the parameters $\mathcal{C}, \textrm{and} \ \sigma$. 

The surface $\bar{Q}=\bar{Q}(\mathcal{C},\sigma)$ that best describes the data has the functional form 
\begin{equation}
\label{eq:q_bar_c_sigma}
\bar{Q}(\mathcal{C},\sigma)=\sum_{n=0}^{5}\sum_{m=0}^{5-n}\hat{a}_{nm} \ \mathcal{C}^n \ \sigma^m.
\end{equation}
Compared to other regression models examined, this is the mathematical model with the best statistical evaluation metric functions at LOOCV. The corresponding results for an indicative list of models are presented in table (\ref{tab:q_c_sigma_tab}).

\begin{table}[!h]
	\small
	\caption{\label{tab:q_c_sigma_tab} Indicative list of LOOCV evaluation metrics 
	for the $\bar{Q}(\mathcal{C},\sigma)=\sum_{n=0}^{\kappa}\sum_{m=0}^{\kappa-n}\hat{a}_{nm} \ \mathcal{C}^n \ \sigma^m$ parameterization, where $\kappa$ is the highest order of the polynomial function.}
	\begin{ruledtabular}
		\begin{tabular}{ccccccc}
			MAE& Max Error& MSE &$d_{\text{max}}$($\%$)& MAPE ($\%$) & Exp Var& $\kappa$ \\
			\hline
			0.2265&2.549& 0.0900&81.853 & 7.321 & 1.0 & 2 \\
			\hline
			  0.092&1.093& 0.0174&36.463& 2.689 & 1.0 & 3 \\
			\hline
			  0.066& 0.771 & 0.0104& 13.011& 1.585 & 1.0& 4 \\
			\hline
			  {\bf 0.063}&{\bf 0.744}& {\bf 0.0097} & {\bf 6.899}&  {\bf 1.491} & {\bf 1.0} & {\bf 5} \\
			\hline
			0.065 &0.786 &0.0099 & 9.538 & 1.575 & 1.0 & 6 \\
			\hline
			 0.062&0.758 & 0.0095&7.173 & 1.455 & 1.0 & 7 
		\end{tabular}
	\end{ruledtabular}
\end{table}
More or less complicated polynomial combinations of $\mathcal{C}, \textrm{and} \ \sigma$ do not improve the fit quality. From the surface-fit evaluation, the polynomial function's (\ref{eq:q_bar_c_sigma}) (best fit) optimizers $\hat{a}_{nm}$ are presented in table (\ref{tab:qbar_c_sigma_optimizers}).
\begin{table}[!h]
	\caption{\label{tab:qbar_c_sigma_optimizers} $\hat{a}_{nm}$ regression optimizers for the $\bar{Q}(\mathcal{C},\sigma)$ parameterization (\ref{eq:q_bar_c_sigma}).}
	\begin{ruledtabular}
		\begin{tabular}{cccc}
			$\hat{a}_{00}\cdot 10^{2}$ & $\hat{a}_{01}\cdot 10^{1}$ & $\hat{a}_{02}\cdot 10^{1}$ & $\hat{a}_{03}\cdot 10^{1}$   \\
			0.6802751 & -4.3993096 & 2.7137482 & -1.0955132  \\
			\hline\hline	
			$\hat{a}_{04}\cdot 10^{-2}$ & $\hat{a}_{05}$ & $\hat{a}_{10}\cdot 10^{3}$ & $\hat{a}_{11}\cdot 10^{2}$  \\
			-9.6154370 & 1.445554 & -1.0699002 & 5.0094955 \\
			\hline\hline
			$\hat{a}_{12}\cdot 10^{2}$ & $\hat{a}_{13}\cdot 10^{1}$ & $\hat{a}_{14}\cdot 10^{1}$ & $\hat{a}_{20}\cdot 10^{3}$ \\
			-2.2554846 & 7.6965321 & -1.3503803 & 7.7681018 \\
			\hline\hline
			$\hat{a}_{21}\cdot 10^{2}$ & $\hat{a}_{22}\cdot 10^{2}$ & $\hat{a}_{23}\cdot 10^{1}$ & $\hat{a}_{30}\cdot 10^{4}$ \\
			5.0094955 & 6.1857273 & -9.3568604 & -3.0342856 \\
			\hline\hline
			$\hat{a}_{31}\cdot 10^{3}$& $\hat{a}_{32}\cdot 10^{2}$ & $\hat{a}_{40}\cdot 10^{4}$ & $\hat{a}_{41}\cdot 10^{3}$ \\
			5.5645259 & -6.0425422 & 6.1273085 & -5.1861927 \\
			\hline\hline
			$\hat{a}_{50}\cdot 10^{4}$& & & \\
			-5.0058667& 
		\end{tabular}
		
	\end{ruledtabular}
\end{table}

\begin{figure}[!h]
	\includegraphics[width=0.24\textwidth]{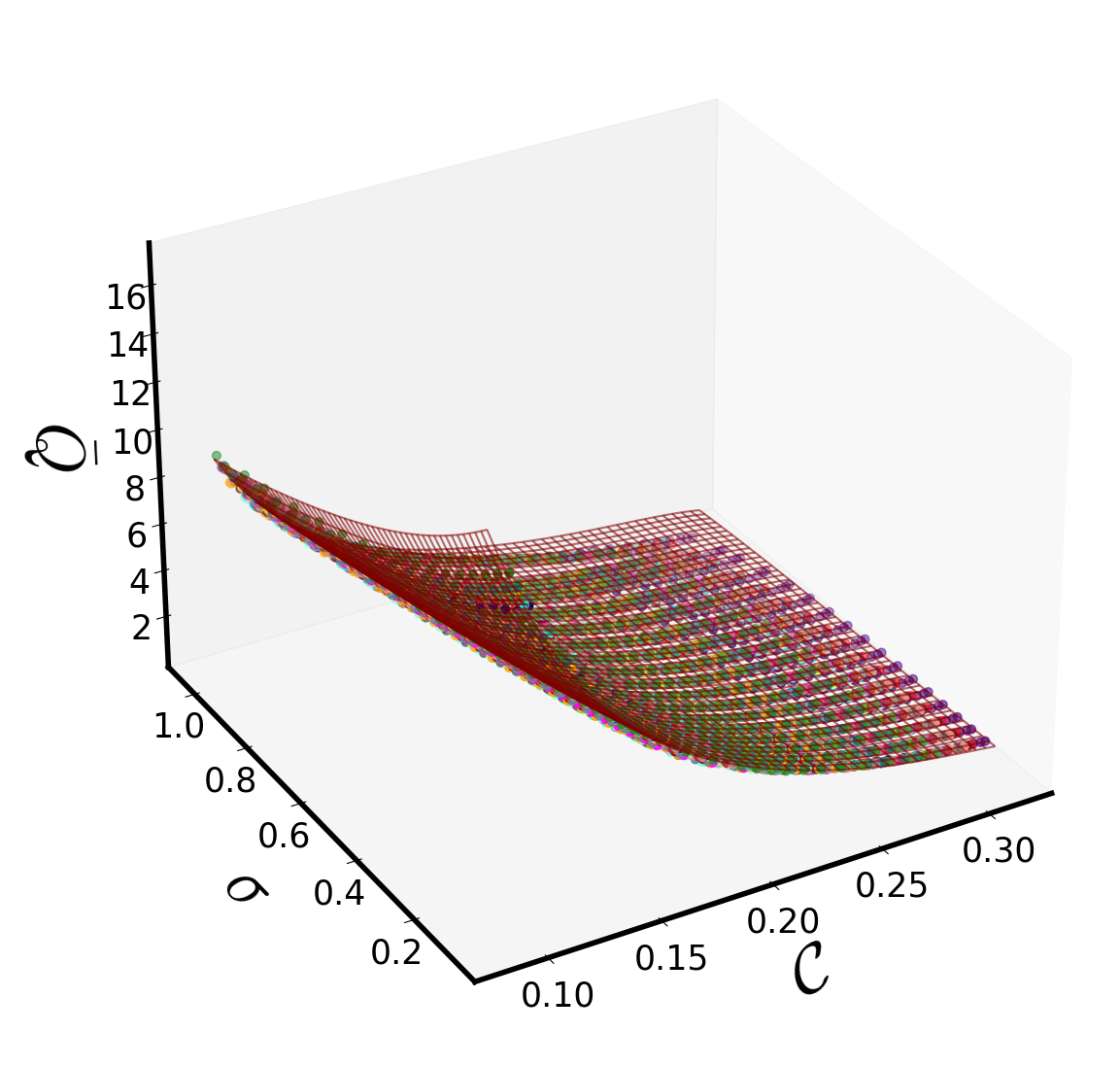}
	\includegraphics[width=0.24\textwidth]{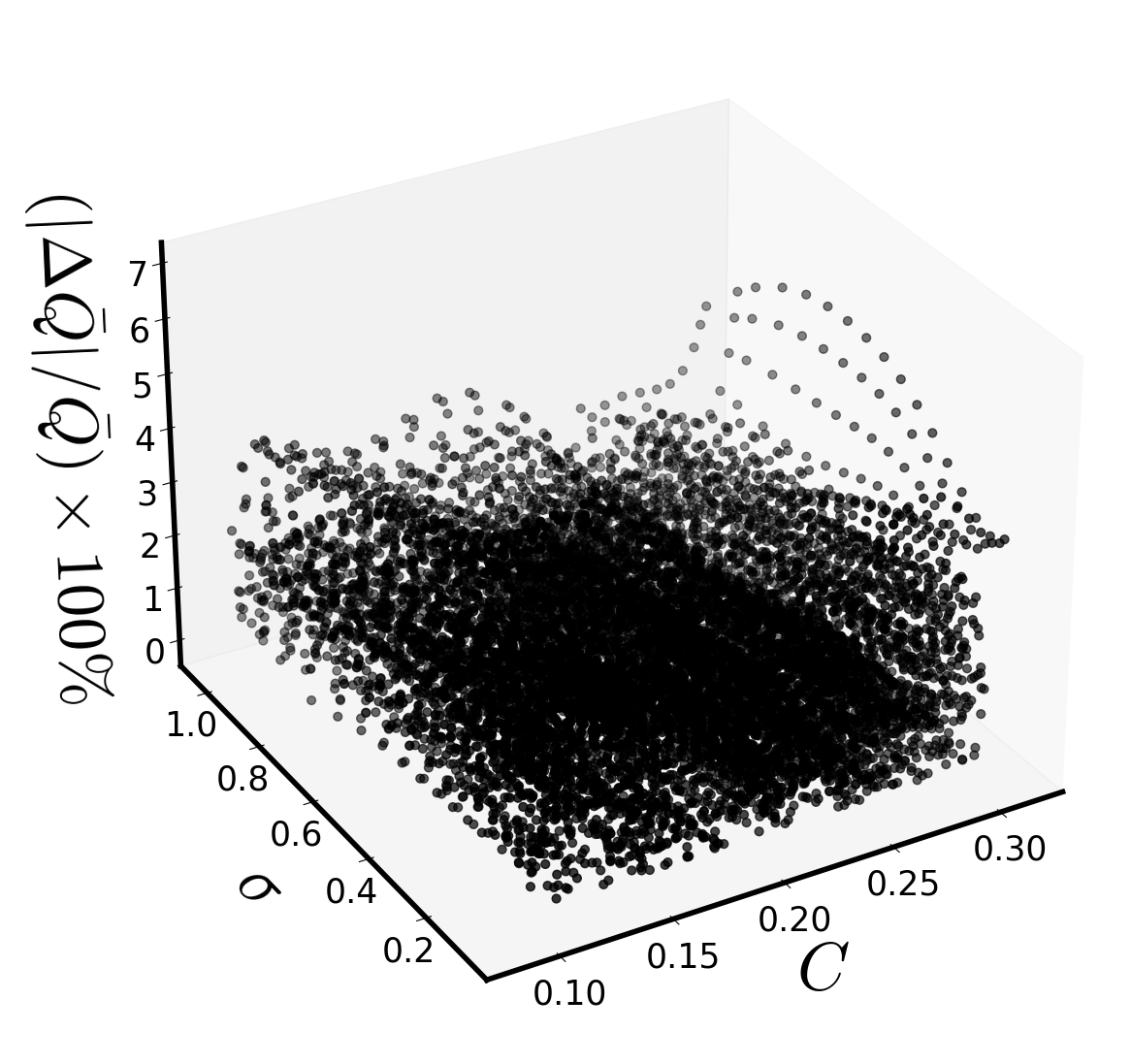}
	\caption{$\bar{Q}$ as a function of the dimensionless parameters $\mathcal{C}, \textrm{and} \ \sigma$ and relative errors. The plotted surface corresponds to the regression polynomial formula (\ref{eq:q_bar_c_sigma}). The relative errors to the fit are given as ($100\%(|\Delta\bar{Q}|/\bar{Q})=100\%|\bar{Q}_{fit}-\bar{Q}|/\bar{Q}$).}
	\label{fig:q_c_sigma_reg_fig}
\end{figure}

The surface fit, and the corresponding relative errors are presented in Fig.\ref{fig:q_c_sigma_reg_fig}.

As we can see in Fig.\ref{fig:q_c_sigma_reg_fig}, with this $\bar{Q}=\bar{Q}(\mathcal{C},\sigma)$-parameterization, the relative errors between the regression formula (\ref{eq:q_bar_c_sigma}) and the observed $\bar{Q}$ are $\lesssim 6.866\%$ for all EoSs and NS models considered. The most significant relative deviations ($\geq 5\%$) are due to the less compact stars with ($0.104 \leq \mathcal{C} \leq 0.162$, $48$ models) and the most compact ones with ($\mathcal{C} \geq 0.264$, $13$ models) regardless of the spin parameter $\sigma$. Consequently, the $\bar{Q}=\bar{Q} (\mathcal{C},\sigma)$ formula  (\ref{eq:q_bar_c_sigma}) corresponds to a well-behaved universal relation for all the rotating NS models considered. 

\begin{figure}[!h]
	\includegraphics[width=0.28\textwidth]{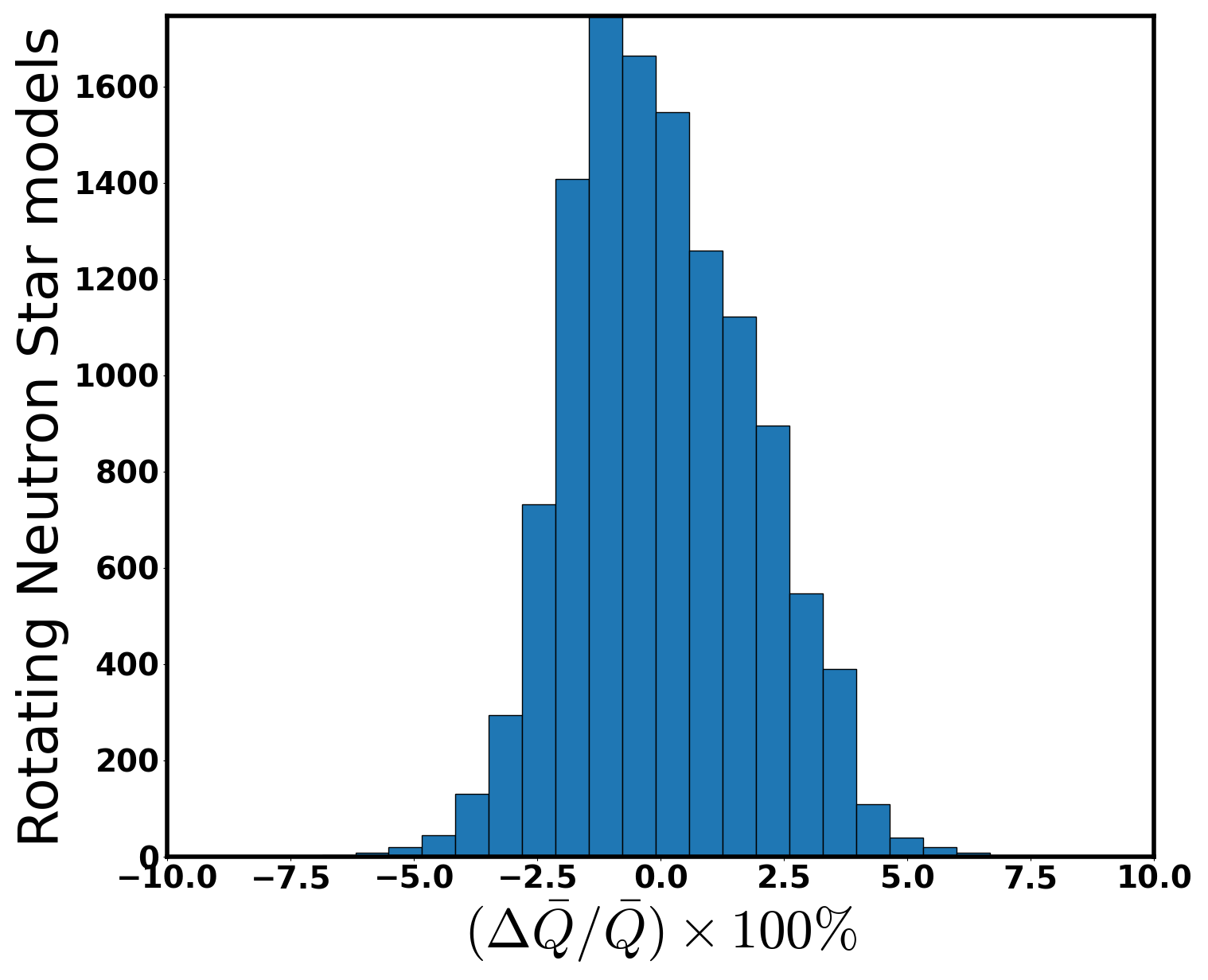}
	\caption{Histogram: Distribution of the number of rotating NS models vs relative errors for the regression formula (\ref{eq:q_bar_c_sigma}).}
	\label{fig:hist_barQ_sigma_c}
\end{figure}

\begin{figure}[!h]
    \includegraphics[width=0.3\textwidth]{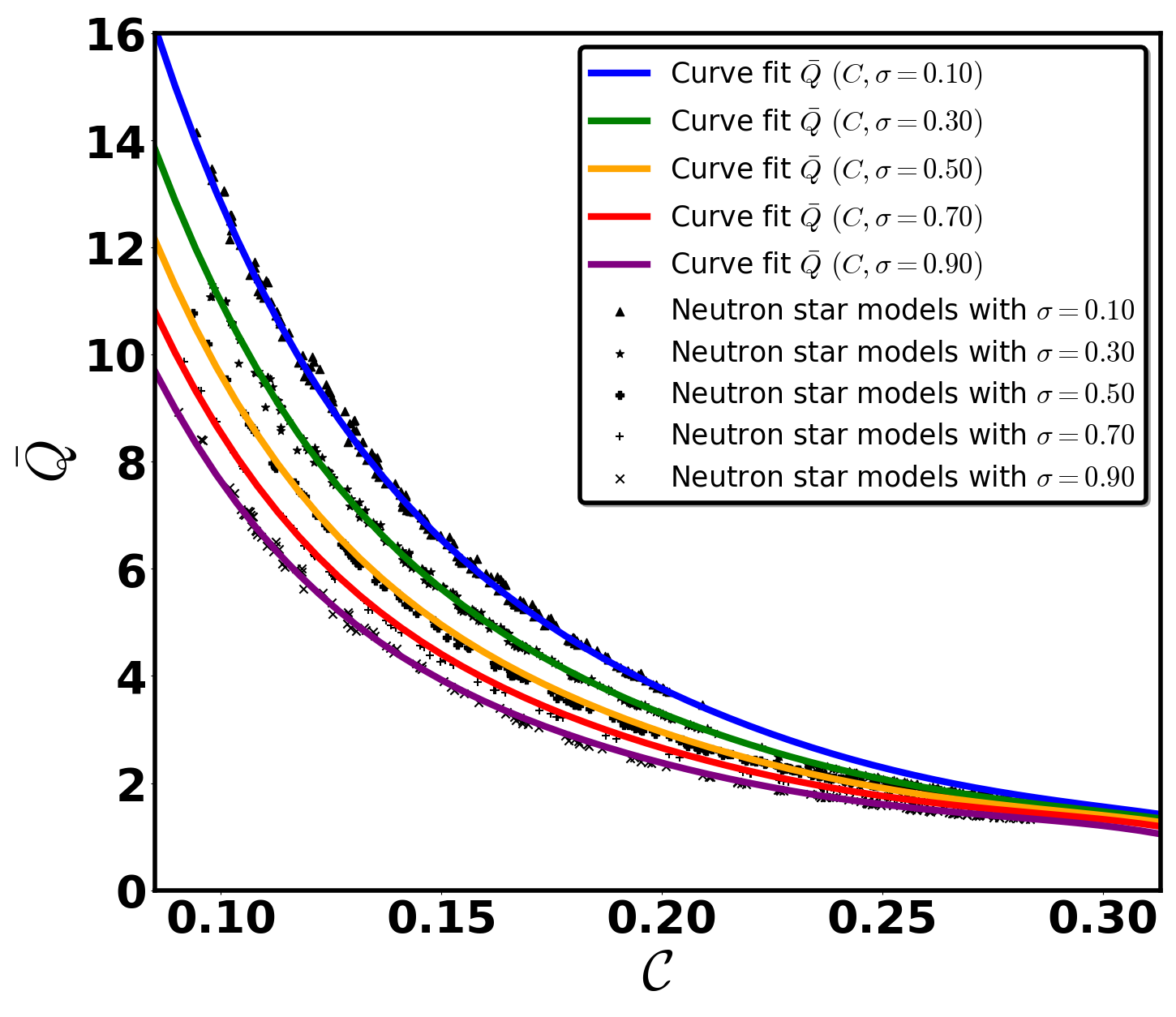}
    \includegraphics[width=0.3\textwidth]{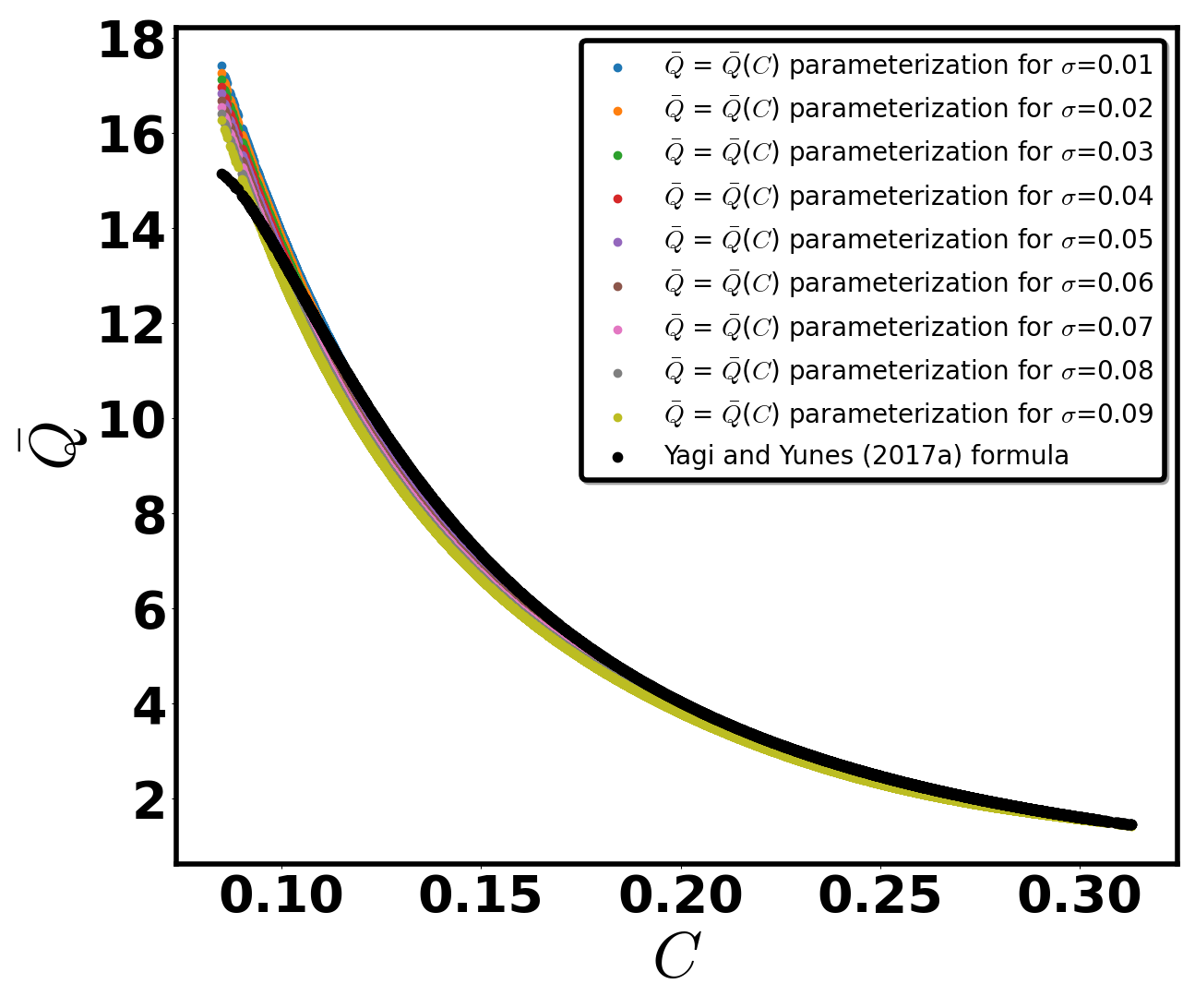}
	\caption{$\bar{Q}=\bar{Q}(\mathcal{C})$ curves for different values of the reduced spin parameter $\sigma$ for rapidly rotating stellar models. Left plot: The black dots correspond to the quantities derived from \textit{RNS} for the total ensemble of EoSs considered, while the colored curves are the prediction according to the formula (\ref{eq:q_bar_c_sigma}).
	Right plot: Prediction of (\ref{eq:q_bar_c_sigma}) in the slowly rotating limit with $\sigma \in[0.01,0.09]$.}
	\label{fig:q_c_sigma_fixed} 
\end{figure}

In the histogram presented in Fig.\ref{fig:hist_barQ_sigma_c}, we show the rotating models' data distribution concerning the relative errors $100\% \times(\Delta\bar{Q}/\bar{Q})$ derived.
It is evident that the formula (\ref{eq:q_bar_c_sigma}) reproduces the vast majority of data values with an error $\lesssim 5\%$. Finally, as an additional demonstration of the success of equation (\ref{eq:q_bar_c_sigma}), we present in Fig.\ref{fig:q_c_sigma_fixed} different curves derived from the relation (\ref{eq:q_bar_c_sigma}), for different values of the rotation spin parameter $\sigma$, plotted against the data from all of our EoSs, for corresponding rotation rates.

As we can see in Fig.\ref{fig:q_c_sigma_fixed} (upper plot), the theoretical prediction coming from (\ref{eq:q_bar_c_sigma}) is in good agreement with the data for different $\sigma$. 
In addition, formula (\ref{eq:q_bar_c_sigma}) also reproduces the universal behavior between the $\bar{Q}$ and $\mathcal{C}$ for the slowly rotating stellar configurations described by the Hartle-Thorne approximation as presented in \cite{yagi2017approximate} (see the bottom plot in Fig.\ref{fig:q_c_sigma_fixed} for $\sigma \in[0.01,0.09]$). However, we have to note that the two formulations fail to converge for the less compact stellar models, which correspond to the higher values of the reduced quadrupole.

\subsection{\label{sec:Universal inverse stellar compactness} 
A universal relation for the inverse stellar compactness $\mathcal{K}=\mathcal{C}^{-1}$}

In the literature, one can find analytic spacetimes that describe the geometry exterior to NSs \cite{Pappas:2015mba, Pappas:2016sye, Maselli:2019qbf} as well as models for the shape of the surface of NSs \cite{Silva:2020oww}, that are all based on the multipole moments of the central object. Therefore, another universal relation that would be useful to investigate is one that links the inverse stellar compactness $\mathcal{K}=R_{eq}/M$ (i.e., the normalized equatorial stellar radius) to the multipole moments, and more specifically to the reduced quadrupole moment $\bar{Q}$ and the dimensionless angular momentum $\chi$. 

\begin{figure}[!h]
	\includegraphics[width=0.26\textwidth]{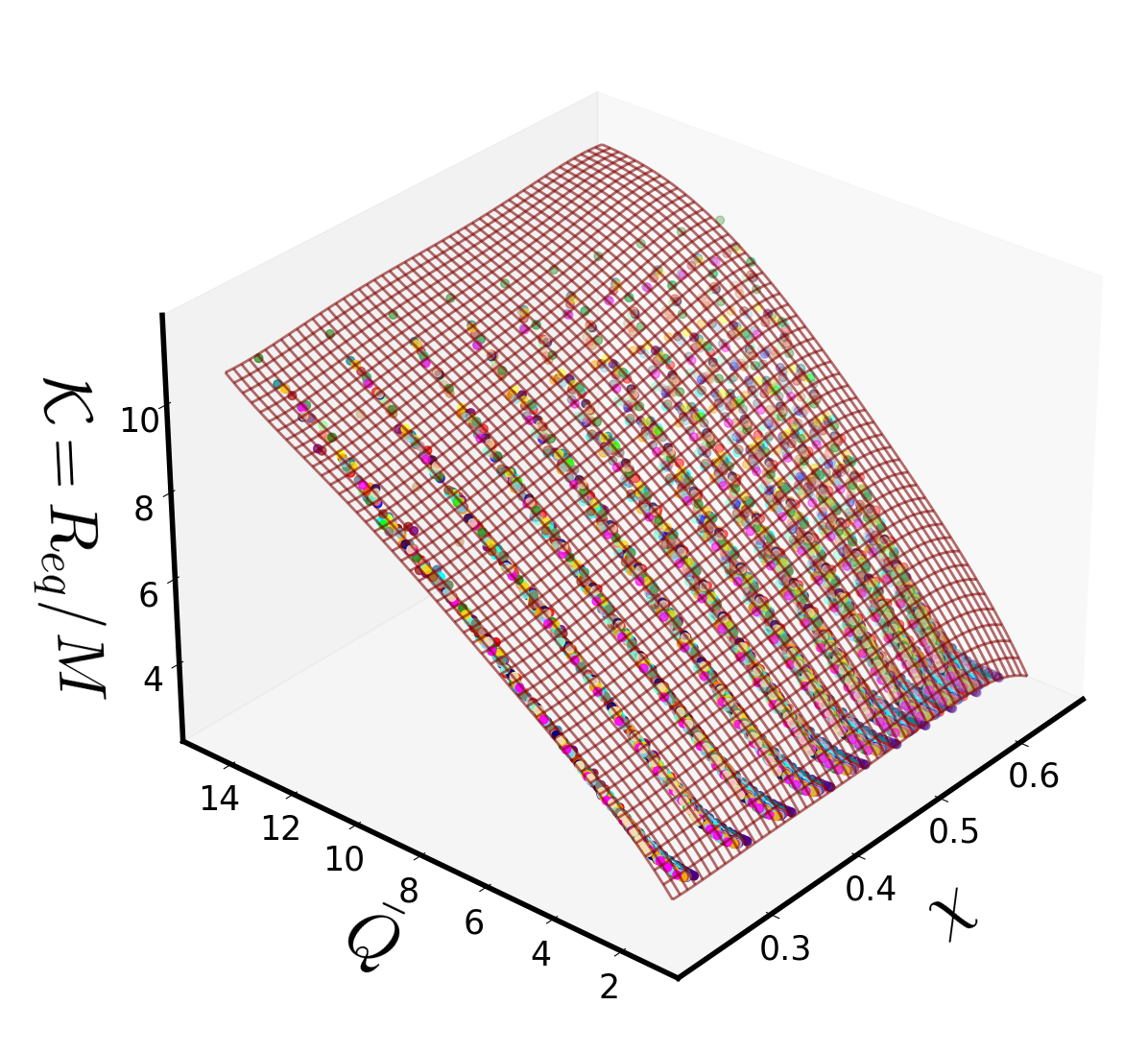}
	\includegraphics[width=0.26\textwidth]{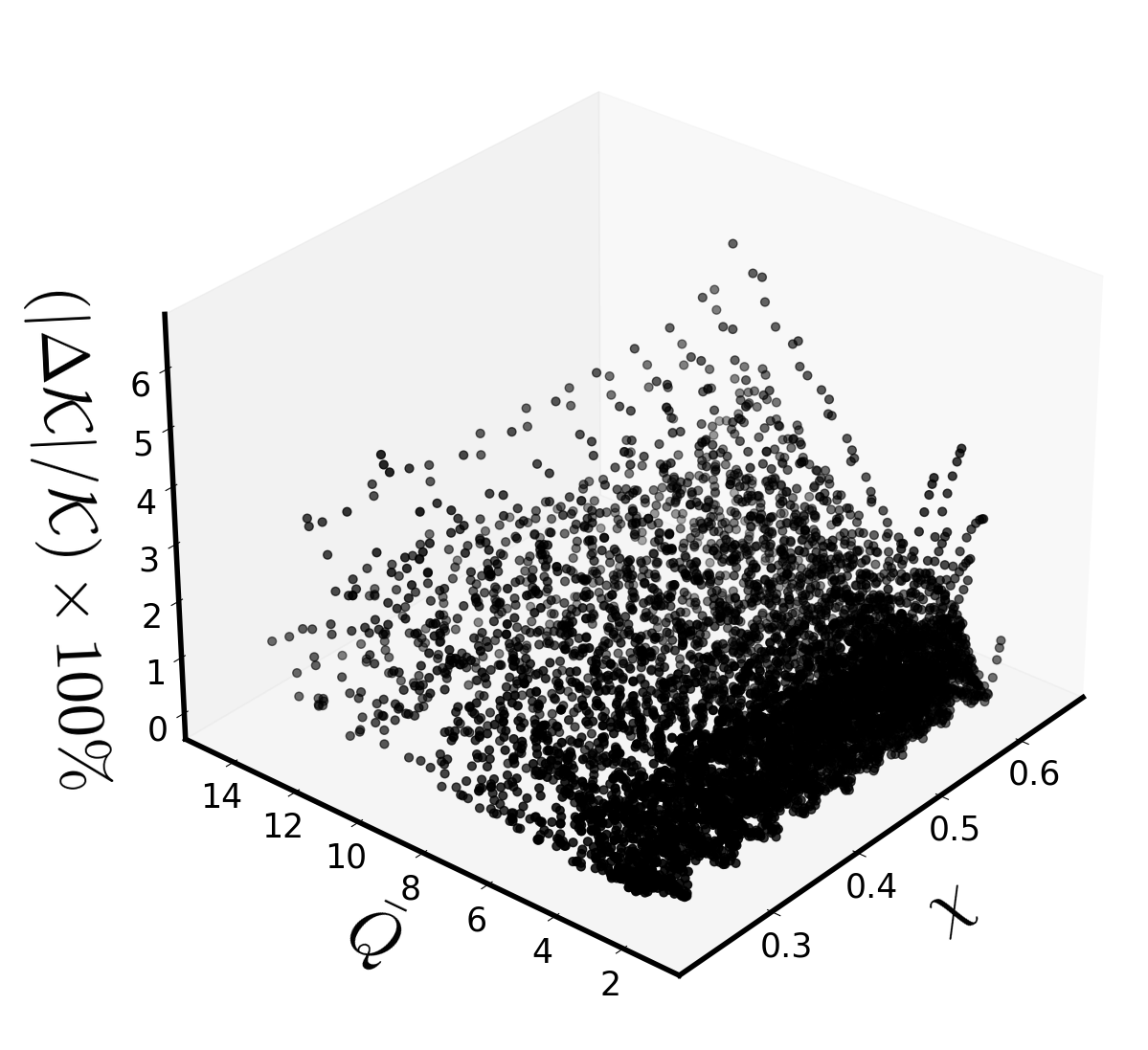}
	\caption{$\mathcal{K}=R_{eq}/M$ as a function of the dimensionless parameters $\chi,\ \bar{Q} $ and relative error distribution. The analysis considers rapidly rotating NS models with frequencies in the range of $0.2278 \ kHz \lesssim f \lesssim 1.7528 \ kHz$. The surface corresponds to the formula (\ref{eq:K_chi_barQ}). The relative errors given as ($100\%(|\Delta\mathcal{K}|/\mathcal{K})=100\%|\mathcal{K}_{fit}-\mathcal{K}|/\mathcal{K}$) are computed between the fit and the data.} 	
	\label{fig:K_chi_q_fig}
\end{figure}

Due to the large deviations introduced by rotating models near the mass shedding limit $f\sim 2 \ kHz$, we limit our analysis to NS configurations that rotate with frequencies in the range of $0.2278 \ kHz \lesssim f \lesssim 1.7528 \ kHz$ and have stellar parameters that range from $0.094 \lesssim \mathcal{C} \lesssim 0.313$, and $0.23 \lesssim \chi \lesssim 0.65$. This ensemble includes 7046 stellar models out of the total 11983 that were used in the previous subsection. The surface-formula $\mathcal{K}(\chi,\bar{Q})$ that best describes the data has the functional form 
\begin{equation}
\label{eq:K_chi_barQ}
\mathcal{K}(\chi,\bar{Q})=\frac{R_{eq}}{M}=\sum_{n=0}^{5}\sum_{m=0}^{5-n}\hat{b}_{nm} \ \chi^n \ \bar{Q}^{m},
\end{equation}
and gives the NS's equatorial radius in terms of the mass, dimensionless spin $\chi$, and $\bar{Q}$. Again, this is the regression model with the optimal statistical evaluation metric functions at LOOCV. 
Other regression functions of $\chi, \textrm{and} \ \bar{Q}$ do not improve the fit quality. The corresponding results for an indicative list of models are presented in table (\ref{tab:K_q_chi_tab}).
\begin{table}[!h]
	\small
	\caption{\label{tab:K_q_chi_tab} Indicative list of LOOCV evaluation metrics for the $\mathcal{K}(\chi,\bar{Q}) = \sum_{n=0}^{\kappa}\sum_{m=0}^{\kappa-n}\hat{b}_{nm} \ \chi^n \ \bar{Q}^{m}$ parameterization.}
	\begin{ruledtabular}
		\begin{tabular}{ccccccc}
			MAE&Max Error& MSE &$d_{\text{max}}$($\%$) & MAPE ($\%$) & Exp Var& $\kappa$ \\
			\hline
			0.088& 0.708 &0.015 & 11.919 & 1.5829 & 1.0 & 2\\
			\hline
			0.077& 0.707&0.012 & 7.190 & 1.329 & 1.0 & 3\\
			\hline
			0.072&0.682&0.011 & 6.930 & 1.221 & 1.0 & 4\\
			\hline
			{\bf 0.070}&{\bf 0.638}&{\bf 0.010} & {\bf 6.480} & {\bf 1.177} & {\bf 1.0} & {\bf 5}\\
			\hline
			0.087&0.993& 0.015& 9.374 & 1.487 & 1.0 & 6\\
			\hline
			 0.070&0.618 &0.010 & 6.688 & 1.179 & 1.0 & 7\\
		\end{tabular}
	\end{ruledtabular}
\end{table}
From the surface-fit evaluation, the best-fit optimizers $\hat{b}_{nm}$ are presented in the table (\ref{tab:k_chi_qbar_optimizers}).
\begin{table}[!h]
	\caption{\label{tab:k_chi_qbar_optimizers} $\hat{b}_{nm}$ regression optimizers for the $\mathcal{K}(\chi,\bar{Q})$ parameterization (\ref{eq:K_chi_barQ}).}
	\begin{ruledtabular}
		\begin{tabular}{cccc}
			$\hat{b}_{00}\cdot10^{1}$ & $\hat{b}_{01}$ & $\hat{b}_{02}\cdot10^{-1}$ & $\hat{b}_{03}\cdot10^{-2}$   \\
			1.0096592 & 1.8238166 & -2.0956902 & 2.7949308   \\
			\hline\hline	
			$\hat{b}_{04}\cdot10^{-3}$ & $\hat{b}_{05}\cdot10^{-5}$ & $\hat{b}_{10}\cdot10^{2}$ & $\hat{b}_{11}$  \\
			-1.9398241 & 4.9502268 & -1.2168276 & -4.9461032 \\
			\hline\hline	
			$\hat{b}_{12}\cdot10^{-1}$ & $\hat{b}_{13}\cdot10^{-2}$ & $\hat{b}_{14}\cdot10^{-4}$ & $\hat{b}_{20}$ \\
			-1.1701846 & 1.9652010, & -4.3842554 & -4.9461032 \\
			\hline\hline
			$\hat{b}_{21}\cdot10^{1}$ & $\hat{b}_{22}\cdot10^{-1}$ & $\hat{b}_{23}\cdot10^{-2}$ & $\hat{b}_{30}\cdot10^{3}$ \\
			2.1398537 & -1.5834957 & -1.2228542 & -1.8417567 \\
			\hline\hline
			$\hat{b}_{31}\cdot10^{1}$& $\hat{b}_{32}\cdot10^{-1}$ & $\hat{b}_{40}\cdot10^{3}$ & $\hat{b}_{41}\cdot10^{1}$ \\
			-3.2814514 & 2.6746975 & 2.4285095 & 1.8288763 \\
			\hline\hline
			$\hat{b}_{50}\cdot10^{3}$ 
			\\
			-1.2443348
		\end{tabular}
		
	\end{ruledtabular}
\end{table}
The surface fit (\ref{eq:K_chi_barQ}) that best reproduces the data and the corresponding relative errors are presented in Fig.\ref{fig:K_chi_q_fig}.

Using this $\mathcal{K}=\mathcal{K}(\chi,\bar{Q})$-parameterization, the relative errors between the fit (\ref{eq:K_chi_barQ}) and the observed $\mathcal{K}$ are $\lesssim 6.443\%$ for all EoSs considered (universality). The cases with relative deviations $\geq 5\%$ correspond to $18$ of the less compact stellar models with $\mathcal{C}\in [0.096,0.124]$, $\bar{Q}\in [6.459,10.564]$, and $\chi \in [0.248,0.602]$. 

\begin{figure}[!h]
	\includegraphics[width=0.28\textwidth]{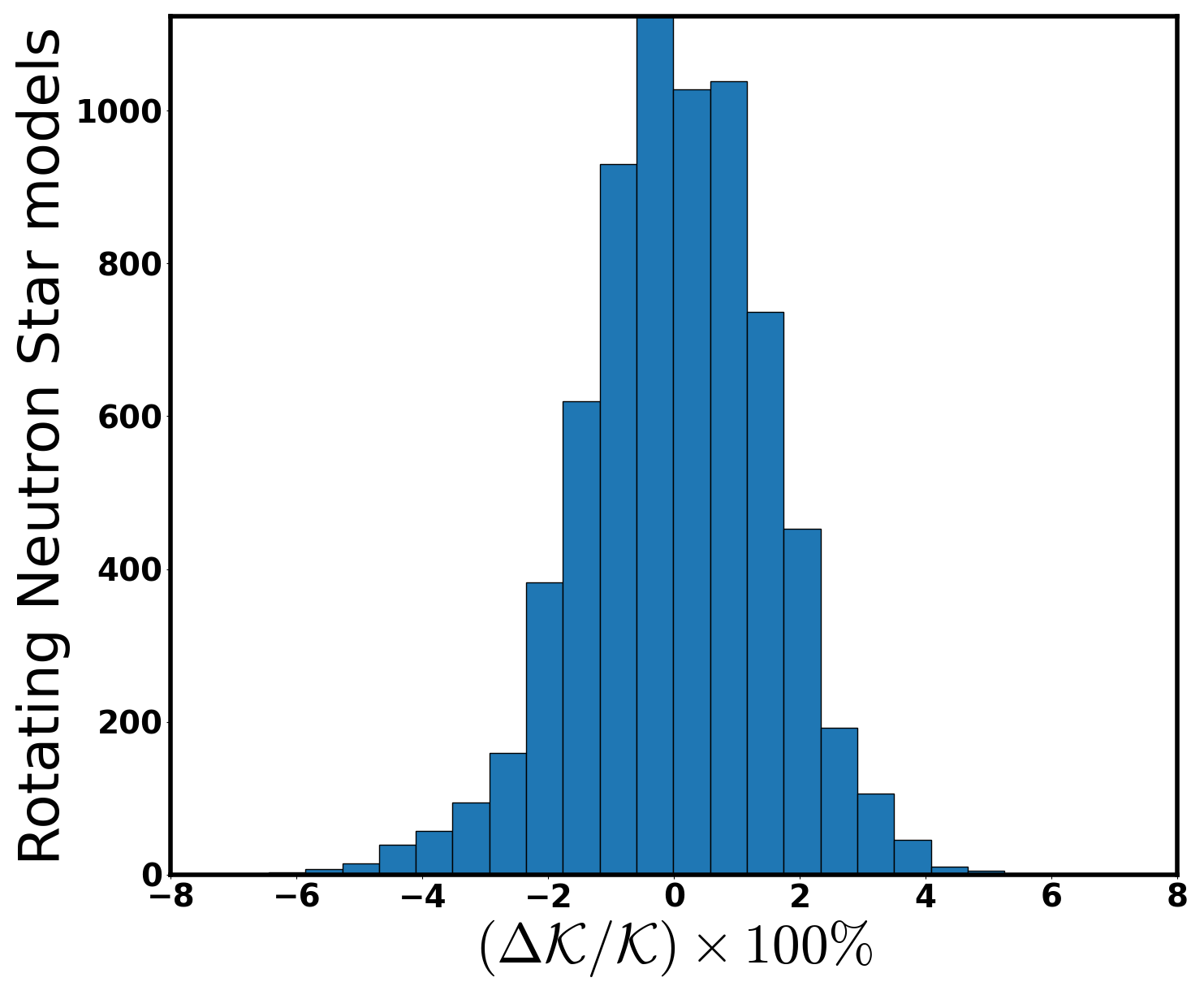}
	\caption{Histogram: The distribution of the 7046 models used vs the relative errors for the formula (\ref{eq:K_chi_barQ}).}
	\label{fig:hist_K_chi_varQ_fig}
\end{figure}

Therefore, the $\mathcal{K}=\mathcal{K}(\chi,\bar{Q})$ %
formula is, to a satisfactory degree, a universal relation for all the rotating stellar models considered. This is further demonstrated in Fig.\ref{fig:hist_K_chi_varQ_fig} where it is evident that (\ref{eq:K_chi_barQ}) reproduces most data values with an error $\lesssim 5\%$. To conclude, the formula (\ref{eq:K_chi_barQ}) represents a useful EoS-insensitive description of the star's equatorial radius $R_{eq}$ in terms of the mass $M$, reduced quadrupole $\bar{Q}$, and spin $\chi$.

\subsection{\label{sec:Universal Relation for the fraction of kinetic to gravitational energy} 
A universal relation for the fraction of kinetic to gravitational energy $\mathcal{E} = T/|W|$}

One could also look for universal relations that can give us properties of NSs that are not directly observable, but are nevertheless important for their structure, with respect to stability considerations, for example. One such interesting quantity would be the dimensionless fraction of kinetic to gravitational energy $T/|W|$, which is related to the change of the equilibrium shape of a rotating fluid configuration. 
Here we explore a universal relations that relates $\bar{Q}, \ \textrm{and} \ \chi$ to $T/|W|$. We note that $T/|W|$ is essentially a different representation of the star's rotation. A star's rotational kinetic energy is expressed as $T=\frac{1}{2}\int \Omega dJ$, while the gravitational energy is given as $W=M_pc^2+T-Mc^2$ \cite{camenzind_compact_2007} ($M_p$ is the star's proper mass, while $M$ is the gravitational mass), and for the NS ensemble of 7046 models that we have used is in the range $0.012\leq T/|W| \leq 0.1020$.

For the universal relation, we have chosen to express $T/|W|$ as a function of the parameters $\chi, \ln(\bar{Q})$. The fitting formula $T/|W|=\mathcal{E}(\chi,\ln(\bar{Q}))$ that optimally reproduces the data has the functional form
\begin{equation}
\label{eq:TW_chi_lnQbar}
\frac{T}{|W|}=\mathcal{E}\left(\chi,\ln(\bar{Q})\right)=\sum_{n=0}^{5}\sum_{m=0}^{5-n}\hat{c}_{nm} \  \chi^n \ (\ln \bar{Q})^{m}.
\end{equation}
This is the simplest (less complicated) regression function of the functions that we tested that gave a satisfactory fit, i.e., there were higher-order ($\kappa > 5$) polynomial functions that gave better statistical evaluation metric functions at LOOCV from those presented in the table (\ref{tab:R_M_chi_lnq_tab}) for $\kappa = 5$.
\begin{table}[!h]
	\small
	\caption{\label{tab:R_M_chi_lnq_tab} Indicative list of LOOCV evaluation metrics for the $\mathcal{E}(\chi,\ln(\bar{Q})) = \sum_{n=0}^{\kappa}\sum_{m=0}^{\kappa-n}\hat{c}_{nm} \  \chi^n \ (\ln \bar{Q})^{m}$ parameterization.}
	\begin{ruledtabular}
		\begin{tabular}{ccccccc}
			MAE& Max Error & MSE & $d_{\text{max}}$($\%$) & MAPE ($\%$) & Exp Var & $\kappa$ \\
            \multicolumn{0}{c}{$(\cdot10^{-3})$}&\multicolumn{1}{c}{$(\cdot10^{-2})$}&\multicolumn{1}{c}{$(\cdot10^{-5})$}\\
			\hline
			0.713&0.406& 0.092&23.377 & 2.089 & 1.0 & 2\\
			\hline
			0.232&0.182& 0.011&11.888 & 0.608 & 1.0 & 3\\
			\hline
			0.151&0.158& 0.006&5.433 & 0.325 & 1.0 & 4\\
			\hline
			{\bf 0.134}&{\bf 0.148}&{\bf 0.005} &{\bf 3.020} & {\bf 0.269}& \bf {1.0} & {\bf 5}\\
			\hline
			0.131 &0.152&0.005 & 2.362 & 0.259  & 1.0 & 6\\
			\hline
			0.129&0.154&0.005 & 2.092 & 0.256 & 1.0 & 7
		\end{tabular}
	\end{ruledtabular}
\end{table}
Nevertheless, selecting functions that were too complicated was not worth the slight improvement of the fit quality. Therefore, from the surface-fit evaluation, the model's optimizers $\hat{c}_{nm}$ are presented in the table (\ref{tab:tw_chi_ln_qbar_optimizers}).
\begin{table}[!h]
	\caption{\label{tab:tw_chi_ln_qbar_optimizers} $\hat{c}_{nm}$ regression optimizers for the $\mathcal{E}(\chi,\ln\bar{Q})$ parameterization (\ref{eq:TW_chi_lnQbar}).}
	\begin{ruledtabular}
		\begin{tabular}{cccc}
			$\hat{c}_{00}\cdot10^{-3}$ & $\hat{c}_{01}\cdot10^{-3}$ & $\hat{c}_{02}\cdot10^{-4}$ & $\hat{c}_{03}\cdot10^{-3}$   \\
			-6.9786155 & -9.5138577 & -5.6766317 & 3.8583851    \\
			\hline\hline	
			$\hat{c}_{04}\cdot10^{-3}$ & $\hat{c}_{05}\cdot10^{-4}$ & $\hat{c}_{10}\cdot10^{-1}$ & $\hat{c}_{11}\cdot10^{-1}$  \\
			-2.7620777& 5.1787682 & 1.2959001 & 1.1731501 \\
			\hline\hline	
			$\hat{c}_{12}\cdot10^{-2}$ & $\hat{c}_{13}\cdot10^{-2}$ & $\hat{c}_{14}\cdot10^{-3}$ & $\hat{c}_{20}\cdot10^{-1}$ \\
			-4.9604453  & 2.8783020 & -4.3501526 & -7.0668356 \\
			\hline\hline
			$\hat{c}_{21}\cdot10^{-1}$ & $\hat{c}_{22}\cdot10^{-1}$ & $\hat{c}_{23}\cdot10^{-3}$ & $\hat{c}_{30}$ \\
			-1.4800551 & -1.1282566 & 5.3987682 & 2.8942703 \\
			\hline\hline
			$\hat{c}_{31}\cdot10^{-1}$& $\hat{c}_{32}\cdot10^{-2}$ & $\hat{c}_{40}$ & $\hat{c}_{41}\cdot10^{-1}$ \\
			4.2370939 & 8.3353042 & -4.1338397 & -4.4797049  \\
			\hline\hline
			$\hat{a}_{50}$ 
			\\
			2.3560209 
		\end{tabular}
		
	\end{ruledtabular}
\end{table}

\begin{figure}[!ht]
	\includegraphics[width=0.26\textwidth]{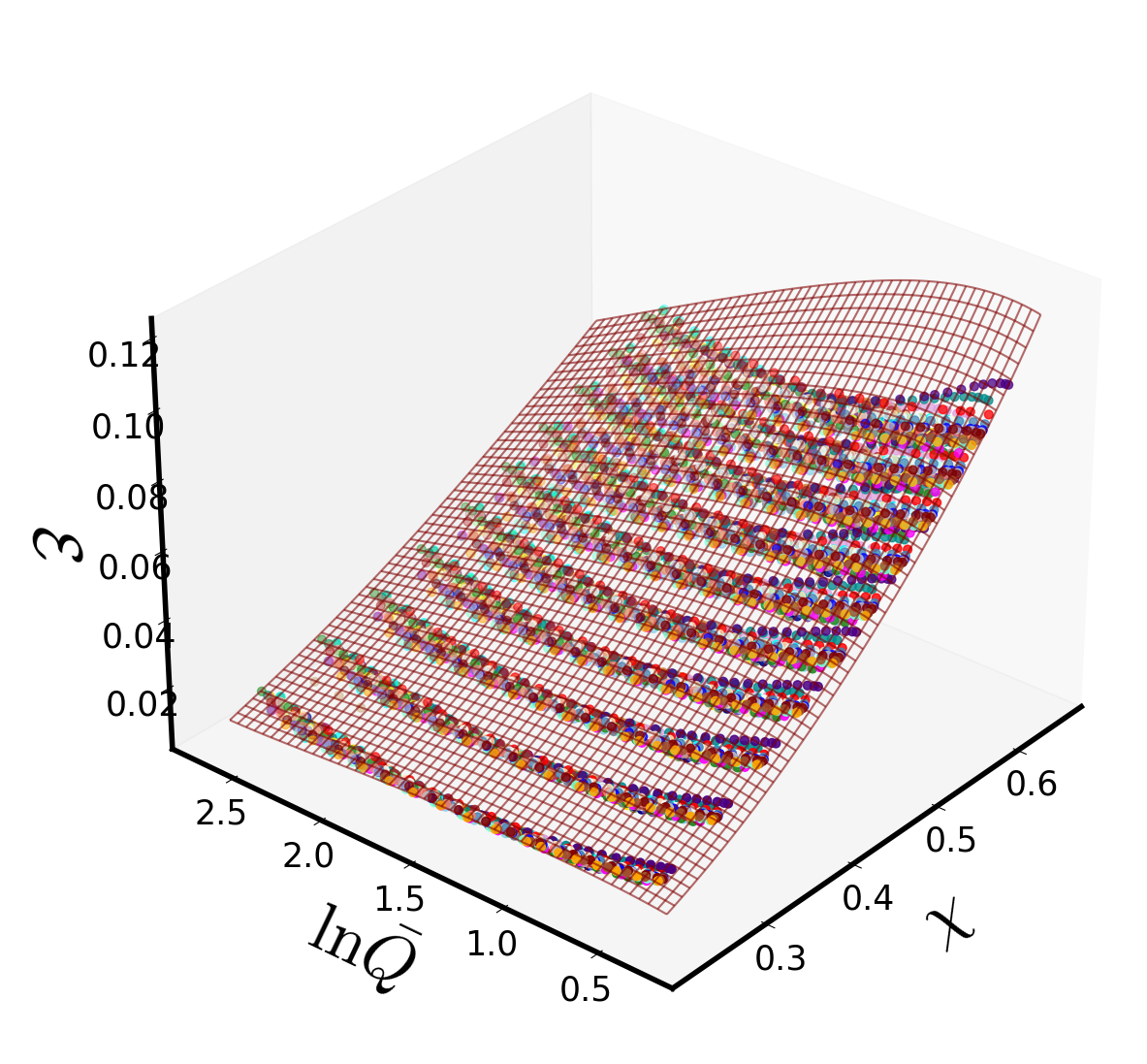}
	\includegraphics[width=0.26\textwidth]{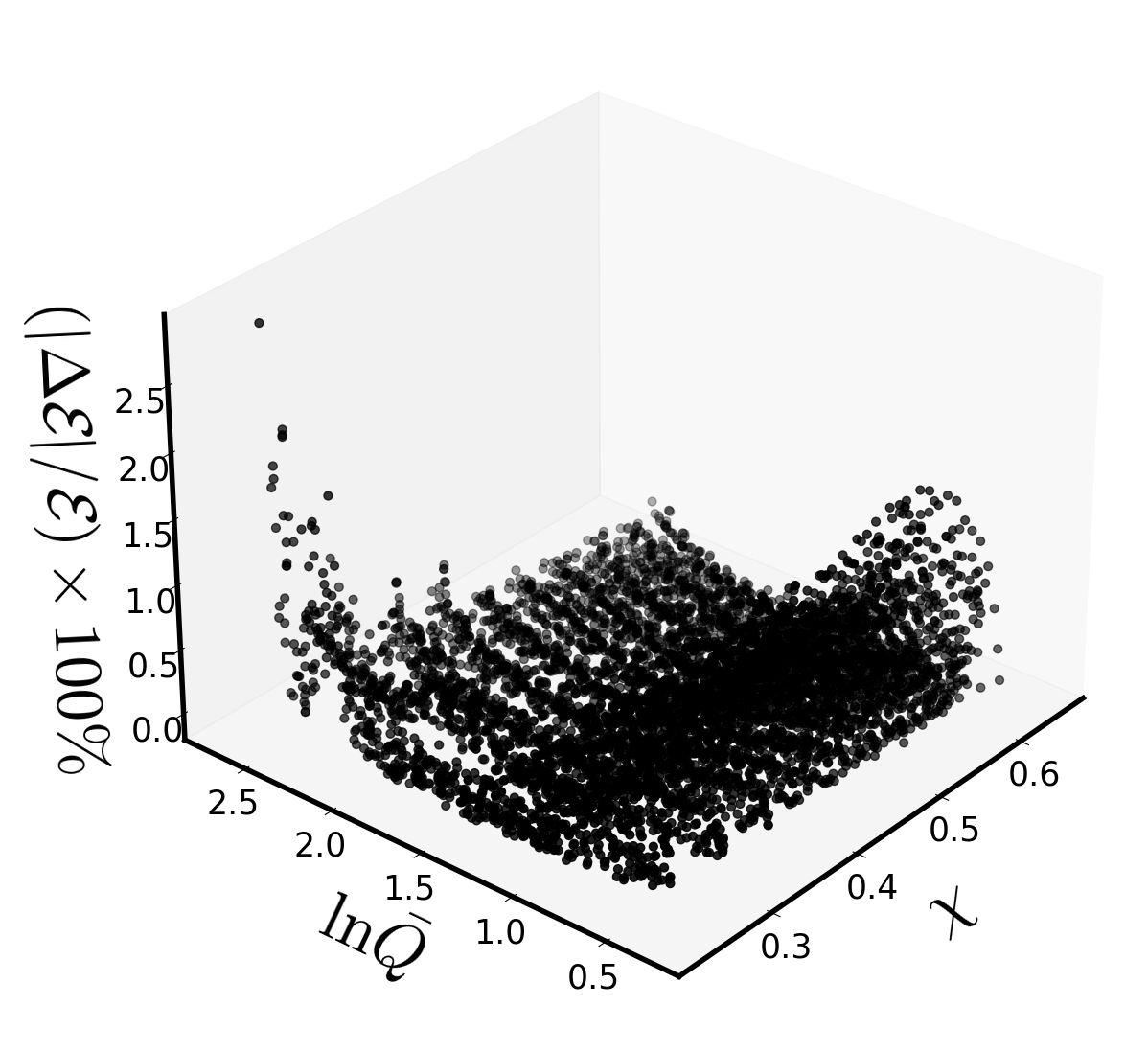}
	%
	\caption{$\mathcal{E}=T/|W|$ as a function of the parameters $\chi,\ \ln(\bar{Q})$ and relative error distribution. The analysis considers rapidly rotating NS models with frequencies in the range of $0.2278 \ kHz \lesssim f \lesssim 1.7528 \ kHz$. The surface corresponds to the formula (\ref{eq:TW_chi_lnQbar}).}
	\label{fig:T_W_chi_lnq_fig}
\end{figure}

\begin{figure}[!ht]
	\includegraphics[width=0.28\textwidth]{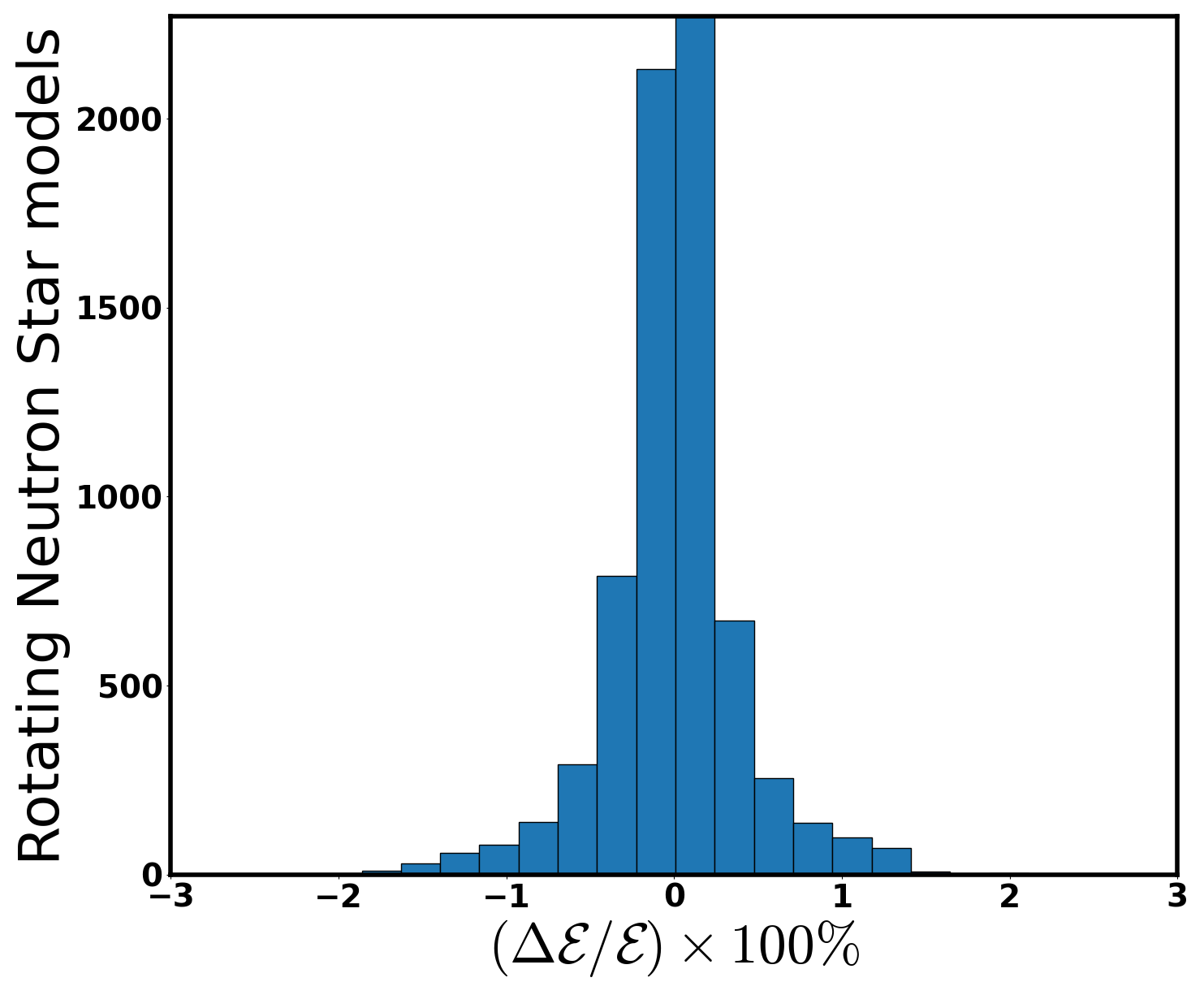}
	\caption{Histogram: Distribution of the 7046 models used vs relative errors from the regression formula (\ref{eq:TW_chi_lnQbar}).}
	\label{fig:hist_T_W_chi_lnbarQ.png}
\end{figure}

The surface evaluation fit (\ref{eq:TW_chi_lnQbar}) and the corresponding relative errors are presented in Fig.\ref{fig:T_W_chi_lnq_fig}.
The $\mathcal{E}(\chi,\ln(\bar{Q}))$-parameterization gives relative deviations between the fit (\ref{eq:TW_chi_lnQbar}) and the data that are $\lesssim 2.815\%$, with only $35$ models out of the $7046$ having $\gtrsim 1.5\%$.  
In Fig.\ref{fig:hist_T_W_chi_lnbarQ.png}, we present the distribution histogram concerning the relative errors $100\% \times(\Delta {\mathcal{E}}/{\mathcal{E}})$ derived.
From Fig.\ref{fig:hist_T_W_chi_lnbarQ.png}, it is evident that the regression formula (\ref{eq:TW_chi_lnQbar}) corresponds to a very good universal relation which gives accurate results for all the stellar models considered, reproducing most data values with an error $\lesssim 1.5\%$.

\subsection{\label{sec:Universal Relations for the frequency} 
A universal relation for the normalised rotational frequency $M\times \tilde{f}$}

An observationally important quantity for a NS, is its rotational frequency $\tilde{f}$. This is mainly due to the fact that in many cases it is the simplest quantity to measure. Therefore it would be useful to have a relation that connects the star's rotation frequency with other parameters. In order to do that, we use the quantity $\mathcal{D} = M \times \tilde{f}/\chi$ instead of $\tilde{f}$ itself, and express it as a function of the dimensionless spin $\chi$ and the reduced quadrupole deformation $\bar{Q}$. 

\begin{figure}[!ht]
	\includegraphics[width=0.26\textwidth]{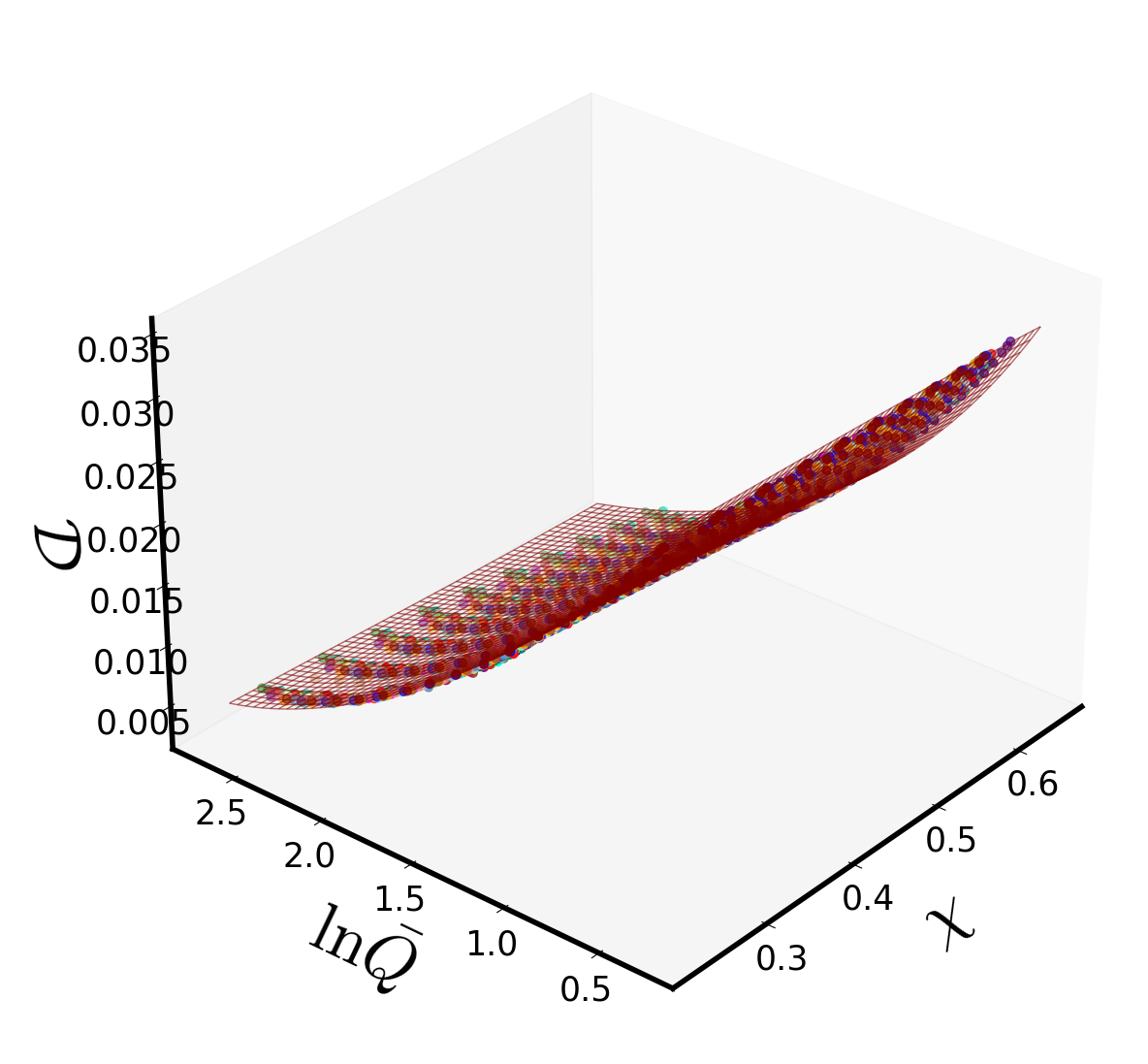}
	\includegraphics[width=0.26\textwidth]{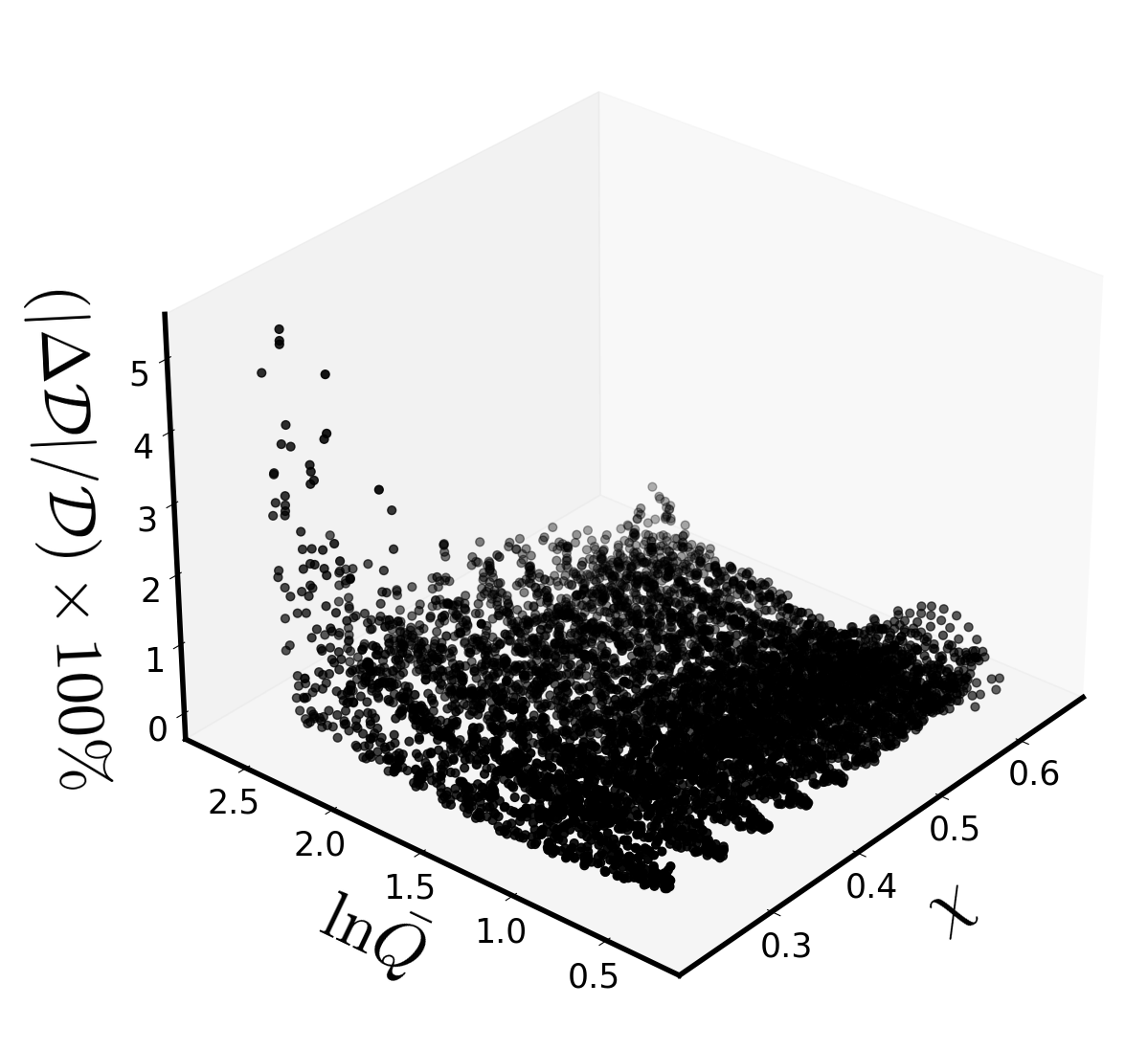}
	\caption{$\mathcal{D} = M\times f/\chi$ as a function of the parameters $\chi,\ \ln(\bar{Q})$ and corresponding relative error distribution. The models used have frequencies in the range of $0.2278 \ kHz \lesssim f \lesssim 1.7528 \ kHz$. The surface correspond to the formula (\ref{eq:Mf/x_chi_lnQbar}), while the relative errors are given as ($100\%(|\Delta \mathcal{D}|/\mathcal{D})=100\%|\mathcal{D}_{fit}-\mathcal{D}|/\mathcal{D}$).} %
	\label{fig:Mf_chi_chi_lnq_fig}
\end{figure}

\begin{figure}[!ht]
	\includegraphics[width=0.28\textwidth]{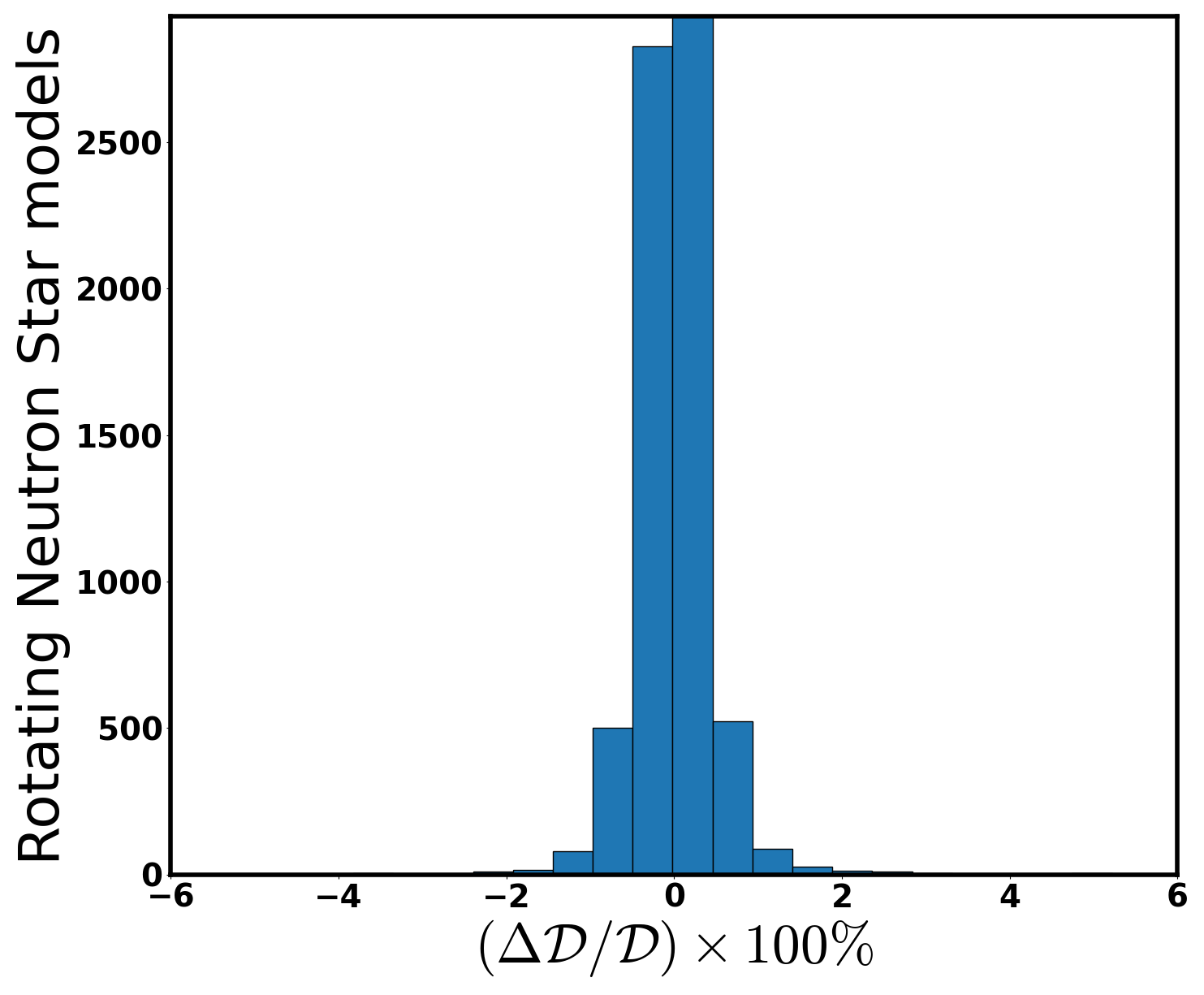}
	\caption{Histogram: Distribution of the 7046 models used vs relative errors from the formula (\ref{eq:Mf/x_chi_lnQbar}).}
	\label{fig:hist_Mf_chi_chi_lnbarQ.png}
\end{figure}

In the definition mentioned above for $\mathcal{D}$, the star's mass is given in geometric units, whereas $\tilde{f}=\Omega/2\pi c$. A similar relation had been proposed in \cite{Pappas:2015mba}, where it was pointed out that $\mathcal{D}$ corresponds to the reciprocal of $\bar{I}$. The surface that best describes the data has the functional form
\begin{equation}
\label{eq:Mf/x_chi_lnQbar}
\mathcal{D}=\sum_{n=0}^{4}\sum_{m=0}^{4-n}\hat{c}_{nm} \  \chi^n \ (\ln \bar{Q})^{m}.
\end{equation}
This is the regression function with the optimal statistical evaluation metrics at LOOCV. The corresponding results for an indicative list of models examined are given in the table (\ref{tab:Mf/x_chi_lnq_tab}), while the regression optimizers $\hat{c}_{nm}$ for the fitting function (\ref{eq:Mf/x_chi_lnQbar}) are presented in the table (\ref{tab:Mf/x_chi_lnqbar_optimizers}).
\begin{table}[!h]
	\small
	\caption{\label{tab:Mf/x_chi_lnq_tab} Indicative list of LOOCV evaluation metrics for the $\mathcal{D} = \sum_{n=0}^{\kappa}\sum_{m=0}^{\kappa-n}\hat{c}_{nm} \  \chi^n \ (\ln \bar{Q})^{m}$ parameterization.}
	\begin{ruledtabular}
		\begin{tabular}{ccccccc}
			MAE &Max Error& MSE &$d_{\text{max}}$($\%$) & MAPE ($\%$) & Exp Var & $\kappa$ \\
           \multicolumn{0}{c}{$(\cdot10^{-3})$}&\multicolumn{1}{c}{$(\cdot10^{-3})$}&\multicolumn{1}{c}{$(\cdot10^{-6})$}\\
			\hline
			0.192&1.352&0.065 &25.341 &1.351 & 1.0 & 2\\
			\hline
			0.082&0.638& 0.012&13.545&0.532 & 1.0 & 3\\
			\hline
			{\bf 0.057}&{\bf 0.439}& {\bf 0.007}&{\bf 5.306} & {\bf 0.322} & {\bf 1.0} & {\bf 4}\\
			\hline
			0.054&0.407&0.006 &5.634& 0.297& 1.0 &  5\\
			\hline
			0.053&0.412& 0.006&5.349 &  0.290 & 1.0 & 6\\
			\hline
			0.053&0.398&0.006 & 4.970 &  0.297& 1.0 & 7
		\end{tabular}
	\end{ruledtabular}
\end{table}
\begin{table}[!h]
	\caption{\label{tab:Mf/x_chi_lnqbar_optimizers} $\hat{c}_{nm}$ regression optimizers for the $\mathcal{D}(\chi,\ln\bar{Q})$ parameterization (\ref{eq:Mf/x_chi_lnQbar}).}
	\begin{ruledtabular}
		\begin{tabular}{cccc}
			$\hat{c}_{00}\cdot10^{-2}$ & $\hat{c}_{01}\cdot10^{-2}$ & $\hat{c}_{02}\cdot10^{-3}$ & $\hat{c}_{03}\cdot10^{-3}$   \\
			 4.0303260& -2.19273872 & 5.0581320 & -1.5218766   \\
			\hline\hline	
			$\hat{c}_{04}\cdot10^{-4}$ & $\hat{c}_{10}\cdot10^{-3}$ & $\hat{c}_{11}\cdot10^{-2}$ & $\hat{c}_{12}\cdot10^{-2}$  \\
			 3.2231043& 2.5119111  & -1.2777319 & 1.1903211 \\
			\hline\hline	
			$\hat{c}_{13}\cdot10^{-3}$ & $\hat{c}_{20}\cdot10^{-3}$ & $\hat{c}_{21}\cdot10^{-2}$ & $\hat{c}_{22}\cdot10^{-3}$ \\
			-2.6225836 & -2.7619227 & -1.7579355 &  3.4932330\\
			\hline\hline
			$\hat{c}_{30}\cdot10^{-2}$ & $\hat{c}_{31}\cdot10^{-3}$ & $\hat{c}_{40}\cdot10^{-2}$ &  \\
			 1.4039214 & 7.5211686 & -1.4772850 & 
		\end{tabular}
	\end{ruledtabular}
\end{table}
The fitting formula (\ref{eq:Mf/x_chi_lnQbar}) that optimally reproduces the data and the corresponding relative deviations are presented in Fig.\ref{fig:Mf_chi_chi_lnq_fig}. Using this $\mathcal{D}=\mathcal{D}(\chi,\ln\bar{Q})$-parameterization the relative errors between the fit (\ref{eq:Mf/x_chi_lnQbar}) and the data are $\lesssim 5.220\%$ (universality). Out of the full data set, only 42 models have relative deviations $\geq 2\%$, corresponding to models with stellar compactness $\mathcal{C}\leq  0.212$ and spin $\chi \in [0.233, 0.3013]$. This behavior is shown in Fig.\ref{fig:hist_Mf_chi_chi_lnbarQ.png} where it is evident that the fitting function (\ref{eq:Mf/x_chi_lnQbar}) reproduces most of the data values with accuracy $\leq 2\%$. This relation provides a useful universal description of rotation frequencies in terms of the NS (and spacetime) parameters $M,\chi$, and $\bar{Q}$.  


\subsection{\label{sec:inertia}Universal relations for the normalized moment of inertia $\bar{I}$}

We now turn our attention to a quantity involved in one of the better-known universal relations, the normalized moment of inertia $\bar{I}$.
The moment of inertia for a rigidly rotating configuration is defined as $I=J/\Omega$. The mass mainly influences the star's moment of inertia in the star's outer regions; therefore, the increases in mass and radius lead generically to larger moments of inertia \cite{breu_maximum_2016}. In geometric units, the NS's moment of inertia is expressed as $I_{geom}=IG/c^2$ with $[I_{geom}]=[km^3]$. The normalized (dimensionless) moment of inertia $\bar{I}$ is defined as $\bar{I}=I_{geom}/M^3$, where $M$ is the mass of the NS in geometric units ($[M]=[km]$). In what follows, we investigate universal relations between the normalized moment of inertia $\bar{I}$ and the dimensionless quantities $\mathcal{C}, \ \chi, \ \sigma,\ \mathcal{E} \ (\textrm{i.e., }T/|W|), \ \textrm{and} \ \bar{Q}$. We remind that, for the total sample of the 11983 NS models considered, for all the rotation rates, these parameters are in the ranges of, $0.085\leq \mathcal{C} \leq 0.313, \ 0.227\leq \chi \leq 0.799,\ 0.067\leq \sigma \leq 1.033,\textrm{and}\ 0.012\leq \mathcal{E} \leq 0.146$. 

The normalized moment of inertia $\bar{I}$ is related to the star's rotation and deformation.
Therefore, it is expected to be related to the parameters $\chi, \sigma, \mathcal{E}$ and the reduced quadrupole deformation $\bar{Q}$. In what follows, we first investigate the correlation between the quantities $\bar{I}$, $\chi$, and $\bar{Q}$, as has already been done in the literature \cite{pappas2014effectively}. The surface ${\bar{I}}=\bar{I}(\chi,\bar{Q})$ that best describes the data for these parameters has the functional form:
\begin{equation}
\label{eq:Ibar_chi_Qbar}
\bar{I}(\chi,\bar{Q})=\sum_{n=0}^{4}\sum_{m=0}^{4-n}\hat{d}_{nm} \ \chi^n \ \bar{Q}^m.
\end{equation}
Again, this is the least complicated function compared to other regression models tested (i.e., higher-order than $\kappa = 4$ polynomial functions), but it is a good enough fit while the improvement from going to higher order is only marginal and not worth the effort. The corresponding results for an indicative list of regression models tested are shown in table (\ref{tab:Ibar_chi_Qbar}).
\begin{table}[!h]
	\small
	\caption{\label{tab:Ibar_chi_Qbar} Indicative list of LOOCV evaluation metrics for the $\bar{I}(\chi,\bar{Q}) = \sum_{n=0}^{\kappa}\sum_{m=0}^{\kappa-n}\hat{d}_{nm} \ \chi^n \ \bar{Q}^m $ parameterization.}
	\begin{ruledtabular}
		\begin{tabular}{ccccccc}
			MAE&Max Error& MSE &$d_{\text{max}}$($\%$) & MAPE ($\%$) & Exp Var & $\kappa$ \\
			\hline
			 0.110&1.808& 0.022& 11.320 & 1.205 & 1.0 & 2\\
			\hline
			 0.048&1.605&0.005 & 5.919 & 0.524 & 1.0 & 3\\
			\hline
			 {\bf0.035}&{\bf 1.522}&{\bf 0.004} & {\bf 5.613} & {\bf 0.360}& {\bf 1.0} & {\bf 4}\\
			\hline
			 0.031&1.482&0.003 & 5.466 & 0.311 & 1.0 & 5\\
			\hline
			 0.030&1.588&0.003 & 5.858 & 0.294 & 1.0 & 6\\
			 \hline
			 0.029&1.549&0.003 & 5.716 & 0.290 & 1.0 & 7\\
			
		\end{tabular}
	\end{ruledtabular}
\end{table}
Therefore, from the surface-fit evaluation, the fitting-optimizers $\hat{d}_{nm}$ are presented in table (\ref{tab:Ibar_chi_qbar_optimizers}).
\begin{table}[!h]
	\caption{\label{tab:Ibar_chi_qbar_optimizers} $\hat{d}_{nm}$ regression optimizers for the $\bar{I}(\chi,\bar{Q})$ parameterization (\ref{eq:Ibar_chi_Qbar}).}
	\begin{ruledtabular}
		\begin{tabular}{cccc}
			$\hat{d}_{00}$ & $\hat{d}_{01}$ & $\hat{d}_{02}\cdot10^{-1}$ & $\hat{d}_{03}\cdot10^{-2}$   \\
			2.0783240 & 2.1236130 & -1.1322365 & 1.4017422   \\
			\hline\hline	
			$\hat{d}_{04}\cdot10^{-4}$ & $\hat{d}_{10}$ & $\hat{d}_{11}\cdot10^{-1}$ & $\hat{d}_{12}\cdot10^{-2}$  \\
			-4.7146766 & -1.0681569 & 3.6822470 & -6.4286919 \\
			\hline\hline	
			$\hat{d}_{13}\cdot10^{-3}$ & $\hat{d}_{20}\cdot10^{-1}$ & $\hat{d}_{21}$ & $\hat{d}_{22}\cdot10^{-3}$ \\
			3.2317173 & 2.7069932 & 2.5042417 & -9.8870123 \\
			\hline\hline
			$\hat{d}_{30}$ & $\hat{d}_{31}\cdot10^{-1}$ & $\hat{d}_{40}$ &  \\
			-5.9022158 & -1.9102353 & 4.1281492 & 
		\end{tabular}
	\end{ruledtabular}
\end{table}

\begin{figure}[!ht]
	\includegraphics[width=0.26\textwidth]{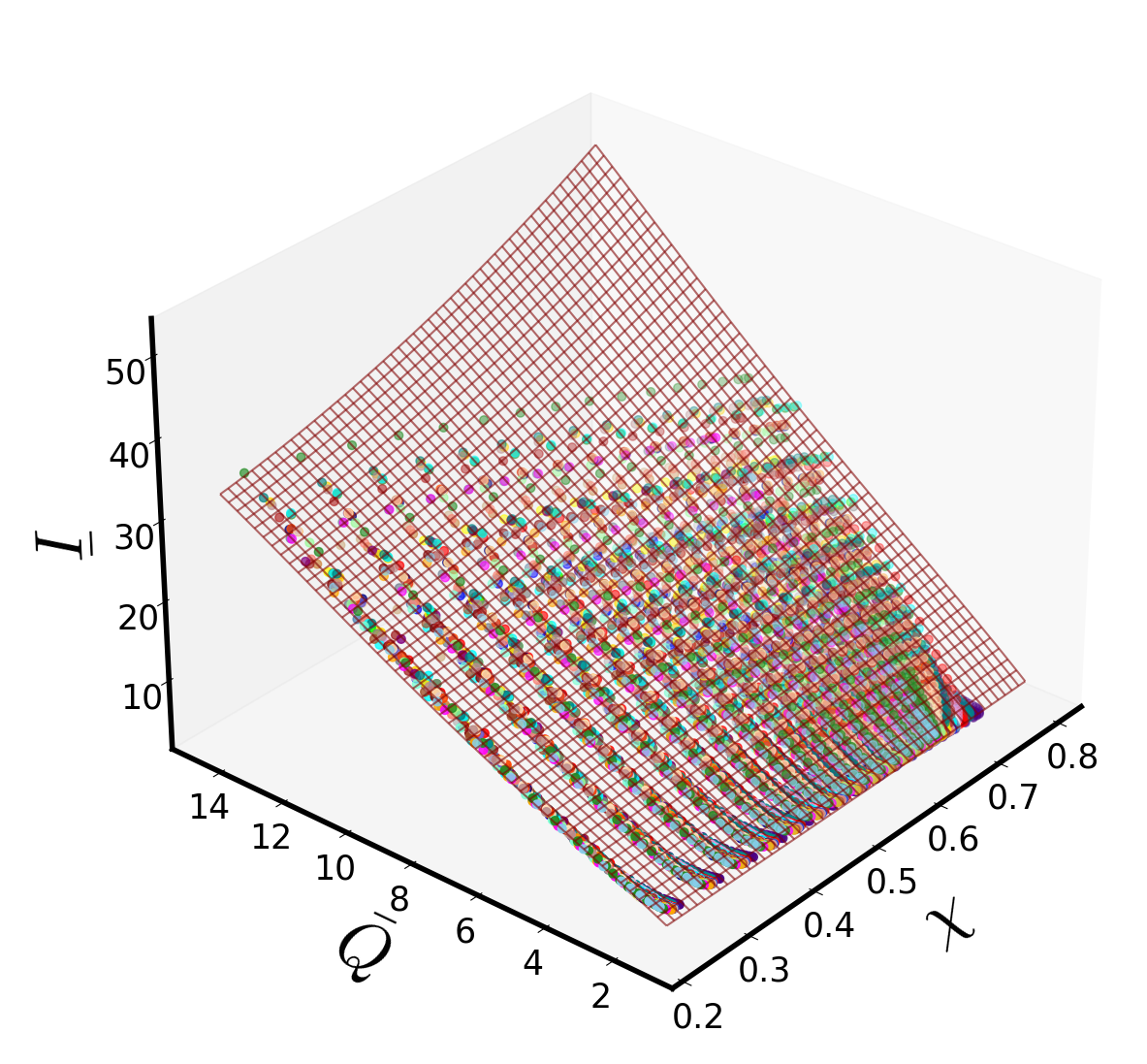}
	\includegraphics[width=0.26\textwidth]{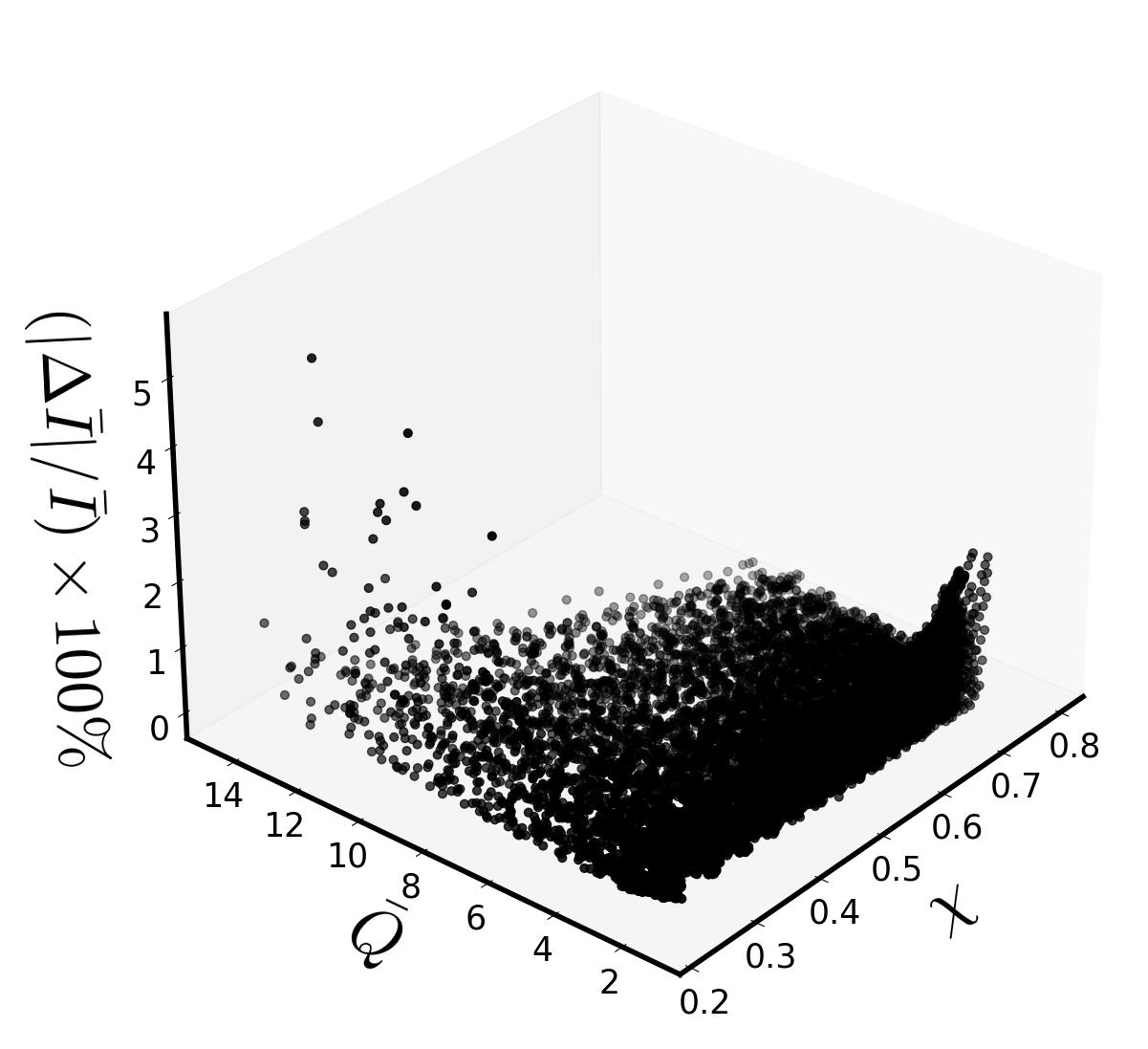}
	%
	\caption{$\bar{I}$ as a function of the dimensionless parameters $\chi,\ \bar{Q}$ and relative error distribution for our sample. The surface corresponds to the formula (\ref{eq:Ibar_chi_Qbar}). The relative errors are given as ($100\%(|\Delta{\bar{I}}|/{\bar{I}})=100\%|{\bar{I}_{fit}}-{\bar{I}}|/{\bar{I}}$).} 
		\label{fig:Ibar_chi_Qbar_fig}
\end{figure}

\begin{figure}[!h]
	\includegraphics[width=0.28\textwidth]{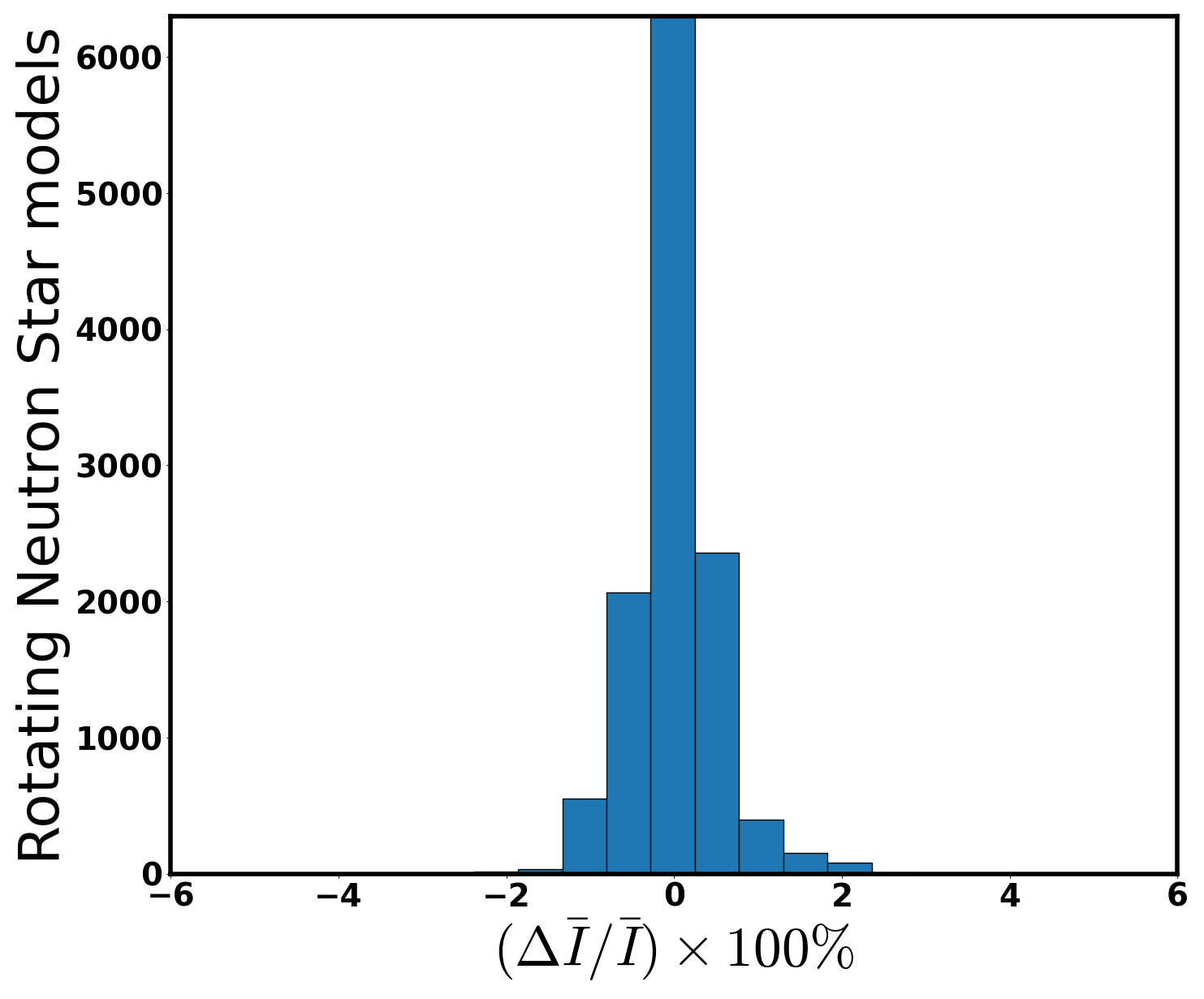}
	\caption{Histogram: Rotating NS models used vs relative errors distribution derived from regression formula (\ref{eq:Ibar_chi_Qbar}).}
	\label{fig:hist_Ibar_chi_barQ_hist}
\end{figure}

\begin{figure}[!h]
    \includegraphics[width=0.30\textwidth]{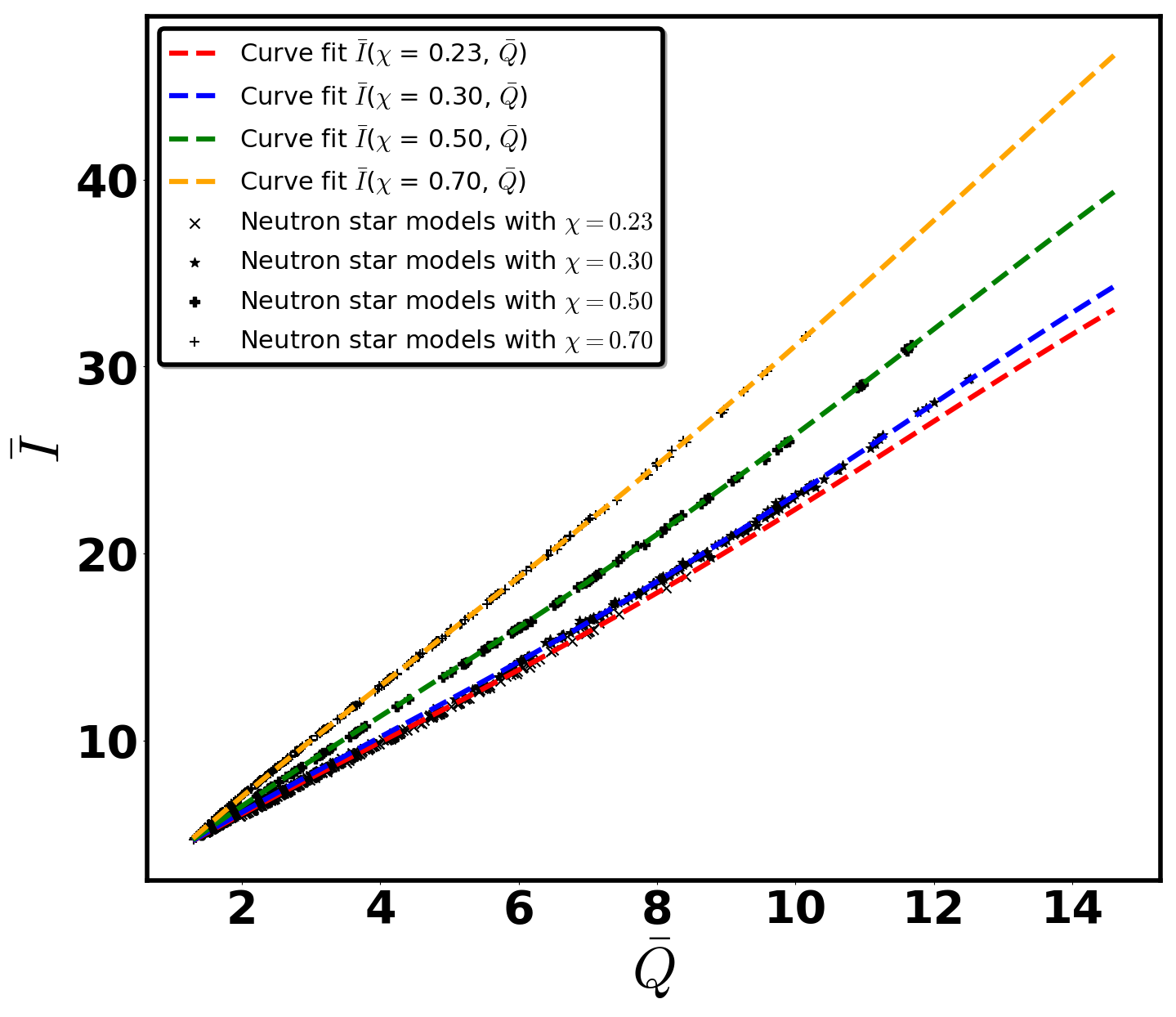}
    \includegraphics[width=0.30\textwidth]{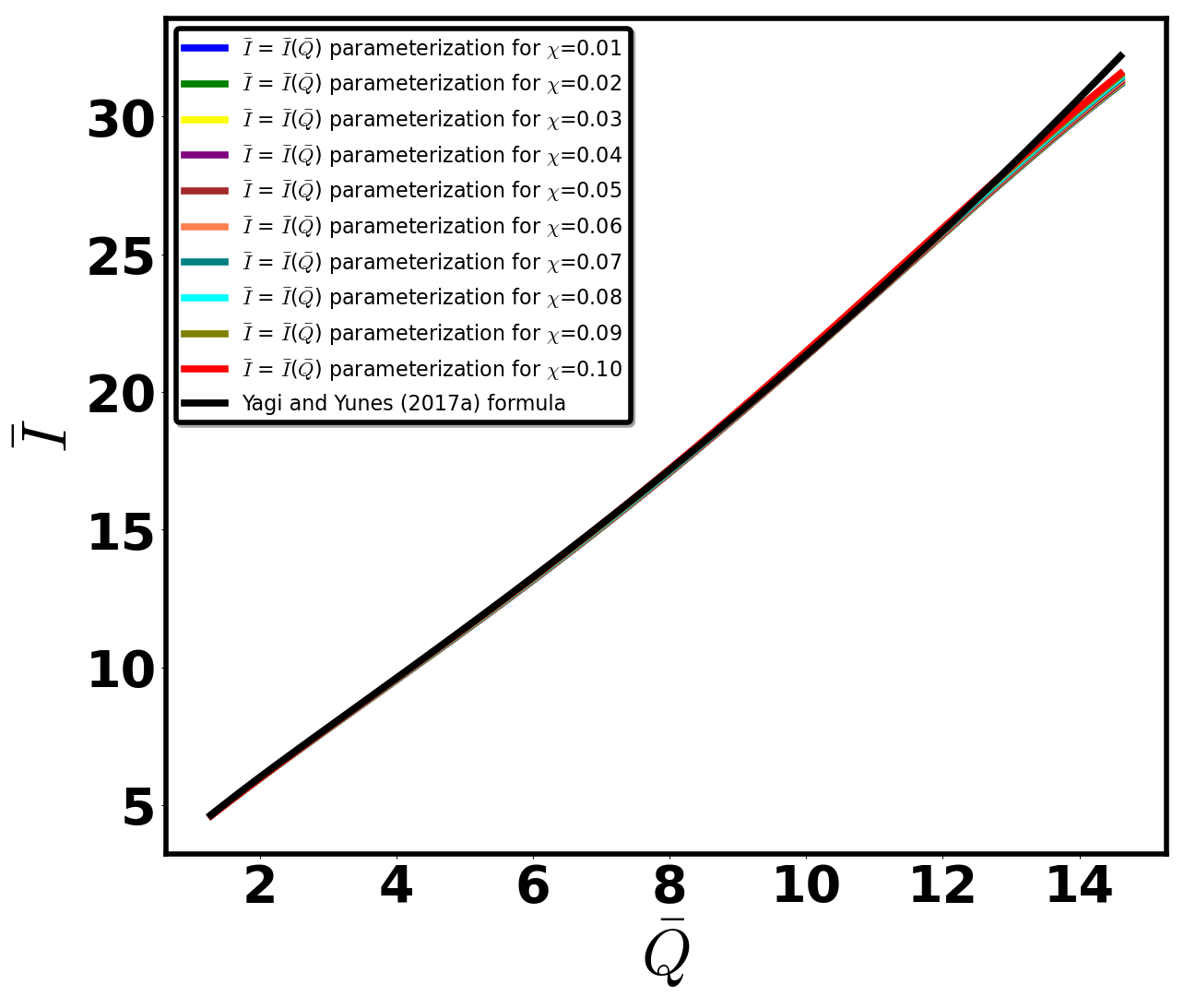}
	\caption{$\bar{I}=\bar{I}(\bar{Q})$ curves for different values of the dimensionless angular momentum $\chi$. Left plot: The black dots correspond to the observable quantities derived from \textit{RNS}, while the colored curves correspond to the theoretical prediction coming from the regression formula (\ref{eq:Ibar_chi_Qbar}). Right plot: Prediction of (\ref{eq:Ibar_chi_Qbar}) in the slowly rotating limit using different values of $\chi \in [0.01,0.10]$ compared against results in the literature \cite{yagi2017approximate}.}
	\label{fig:Ibar_chi_fixed_qbar.png}
\end{figure}

The best fit (\ref{eq:Ibar_chi_Qbar}) to the data and the corresponding relative errors are presented in Fig.\ref{fig:Ibar_chi_Qbar_fig}.
The relative errors between the fit (\ref{eq:Ibar_chi_Qbar}) and the observed $\bar{I}$ are $\lesssim 5.515\%$, while only $88$ stellar models out of the total have relative deviation $\geq 2\%$. %
The histogram in Fig.\ref{fig:hist_Ibar_chi_barQ_hist} presents the relative errors distribution. %
It is clear from Fig.\ref{fig:hist_Ibar_chi_barQ_hist} that the regression formula (\ref{eq:Ibar_chi_Qbar}) corresponds to a well-behaved EoS-independent relation which gives good results for all the rotating models considered. Moreover, it reproduces most data values with an error $\lesssim2\%$. Consequently, it is a useful relation between $\bar{I}$ and the parameters $\chi$ and $\bar{Q}$.  Additionally, in Fig.\ref{fig:Ibar_chi_fixed_qbar.png}, we present curves that are derived from the universal relation (\ref{eq:Ibar_chi_Qbar}) for different values of the dimensionless angular momentum $\chi$. 

As we can see from Fig.\ref{fig:Ibar_chi_fixed_qbar.png} 
(upper plot) as well, the theoretical prediction given by (\ref{eq:Ibar_chi_Qbar}) verifies the data for different $\chi$ values quite accurately. Also, for relatively "slow" rotating stellar models with $\chi$-values $0.23$ and $0.30$, the corresponding curves from the $\bar{I}-\bar{Q}$ relation tend to coincide, which leads to the well-known universal behavior for slowly rotating NSs in the Hartle-Thorne approximation (bottom plot in Fig.\ref{fig:Ibar_chi_fixed_qbar.png}).

\subsection{Universal relations for the reduced spin octupole moment $\bar{S_3}$}
\label{sec:Universal Relations with Octupole Moment}

Apart from the mass quadrupole moment, another multipole moment that can be useful for NSs is the spin octupole moment $S_3$, which is the next order contribution in the spin moments after the angular momentum. It has been shown that NSs seem to observe a universal three-hair property, where the moments higher than the quadrupole depend on the first three, i.e., the mass, the dimensionless angular momentum, and the quadrupole \cite{pappas2014effectively, stein2014three, yagi2014effective}.

In this section, we revisit the universal relation for the Geroch-Hansen octupole moment $S_3$ in terms of the quadrupole and look for improvements, but also explore other possibilities as well. The $S_3$ is computed in geometric units ($[S_3]=[km^4]$), while the reduced octupole moment $\bar{S}_3$ is defined as $\bar{S}_3=-S_3/(\chi^3M^4)$. 

We first look for a relation similar to the one for the quadrupole and therefore investigate some relation that connects the reduced octupole moment $\bar{S_3}$ with the star's stellar compactness $\mathcal{C}$ and the rotation parameters $\chi,\sigma$. However, as we can see from tables (\ref{tab:S3_chi_C_tab},\ref{tab:S3_sigma_C_tab}), the $\bar{S_3} = \bar{S_3}(\chi,\mathcal{C})$ and the $\bar{S_3} = \bar{S_3}(\sigma,\mathcal{C})$ parameterizations do not give satisfactory results at the LOOCV evaluation test for the indicative class of regression models examined.
\begin{table}[!h]
	\small
	\caption{\label{tab:S3_chi_C_tab} Indicative list of LOOCV evaluation metrics for the $\bar{S_3}(\chi,\mathcal{C}) = \sum_{n=0}^{\kappa}\sum_{m=0}^{\kappa-n}\hat{\tilde{a}}_{nm} \  \chi^n \mathcal{C}^m$ parameterization.}
	\begin{ruledtabular}
		\begin{tabular}{ccccccc}
			MAE&Max Error&MSE & $d_{\text{max}}$($\%$) & MAPE ($\%$) & Exp Var & $\kappa$ \\
			\hline
			 0.602&5.383& 0.703&134.168 & 10.788 & 1.0 & 2\\
			\hline
			0.363&3.925&0.326 &65.493 & 5.193 & 1.0 & 3\\
			\hline
			0.321&4.169 & 0.281&25.362 & 4.031 & 1.0 & 4\\
			\hline
			0.310&4.059&0.268 & 24.472 & 3.774 & 1.0 & 5\\
			\hline
			0.310&3.937& 0.267& 24.087  & 3.810 & 1.0 & 6\\
			\hline
			0.305&3.631&  0.262&23.269 & 3.692 & 1.0 & 7
		\end{tabular}
	\end{ruledtabular}
\end{table}

\begin{table}[!h]
	\small
	\caption{\label{tab:S3_sigma_C_tab} Indicative list of LOOCV evaluation metrics for the $\bar{S_3}(\sigma,\mathcal{C}) = \sum_{n=0}^{\kappa}\sum_{m=0}^{\kappa-n}\hat{\tilde{a}}_{nm} \  \sigma^n \mathcal{C}^m$ parameterization.}
	\begin{ruledtabular}
		\begin{tabular}{ccccccc}
			MAE&Max Error&MSE & $d_{\text{max}}$($\%$) & MAPE ($\%$) & Exp Var & $\kappa$ \\
			\hline
			0.522&5.726& 0.4736& 143.273 & 10.061 & 1.0 & 2\\
			\hline
			0.212&2.480&0.0924 &64.275 & 3.578 & 1.0 & 3\\
			\hline
			0.152&2.030&0.0544 &23.279 & 1.986 & 1.0 & 4\\
			\hline
			0.147&1.975&0.0506 &11.532 & 1.853 & 1.0 & 5\\
			\hline
			0.155&2.128 &  0.0531&23.334 & 2.129 & 1.0 & 6\\
			\hline
			0.143&2.017&0.0491 &10.040 & 1.763 & 1.0 & 7
		\end{tabular}
	\end{ruledtabular}
\end{table}
Therefore, we don to pursue this avenue further. 

We continue with an attempt to improve the already-established relation between the reduced octupole moment $\bar{S}_3$ and the star's reduced quadrupole deformation $\bar{Q}$. In order to find a better parameterization, we first investigate the correlation between $\bar{S}_3$ and $\ln\bar{Q}$, as an alternative to the usual approach of $\bar{S_3}=\bar{S_3}(\bar{Q})$ considered so far in the literature \cite{pappas2014effectively,yagi2014effective,yagi2017approximate}. The representation that best describes the data has a functional form
\begin{equation}
\label{eq:S3bar_lnQbar}
\bar{S}_3(\bar{Q})=\sum_{n=0}^{4}\hat{\tilde{b}}_{n} \left(\ln\bar{Q}\right)^n.
\end{equation}
This is the mathematical model with the best statistical evaluation metric functions at LOOCV compared to other regression models examined. An indicative list of regression models tested and the corresponding results are presented in table (\ref{tab:s3_lnq_tab}). 
\begin{table}[!h]
	\small
	\caption{\label{tab:s3_lnq_tab} Indicative list of LOOCV evaluation metrics for the $\bar{S}_3(\bar{Q}) = \sum_{n=0}^{\kappa}\hat{\tilde{b}}_{n} \  \left(\ln\bar{Q}\right)^n$ parameterization.}
	\begin{ruledtabular}
		\begin{tabular}{ccccccc}
			MAE&Max Error&MSE &$d_{\text{max}}$($\%$) & MAPE ($\%$) & Exp Var & $\kappa$ \\
			\hline
			0.2401&2.842& 0.0983&45.553 & 4.806 & 1.0 & 2\\
			\hline
			0.0942&0.912&0.0176 & 6.454 & 1.438 & 1.0 & 3\\
			\hline
			{\bf 0.090}&{\bf 0.893}&{\bf 0.0167}& {\bf 4.847} & {\bf 1.308} & {\bf 1.0} & {\bf4}\\
			\hline
			0.0904&0.889& 0.0167& 4.922 & 1.311 & 1.0 & 5\\
			\hline
			 0.0904&0.886& 0.0167& 4.913 & 1.308 & 1.0 & 6\\
			\hline
			0.0904&0.886&  0.0167& 4.910 & 1.309 & 1.0 & 7
			
		\end{tabular}
	\end{ruledtabular}
\end{table}
From the curve fit evaluation, the fitting optimizers $\hat{\tilde{b}}_n$ are given in the table (\ref{tab:S3bar_lnqbar_optimizers}).

\begin{table}[!h]
	\caption{\label{tab:S3bar_lnqbar_optimizers} $\hat{\tilde{b}}_n$ regression optimizers for the parameterization \ref{eq:S3bar_lnQbar}.}
	\begin{ruledtabular}
		\begin{tabular}{ccccc}
			$\hat{\tilde{b}}_{0}$ & $\hat{\tilde{b}}_{1}$ & $\hat{\tilde{b}}_{2}$ & $\hat{\tilde{b}}_{3}\cdot 10^{-2}$& $\hat{\tilde{b}}_{4}\cdot 10^{-1}$  \\
			1.0201214& 2.2481059& 1.36247244 & 3.7110007 & 2.7018181  \\
		\end{tabular}
		
	\end{ruledtabular}
\end{table}

\begin{figure}[!ht]
	\includegraphics[width=0.25\textwidth]{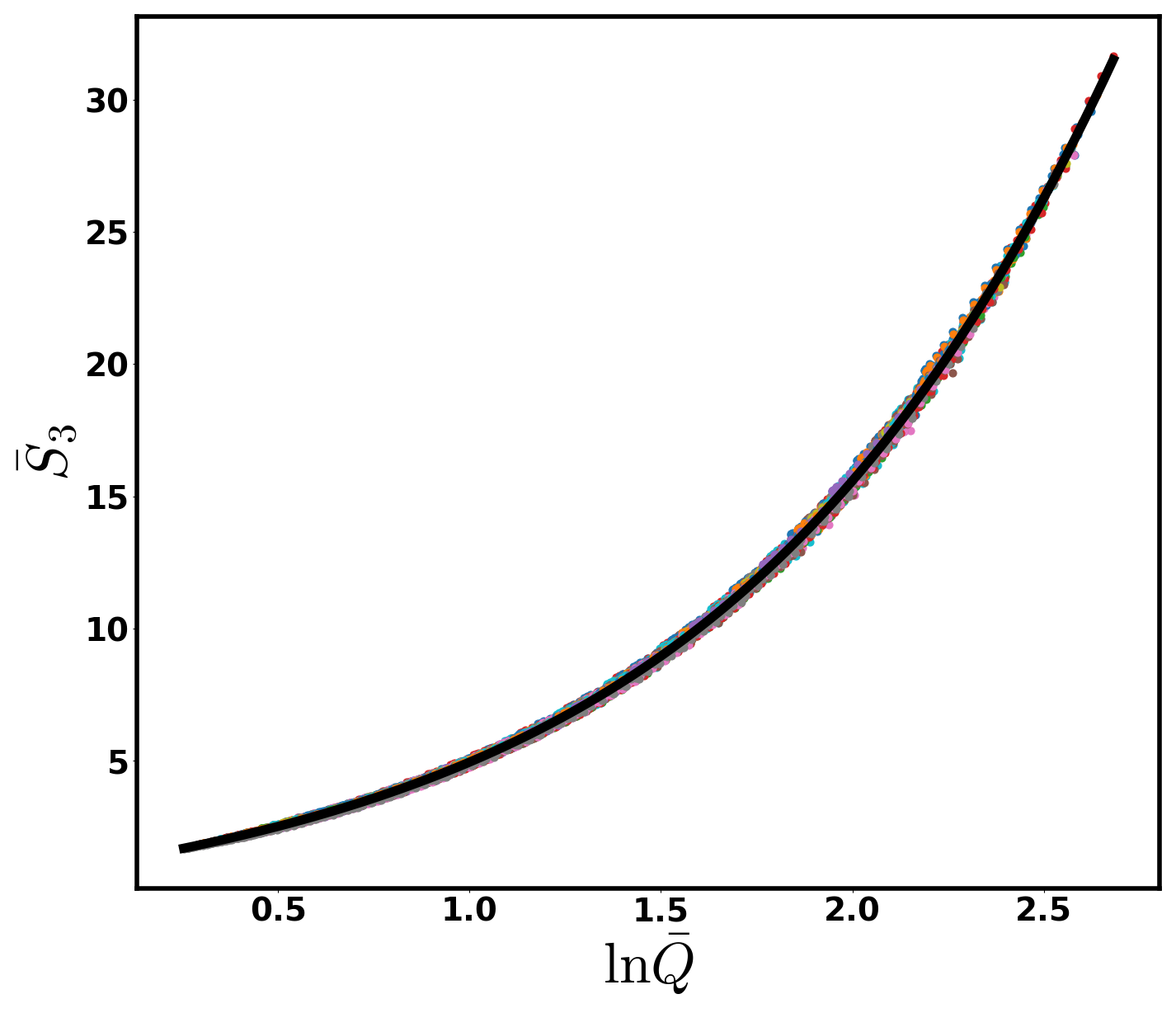}
	\includegraphics[width=0.25\textwidth]{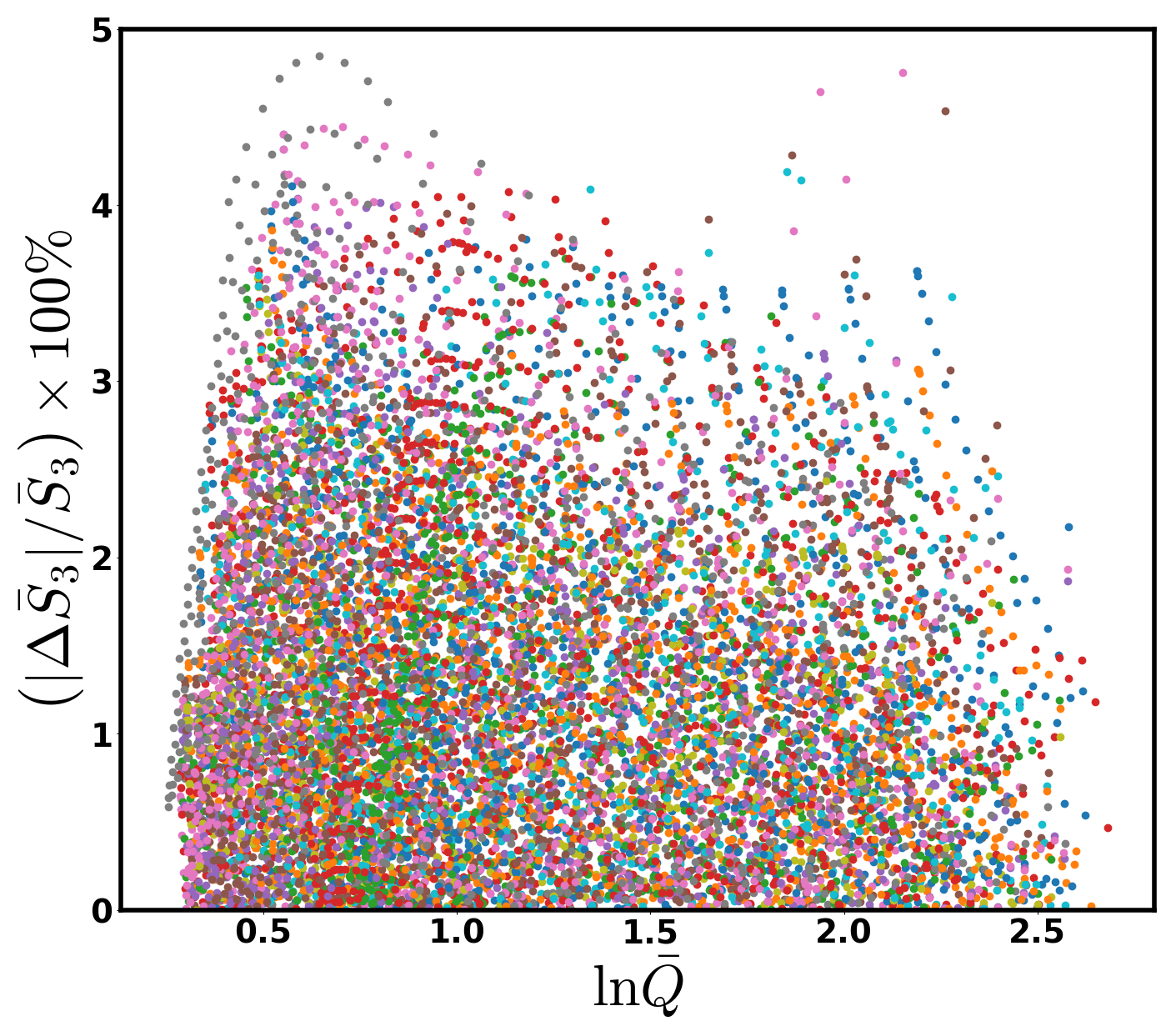}
	\caption{Dimensionless octupole moment $\bar{S}_3$ as a function of the dimensionless $\ln \bar{Q}$ parameter and relative errors for each EoS included in our sample. The black curve corresponds to the regression formula (\ref{eq:S3bar_lnQbar}) used in order to reproduce the data. The relative errors given as $100\%(|\Delta \bar{S}_3|/\bar{S}_3)=100\%|\bar{S}_{3,fit}-\bar{S}_3|/\bar{S}_3$ are computed between the $\bar{S}_3$ and its estimate coming from the regression formula (\ref{eq:S3bar_lnQbar}).}
	\label{fig:S_3_bar_lnQbar_fig}
\end{figure}

The curve evaluation fit (\ref{eq:S3bar_lnQbar}) that best reproduces the data values and the corresponding relative errors are presented in Fig.\ref{fig:S_3_bar_lnQbar_fig}.

\begin{figure}[!h]
	\includegraphics[width=0.28\textwidth]{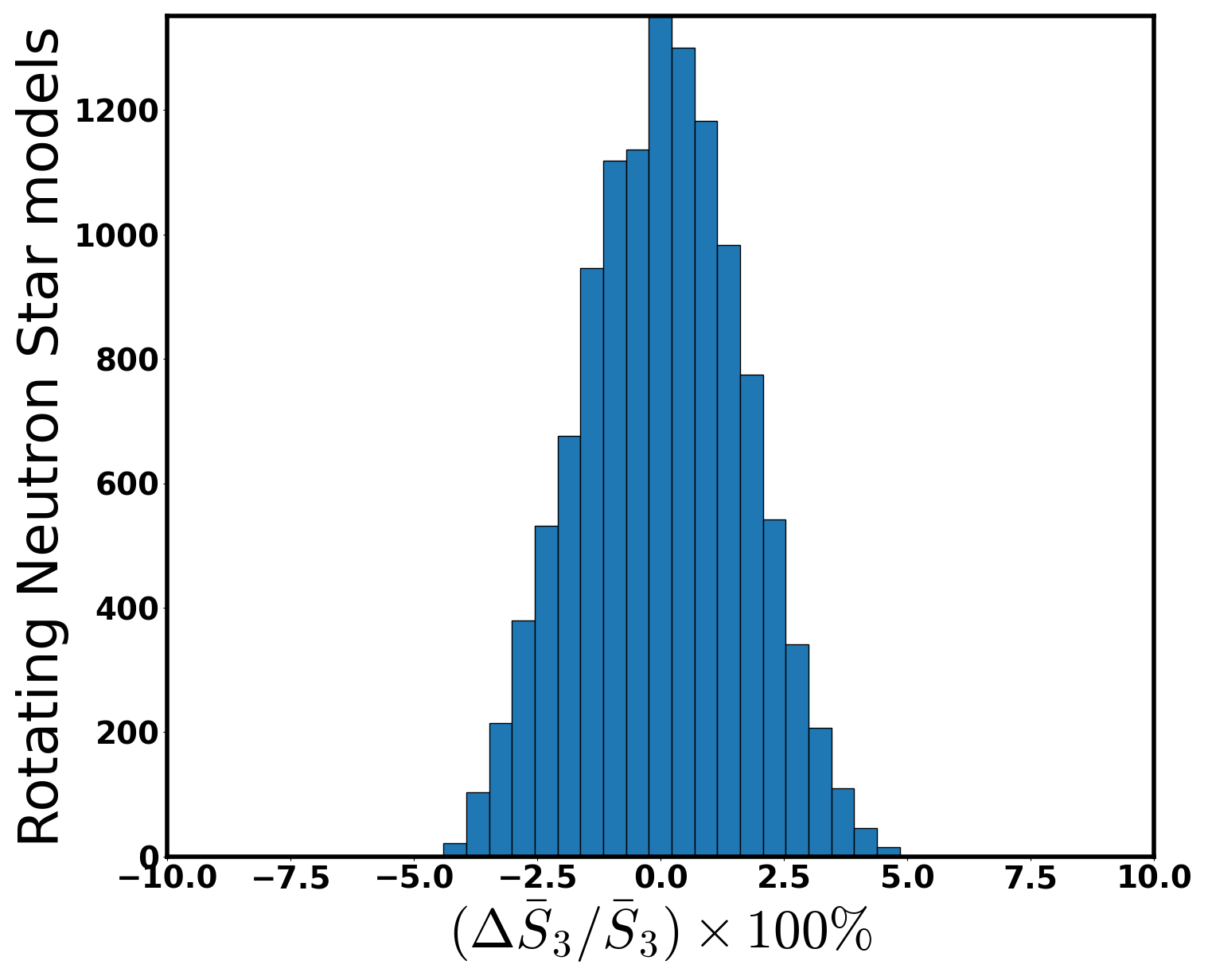}
	\caption{Histogram: distribution of models vs relative errors derived from the regression formula (\ref{eq:S3bar_lnQbar}).}
	\label{fig:hist_S3bar_lnQbar_test_fig}
\end{figure}

Using this ${\bar{S}_3}=\bar{S}_3(\ln \bar{Q})$-parameterization the relative errors between the fit (\ref{eq:S3bar_lnQbar}) and the observed $\bar{S}_3$ are better than $ 4.845\%$ for all EoSs and NS models considered. In addition, this EoS-insensitive formula is slightly better than those presented in \cite{pappas2014effectively,yagi2014effective,yagi2017approximate}. Therefore, this very accurate universal relation can provide the spacetime's octupole moment when the star's quadrupole deformation $\bar{Q}$ is known. In Fig.\ref{fig:hist_S3bar_lnQbar_test_fig}, we present the models' histogram distribution of relative errors $100\% \times(\Delta {\bar{S}_3}/{\bar{S}_3})$.
Therefore, as we can see from the histogram presented in Fig.\ref{fig:hist_S3bar_lnQbar_test_fig}, the regression formula (\ref{eq:S3bar_lnQbar}) corresponds to an accurate EoS-insensitive relation which gives good results for all the rotating models considered. Moreover, it reproduces most of the corresponding data with an error $\lesssim 4\%$.

It is worth attempting to improve the previous relation by including some measure of the rotation of the star as an additional parameter. We therefore explore a relation for $\bar{S_3}$ as a function of one of the star's possible spin parameterizations as well as $\bar{Q}$. We first look into the correlation between the quantities $(\bar{S_3},M\times \tilde{f}, \bar{Q})$, and then we also explore the combination $(\bar{S_3},\sigma, \bar{Q})$. We remind here that the quantities $M\times \tilde{f}$ as well as $\sigma$, are both dimensionless spin parameterizations.

\begin{figure}[!ht]
	\includegraphics[width=0.25\textwidth]{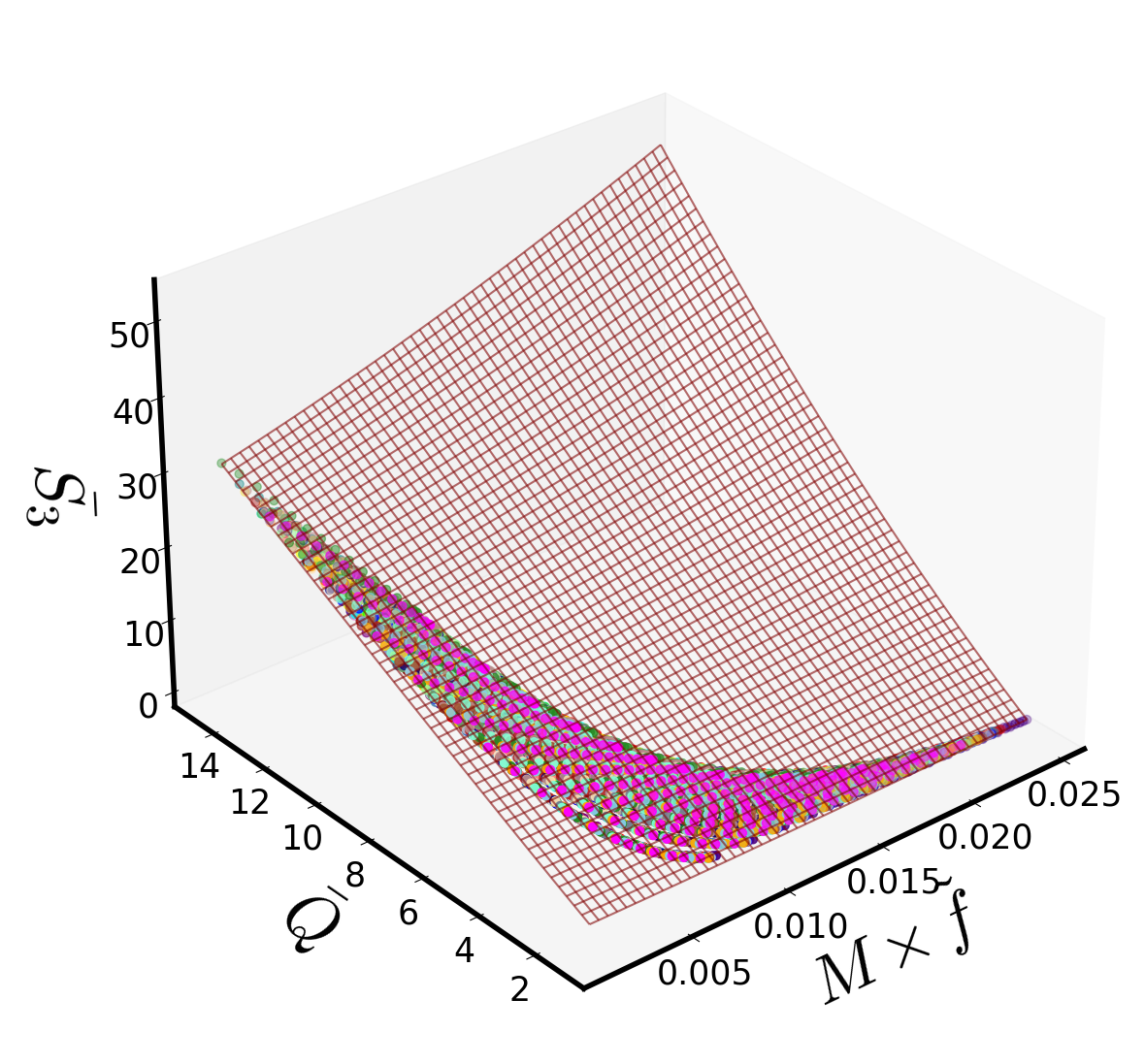}
	\includegraphics[width=0.25\textwidth]{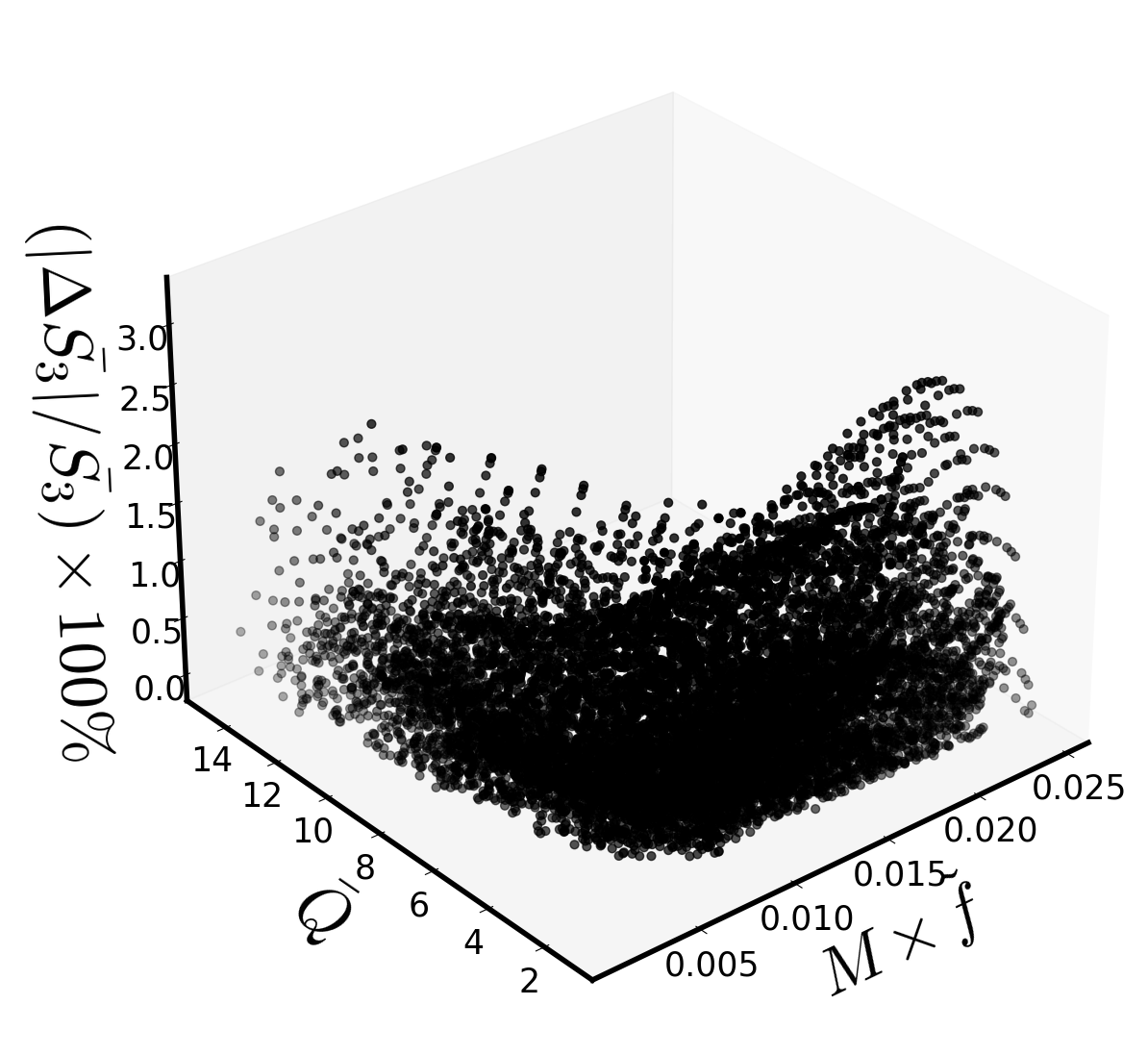}
	\caption{${\bar{S3}}$ as a function of the dimensionless parameters $ M\times \tilde{f},\bar{Q}$ and relative errors. The surface corresponds to the regression formula (\ref{eq:S3bar_Mf_Qbar}). The relative errors are given as ($100\%(|\Delta {\bar{S_3}}|/{\bar{S_3}})=100\%|{\bar{S}}_{3,fit}-{\bar{S_3}}|/{\bar{S_3}}$).}%
		\label{fig:S3bar_Mf_Qbar_fig}
\end{figure}

\begin{figure}[!ht]
	\includegraphics[width=0.25\textwidth]{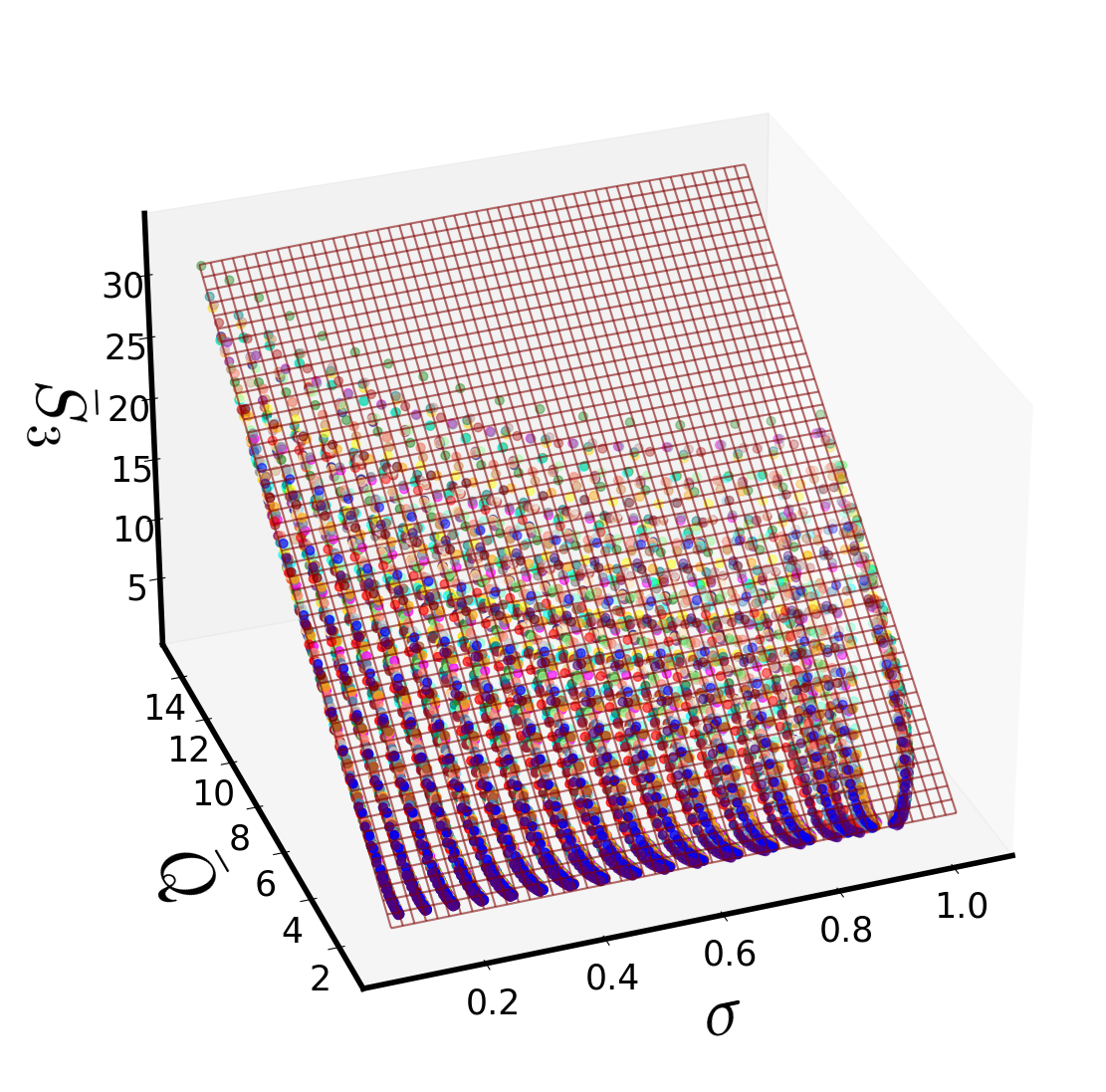}
	\includegraphics[width=0.25\textwidth]{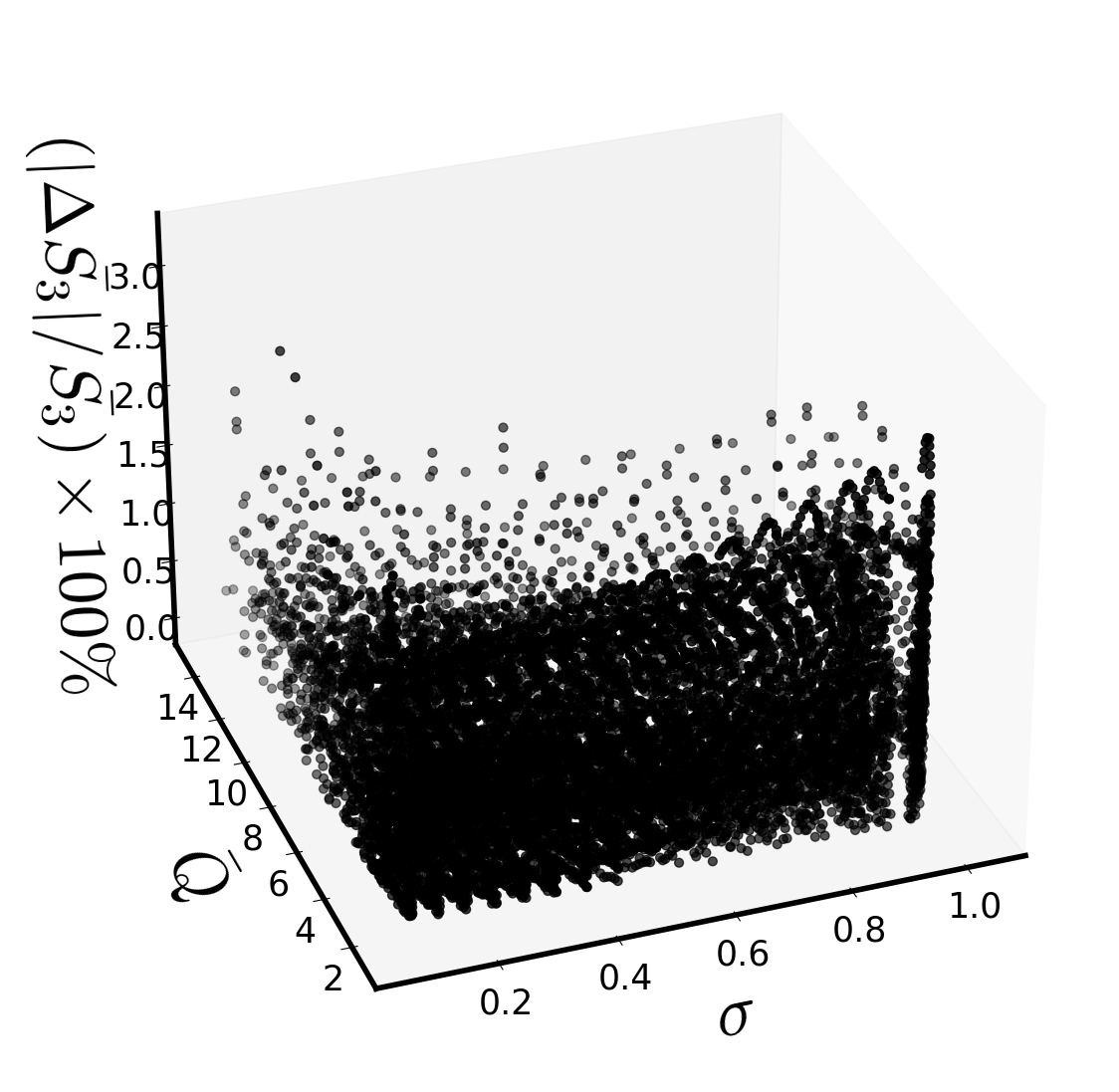}
	\caption{${\bar{S_3}}$ as a function of the dimensionless parameters $\sigma,\bar{Q}$ and relative error distribution. The surface corresponds to the regression formula (\ref{eq:S3bar_sigma_Qbar}). The relative errors are given as ($100\%(|\Delta {\bar{S_3}}|/{\bar{S_3}})=100\%|{\bar{S}}_{3,fit}-{\bar{S_3}}|/{\bar{S_3}}$).}%
		\label{fig:S3_sigma_Qbar_opt_fig}
\end{figure}

\begin{figure}[!h]
	\includegraphics[width=0.235\textwidth]{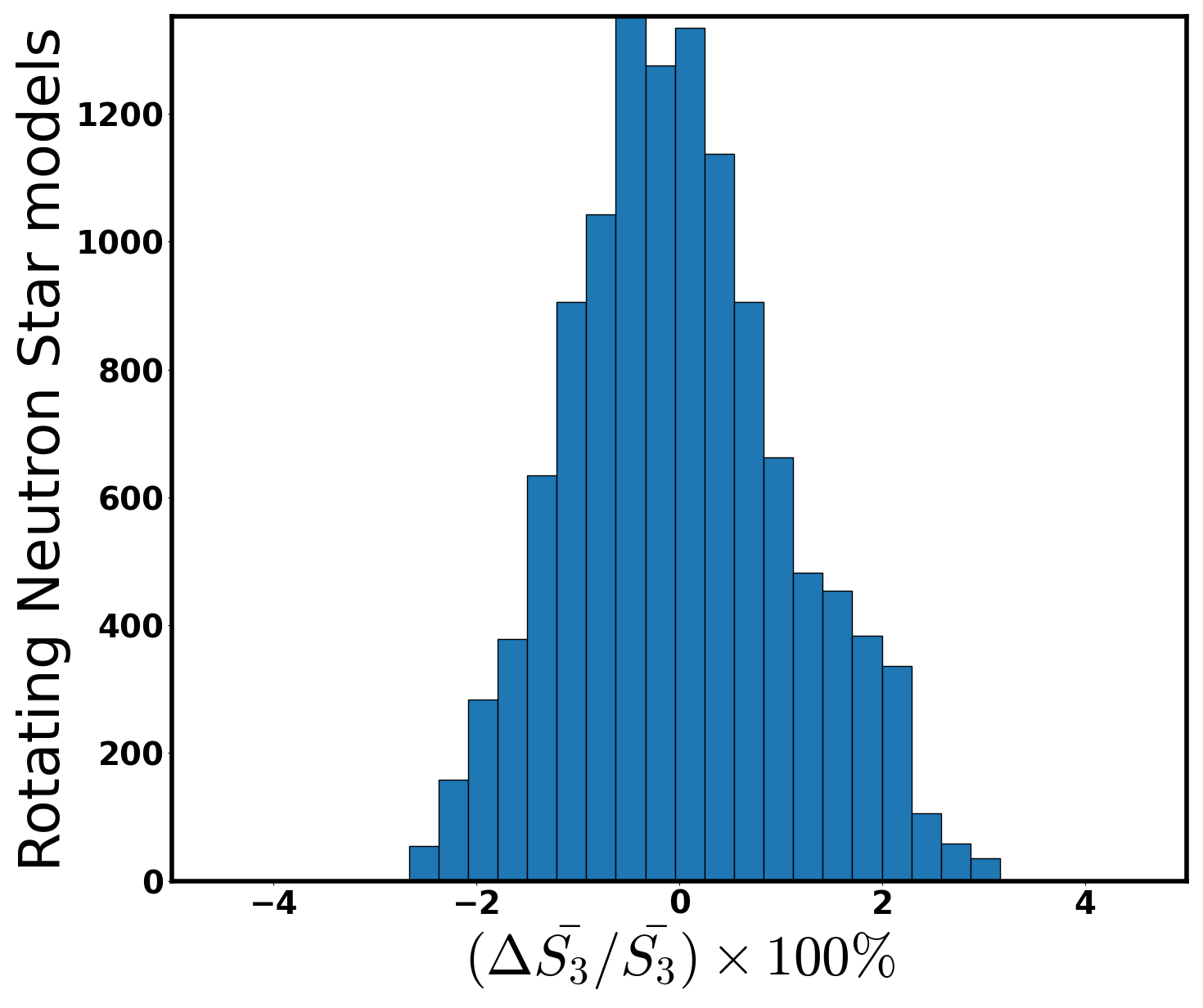}
	\includegraphics[width=0.235\textwidth]{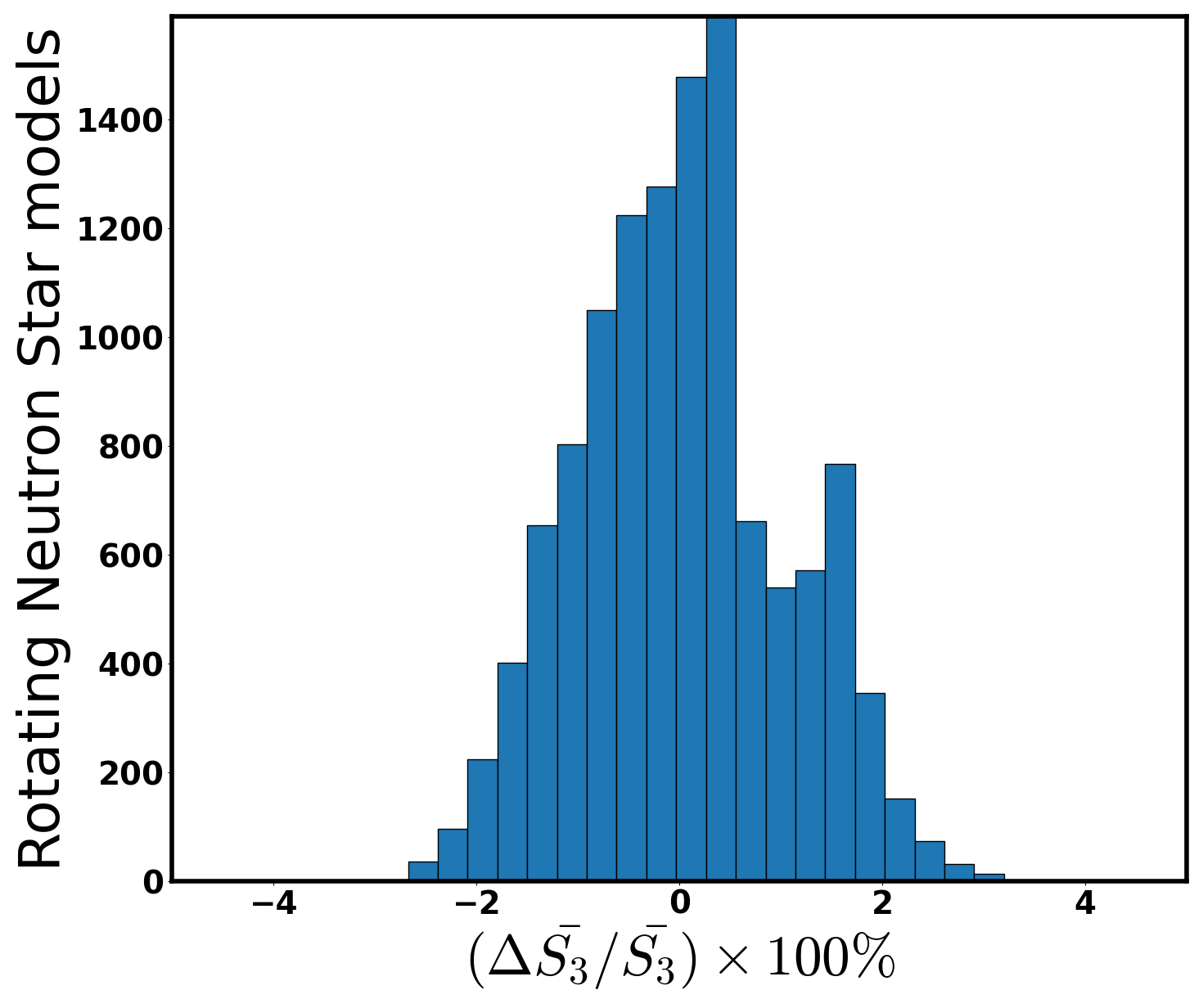}
	\caption{Left plot: Relative errors distribution derived from the regression formula (\ref{eq:S3bar_Mf_Qbar}).
	Right plot: Relative errors distribution derived from the regression formula (\ref{eq:S3bar_sigma_Qbar}).}
	\label{fig:hist_S3bar}
\end{figure}

The surfaces $\bar{S_3}=\bar{S_3}(M\times \tilde{f},\bar{Q})$ and $\bar{S_3}=\bar{S_3}(\sigma,\bar{Q})$ that best reproduces the data have functional forms 
\begin{equation}
\label{eq:S3bar_Mf_Qbar}
\bar{S_3}(M\times \tilde{f},\bar{Q})= \sum_{n=0}^{3}\sum_{m=0}^{3-n}\hat{\tilde{c}}_{nm} \  (M\times \tilde{f})^n \bar{Q}^m,
\end{equation}
\begin{equation}
\label{eq:S3bar_sigma_Qbar}
\bar{S}_3(\sigma,\bar{Q})=\sum_{n=0}^{3}\sum_{m=0}^{3-n}\hat{\tilde{d}}_{nm} \  \sigma^n\bar{Q}^m.
\end{equation}
Compared to other functional forms tested, these are the regression models with the most satisfactory statistical evaluation metric functions at the LOOCV evaluation test. The corresponding results for an indicative list of models are presented in tables (\ref{tab:S3_Mf_qbar_tab}) and (\ref{tab:S3_sigma_Qbar_tab}).
\begin{table}[!h]
	\small
	\caption{\label{tab:S3_Mf_qbar_tab} Indicative list of LOOCV evaluation metrics for the $\bar{S_3}(M\times f/c,\bar{Q}) = \sum_{n=0}^{\kappa}\sum_{m=0}^{\kappa-n}\hat{\tilde{c}}_{nm} \  (M\times \tilde{f})^n \bar{Q}^m$ parameterization.}
	\begin{ruledtabular}
		\begin{tabular}{ccccccc}
			MAE&Max Error&MSE & $d_{\text{max}}$($\%$) & MAPE ($\%$) & Exp Var & $\kappa$ \\
			\hline
			0.0650&0.759&0.0091 & 4.223 & 1.009 & 1.0 & 2\\
			\hline
			{\bf 0.0599}&{\bf 0.576}&{\bf 0.0080} & {\bf 3.161} & {\bf 0.857} & {\bf 1.0} & {\bf3}\\
			\hline
			0.0586&0.647& 0.0076&3.286 & 0.848 & 1.0 & 4\\
			\hline
			0.0579&0.690& 0.0076& 3.508 & 0.824 & 1.0 & 5\\
			\hline
			0.0578&0.664& 0.0075&3.374 & 0.822 & 1.0 & 6 \\
			\hline
			0.0578&0.657& 0.0075&3.336  & 0.822 & 1.0 & 7			
		\end{tabular}
	\end{ruledtabular}
\end{table}
\begin{table}[!h]
	\small
	\caption{\label{tab:S3_sigma_Qbar_tab} Indicative list of LOOCV evaluation metrics for the $\bar{S_3}(\sigma,\bar{Q}) = \sum_{n=0}^{\kappa}\sum_{m=0}^{\kappa-n}\hat{\tilde{d}}_{nm} \  \sigma^n\bar{Q}^m$ parameterization.}
	\begin{ruledtabular}
		\begin{tabular}{ccccccc}
			MAE&Max Error&MSE & $d_{\text{max}}$($\%$) & MAPE ($\%$) & Exp Var & $\kappa$ \\
			\hline
			0.0570&0.662& 0.0071& 3.362 & 0.850 & 1.0 & 2\\
			\hline
			{\bf 0.0560}&{\bf 0.647}& {\bf  0.0070}& {\bf 3.208} & {\bf 0.812} & {\bf 1.0} & {\bf 3}\\
			\hline
			0.0554&0.648 &0.0069 &3.290 & 0.791 & 1.0 & 4\\
			\hline
			0.0554&0.659 &0.0069 &3.349 & 0.794 & 1.0 & 5\\
			\hline
			0.0556&0.687& 0.0070&3.489 & 0.802 & 1.0 & 6\\
			\hline
			0.0553&0.660& 0.0069&3.354 & 0.788 & 1.0 & 7
		\end{tabular}
	\end{ruledtabular}
\end{table}
From the surface-fit evaluation in each case examined, the model-optimizers $\hat{\tilde{c}}_{nm}$, and $\hat{\tilde{d}}_{nm}$ are presented in tables (\ref{tab:S3_Mf_Qbar_opt_tab}), and (\ref{tab:S3_sigma_Qbar_opt_tab}).
\begin{table}[!h]
	\caption{\label{tab:S3_Mf_Qbar_opt_tab} $\hat{\tilde{c}}_{nm}$ regression optimizers for the $\bar{S_3}(M\times \tilde{f},\bar{Q})$ parameterization (\ref{eq:S3bar_Mf_Qbar}).}
	\begin{ruledtabular}
		\begin{tabular}{cccc}
			$\hat{\tilde{c}}_{00}\cdot10^{-1}$ & $\hat{\tilde{c}}_{01}$ & $\hat{\tilde{c}}_{02}\cdot10^{-2}$ & $\hat{\tilde{c}}_{03}\cdot10^{-4}$   \\
			8.1596088 & 2.0515960 & 1.3765340 & -4.1784318 \\
			\hline\hline	
			$\hat{\tilde{c}}_{10}\cdot10^{1}$ & $\hat{\tilde{c}}_{11}$ & $\hat{\tilde{c}}_{12}$ & $\hat{\tilde{c}}_{20}\cdot10^{3}$  \\
			-5.7729297 & 5.6987040 & 2.7351080 & 1.8588174 \\
			\hline\hline	
			$\hat{\tilde{c}}_{21}\cdot10^{2}$ & $\hat{\tilde{c}}_{30}\cdot10^{4}$ &  &  \\
			6.3150248& -4.5052727 &  & \\
		\end{tabular}
	\end{ruledtabular}
\end{table}
\begin{table}[!h]
	\caption{\label{tab:S3_sigma_Qbar_opt_tab} $\hat{\tilde{d}}_{nm}$ regression optimizers for the $\bar{S_3}(\sigma,\bar{Q})$ parameterization (\ref{eq:S3bar_sigma_Qbar}).}
	\begin{ruledtabular}
		\begin{tabular}{cccc}
			$\hat{\tilde{d}}_{00}$ & $\hat{\tilde{d}}_{01}$ & $\hat{\tilde{d}}_{02}\cdot10^{-4}$ & $\hat{\tilde{d}}_{03}\cdot10^{-4}$   \\
			-1.2441831 & 2.2243054 & -1.9394310 & 1.4626058 \\
			\hline\hline	
			$\hat{\tilde{d}}_{10}\cdot10^{-2}$ & $\hat{\tilde{d}}_{11}\cdot10^{-1}$ & $\hat{\tilde{d}}_{12}\cdot10^{-3}$ & $\hat{\tilde{d}}_{20}\cdot10^{-1}$  \\
			-9.2998663 & 1.6736800 & -2.2462191 & 1.0977390 \\
			\hline\hline	
			$\hat{\tilde{d}}_{21}\cdot10^{-2}$ & $\hat{\tilde{d}}_{30}\cdot10^{-2}$ &  &  \\
			-6.3552382 & -4.7295677 &  & \\
		\end{tabular}
	\end{ruledtabular}
\end{table}
The fits (\ref{eq:S3bar_Mf_Qbar}),(\ref{eq:S3bar_sigma_Qbar}) that best describe the data and the corresponding relative deviations are shown in Figs.\ref{fig:S3bar_Mf_Qbar_fig} and \ref{fig:S3_sigma_Qbar_opt_fig} respectively. 
The formulae (\ref{eq:S3bar_Mf_Qbar}), and (\ref{eq:S3bar_sigma_Qbar}) describe the data very well with relative errors $\lesssim 3.159\%$ and $\lesssim 3.198\%$ respectively. Considering the $\bar{S_3}(M\times \tilde{f},\bar{Q})$-parameterization, we note that only $19$ models have relative deviations $\sim 3\%$. These are compact stellar configurations with $\mathcal{C}\in[0.230,0.270]$, quadrupole deformation $\bar{Q}\in [1.547, 1.886]$ and high spin rates $\chi\in[0.639, 0.719]$, and $\sigma \in [0.605,0.962]$. For the $\bar{S_3}(\sigma,\bar{Q})$-parameterization on the other hand, all relative deviations are $< 3\%$. In Fig.\ref{fig:hist_S3bar}, the left plot shows the distribution of errors for the $\bar{S_3}(M\times \tilde{f},\bar{Q})$ universal formula, while the right shows the distribution for the $\bar{S_3}(\sigma,\bar{Q})$ formula.

Consequently, the aforementioned universal relations (\ref{eq:S3bar_Mf_Qbar}) and (\ref{eq:S3bar_sigma_Qbar}) can accurately reproduce the reduced octupole moment $\bar{S_3}$ when the parameters $(M,\tilde{f}, \bar{Q})$ or $(\sigma,\bar{Q})$ are known. It is worth noting that such improvements to the original 3-hair relations have been achieved before in the literature, following though a different approach and parameterization \cite{Majumder:2015kfa}. It is then a matter of specific application or convenience which relation one would prefer to use.  Similar results are obtained if someone uses the dimensionless quantities $\chi$ and $\mathcal{E}$ to parameterize rotation. The corresponding regression models have functional forms $\bar{S_3}(\chi,\bar{Q}) = \sum_{n=0}^{3}\sum_{m=0}^{3-n}\hat{\tilde{e}}_{nm} \  \chi^n\bar{Q}^m$ and $\bar{S_3}(\mathcal{E},\bar{Q}) = \sum_{n=0}^{2}\sum_{m=0}^{2-n}\hat{\tilde{f}}_{nm} \  \mathcal{E}^n\bar{Q}^m$ respectively. A Jupyter notebook implementing these optimizers, as well as the relative plots, can be found in \href{https://github.com/gregoryPapi/UR-for-rotating-NS-using-ML-}{a dedicated GitHub repository}. To conclude, it is worth noting that since the quantities $\bar{S_3}$, $\bar{Q}$ have a direct universal correlation between them, the EoS-insensitive relations involving combinations of $\bar{Q}$ with the ($M\times \tilde{f},\sigma,\chi,\mathcal{E}$) parameters essentially correspond to higher-order corrections to the $\bar{S_3}-\bar{Q}$ base relations.

\section{\label{sec:conclusions} Summary, Discussion and Conclusions}

In this work, we conduct a systematic investigation of EoS-insensitive relations for rapidly rotating NSs in $\beta$-equilibrium using an ensemble of 38 tabulated cold EoSs. The analysis is performed by utilizing statistical evaluation tests, in order to highlight possible relations between various NS structure parameters (section \ref{sec:cor_matrix}), as well as supervised machine learning methods, such as Cross-Validation (section \ref{sec:cross_validation}) and Least-Squares Regression (section \ref{sec:regression}), in order to produce the best possible functions that describe the aforementioned relations. It is clear from the current work as well, that the universality is sensitive to the choice of parameters one makes in describing the star's global properties. Furthermore, the presented analysis constitutes a systematic algorithm for discovering and tuning such universal relations.

Here we briefly summarise the main findings, presented in section \ref{sec:results} as well as in appendix \ref{app:rest_universal}.

First, we proposed a universal relation for the star's reduced quadrupole moment $\bar{Q}$ 
in terms of the stellar compactness $\mathcal{C}=M/R_{eq}$ and reduced spin $\sigma = \Omega^2R^{3}_{eq}/GM$. The resulting %
formula $\bar{Q}(\mathcal{C},\sigma)$ (\ref{eq:q_bar_c_sigma}) describes the data with accuracy $\lesssim 6.866\%$ for all EoSs considered, while it reproduces most data values with an error $\lesssim 5\%$. 

We next explored the possibility of determining the star's equatorial radius universally. For this, we investigated the inverse stellar compactness $\mathcal{K}=R_{eq}/M$ as a function of $\bar{Q}$ and the dimensionless angular momentum, $\chi=J/M^2$. Due to the large deviations introduced by rotating models near the mass shedding limit $f\sim 2 \ kHz$, we decided to limit our analysis to stellar models that rotate with frequencies $f \lesssim 1.7528 \ kHz$, a rotation rate that is still quite high and much higher than the most rapidly rotating NS detected so far. 
This particular ensemble included 7046 rapidly rotating NS models. The evaluated fitting formula  $\mathcal{K}(\chi,\bar{Q})$ (\ref{eq:K_chi_barQ}) reproduces the data with accuracy $\lesssim 6.443\%$ for all EoSs considered, while most of the models are given with a relative error that is better than $5\%$. Deviations greater than $5\%$ were introduced only by the less compact stellar models with $\mathcal{C}\in[0.096,0.124]$. 

We looked for a universal relation between the rotation frequency of the star, in the form of the dimensionless expression $\mathcal{D}=M\times \tilde{f}/\chi$, and the dimensionless angular momentum, $\chi$ and $\ln\bar{Q}$. We found a relation of the form of Eq. (\ref{eq:Mf/x_chi_lnQbar}) that is accurate with an error $\lesssim 5.220\%$ for all the data and better than $2\%$ for most of the data. 

We also used the fraction of kinetic to gravitational energy $\mathcal{E} = T/|W|$ as an additional spin parameter and looked for a relation between $\mathcal{E}$ and the quantities $\chi,\ \textrm{and} \ \ln\bar{Q}$. The estimated regression formula $\mathcal{E}(\chi,\ln\bar{Q})$ (\ref{eq:TW_chi_lnQbar}) had very good accuracy $\lesssim 2.815\%$ for all EoSs considered, while most data values are reproduced with accuracy $\lesssim 1.5\%$.

We then turned our attention to universal relations for the normalized moment of inertia $\bar{I}$. We have considered $\bar{I}$ as a function of $\bar{Q}$ and the various rotation parameters $\chi,\sigma, \ \textrm{and} \ \mathcal{E}$. We first examined the relation between $\bar{I}$ and the parameters $\chi, \bar{Q}$, that has been previously explored in the literature. The fitting formula $\bar{I}(\chi,\bar{Q})$ (\ref{eq:Ibar_chi_Qbar}) extracted reproduces the data with a relative error $\lesssim 5.515\%$, while for most of the data has an accuracy that is better than $2\%$. 
Then, we considered a relation between $\bar{I}$ and the parameters $\sigma$ and $\bar{Q}$. In this case, the evaluated fitting formula $\bar{I}(\sigma,\bar{Q})$ (\ref{eq:Ibar_sigma_Qbar}) has relative errors that are $\lesssim 6.627\%$. Compared to the $\bar{I}(\chi,\bar{Q})$-parameterization, the $\bar{I}(\sigma,\bar{Q})$ formula performs slightly worse having a little higher relative errors but always being $\lesssim 4\%$ for most of the models, while there are some models that have higher than that.
The third relation we considered is between $\bar{I}$ and the parameters $\mathcal{E}$ and $\bar{Q}$. The derived fitting formula $\bar{I}(\mathcal{E},\bar{Q})$ (\ref{eq:Ibar_T_W_Qbar}) has a relative error that is $\lesssim 5.634\%$. Most significant deviations ($>2\%$) correspond to only $50$ stellar models out of the full set. 
Lastly, we considered a relation between $\bar{I}$ and the parameter $R_{eq}\times \tilde{f}$ (another of the spin parameters) and $\bar{Q}$. The derived $\bar{I}(R_{eq}\times \tilde{f}, \bar{Q})$ (\ref{eq:Ibar_Rf_Qbar}) formula does slightly worst than the previous ones with some relative errors up to $7.301 \%$, but with most relative errors being ($\leq 5\%$).

Finally, we investigated various universal relations for the reduced octupole moment $\bar{S_3}$. We first looked for a universal relation that would express $\bar{S_3}$ in terms of $\mathcal{C}$ and one of the spin parameters, as we did for the quadrupole. However, in this case, it was not possible to find an acceptable relation of this sort.
We then revisited another of the universal relations that have already been established in the literature, i.e., the one between $\bar{S}_3$ and $\bar{Q}$, where in this work, we use $\ln {\bar{Q}}$ as a parameter instead of $\bar{Q}$ itself. The resulting fit (\ref{eq:S3bar_lnQbar}) reproduces the data with a relative error $\lesssim 4.845\%$ for all EoSs considered, with accuracy better than $4\%$ for the majority of the models. 
This is a slightly more accurate EoS-insensitive formula compared to those presented in the literature for the $\bar{S_3}=\bar{S_3}(\bar{Q})$ parameterization. In our attempt to improve the accuracy of the universal relation, we decided to explore EoS-insensitive relations that also include one of the spin parameters. 
The derived $\bar{S_3}(M\times \tilde{f},\bar{Q})$ (\ref{eq:S3bar_Mf_Qbar}) and $\bar{S_3}(\sigma,\bar{Q})$ (\ref{eq:S3bar_sigma_Qbar}) fitting functions satisfactorily described the data with relative deviations better than $3.159 \%$ and $3.198 \%$ respectively, improving on the previous result. 
Similar results hold if one uses the dimensionless quantities $\chi$ and $\mathcal{E}$ as spin parameters. 
Lastly, we make one more attempt to find a new universal relation by replacing $\bar{Q}$ with $\bar{I}$ as parameter, having therefore a relation between $\bar{S_3}$ and the parameters $\chi$ and $\bar{I}$. The evaluated fitting surface $\bar{S_3}(\chi,\bar{I})$ (\ref{eq:S3bar_chi_Ibar}) reproduces the data with relative deviations $\lesssim 9.328\%$ for all EoSs considered. However, only 30 stellar models have significant deviations ($>5\%$). Therefore the resulting relation while not one of the best possible universal relations, is not quite that bad.

This work presents several universal relations between parameters that characterize NSs. Some of them are known relations that have been revisited and verified or improved, while others are being considered for the first time. But more importantly, this work presents a systematic algorithm and a framework for improving old or producing new relations of this sort. Therefore it is essentially a proof-of-concept demonstration in addition to being an (incomplete) catalog of relations. 
Nevertheless, these relations can be useful for informing our analysis of both GW signals \cite{Agathos:2015uaa,Chatziioannou:2015uea,Paschalidis:2017qmb,Chatziioannou:2018vzf,LIGOScientific:2018cki,LIGOScientific:2018dkp,Kumar:2019xgp,Carson:2019rjx,Chatziioannou:2020pqz,Tan:2020ics,Narikawa:2021pak} from NS mergers, as well as electromagnetic observations of rotating NSs \cite{baubock2013relations,Pappas:2015mba,AlGendy:2014eua,Nattila:2017hdb,Silva:2020acr,Al-Mamun:2020vzu,Tan:2021ahl,Riley:2021pdl,Bogdanov:2021yip,Yunes:2022ldq}.  

There are several possible directions one can take from here. One could try, for example, to improve on the description of the surface of NSs \cite{Morsink:2007tv,Cadeau:2006dc,baubock2013relations,AlGendy:2014eua,Watts:2016uzu,Miller:2019cac,Riley:2019yda,Silva:2020oww}, that would be useful for modeling the pulse profiles observed by NICER. Or in a different direction, one could try to include parameters that are more directly related to nuclear physics, thus looking for universal relations connecting actual observables to ``hidden'' nuclear physics parameters that are not directly measurable. In this way, we could try to tackle the inverse problem of constraining the EoS in a more direct way with the help of universal relations. These are left for future work.

\begin{acknowledgments}
        
        This work has not received any funding. The EoSs used in this work come from the \href{https://compose.obspm.fr/home}{comPOSE} database. The notebooks with the calculations presented here can be found in the GitHub repository \href{https://github.com/gregoryPapi/UR-for-rotating-NS-using-ML-}{https://github.com/gregoryPapi/UR-for-rotating-NS-using-ML-}%
        
\end{acknowledgments}

\section*{Appendixes}

\appendix

\section{\label{app:rest_universal}Additional Universal relations}

In this appendix, we provide some additional universal relations regarding the reduced moment of inertia $\bar{I}$ and the reduced spin octupole $\bar{S}_3$. 

\subsection{\label{sec:inertiaAppendix}Additional universal relations for the normalized moment of inertia $\bar{I}$}

We start with the relations %
for the moment of inertia. First, we consider the dimensionless spin $\sigma$ as the parameter that characterizes rotation and investigate a relation that links $\bar{I}$ with the quantities $\sigma, \bar{Q}$. The ${\bar{I}}=\bar{I}(\sigma,\bar{Q})$ representation that best reproduces the data has the functional form 
\begin{equation}
\label{eq:Ibar_sigma_Qbar}
\bar{I}(\sigma,\bar{Q})=\sum_{n=0}^{4}\sum_{m=0}^{4-n}\hat{e}_{nm} \ \sigma^n \ \bar{Q}^m.
\end{equation}
This is the regression model with the best statistical evaluation metric functions at LOOCV. 
From the surface-fit evaluation, the fitting optimizers $\hat{e}_{nm}$ are presented in the table (\ref{tab:Ibar_sigma_qbar_optimizers}).
\begin{table}[!h]
	\caption{\label{tab:Ibar_sigma_qbar_optimizers} $\hat{e}_{nm}$ regression optimizers for the $\bar{I}(\sigma,\bar{Q})$ parameterization (\ref{eq:Ibar_sigma_Qbar}).}
	\begin{ruledtabular}
		\begin{tabular}{cccc}
			$\hat{e}_{00}$ & $\hat{e}_{01}$ & $\hat{e}_{02}\cdot10^{-2}$ & $\hat{e}_{03}\cdot10^{-3}$   \\
			2.1883709 & 1.9979686 & -6.7253986 & 8.4152447   \\
			\hline\hline	
			$\hat{e}_{04}\cdot10^{-4}$ & $\hat{e}_{10}$ & $\hat{e}_{11}$ & $\hat{e}_{12}\cdot10^{-2}$  \\
			-2.5907891 & -2.7558192 & 2.0937835 & -9.1147011 \\
			\hline\hline	
			$\hat{e}_{13}\cdot10^{-3}$ & $\hat{e}_{20}\cdot10^{-1}$ & $\hat{e}_{21}\cdot10^{-1}$ & $\hat{e}_{22}\cdot10^{-2}$ \\
			8.1375930 & 8.6762800 & 3.1569363 & -1.5332768 \\
			\hline\hline
			$\hat{e}_{30}$ & $\hat{e}_{31}\cdot10^{-1}$ & $\hat{e}_{40}$ &  \\
			-1.0039675 & -7.9469215 & 1.0162539  & 
		\end{tabular}		
	\end{ruledtabular}
\end{table}

\begin{figure}[!ht]
	\includegraphics[width=0.26\textwidth]{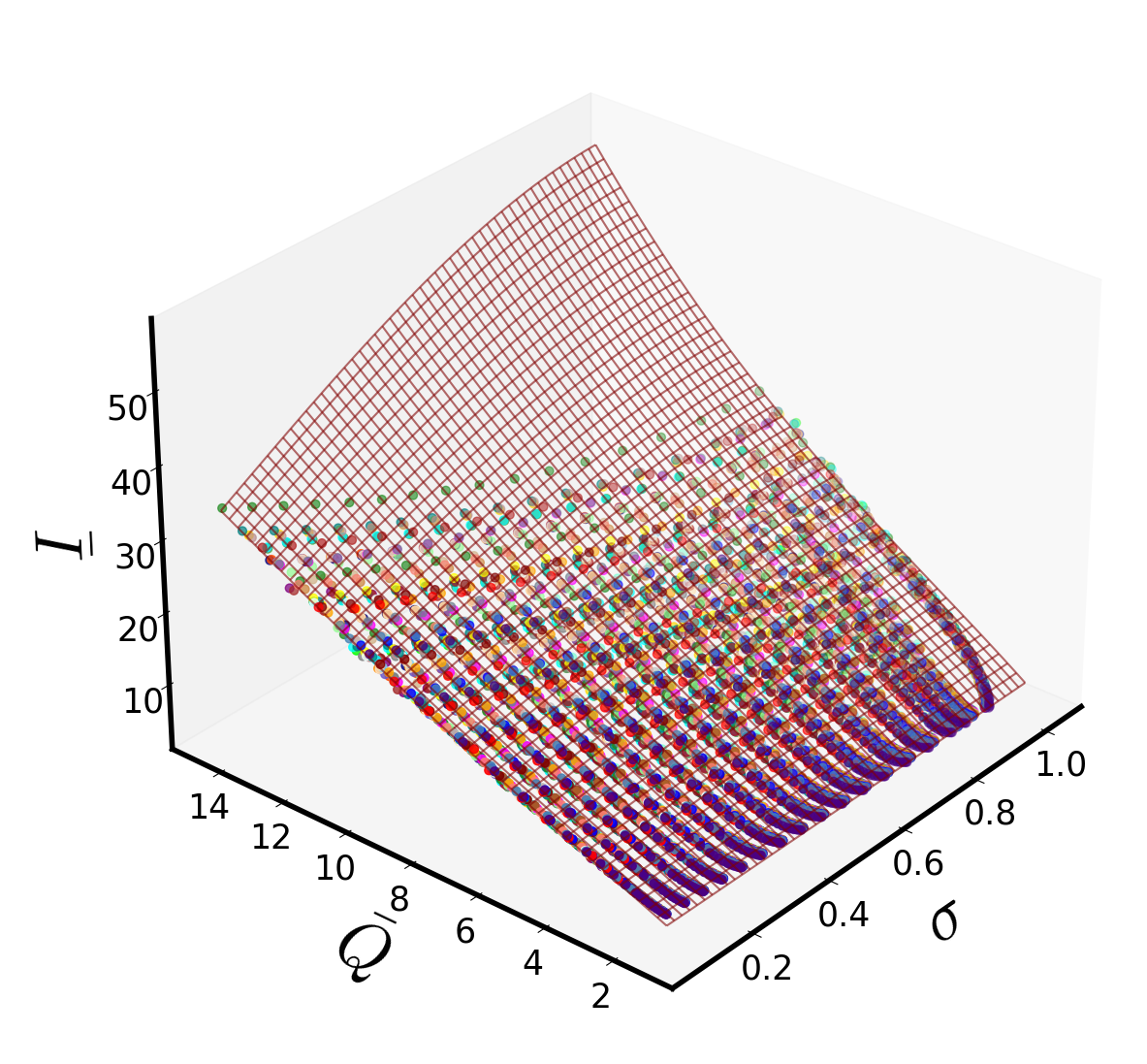}
	\includegraphics[width=0.26\textwidth]{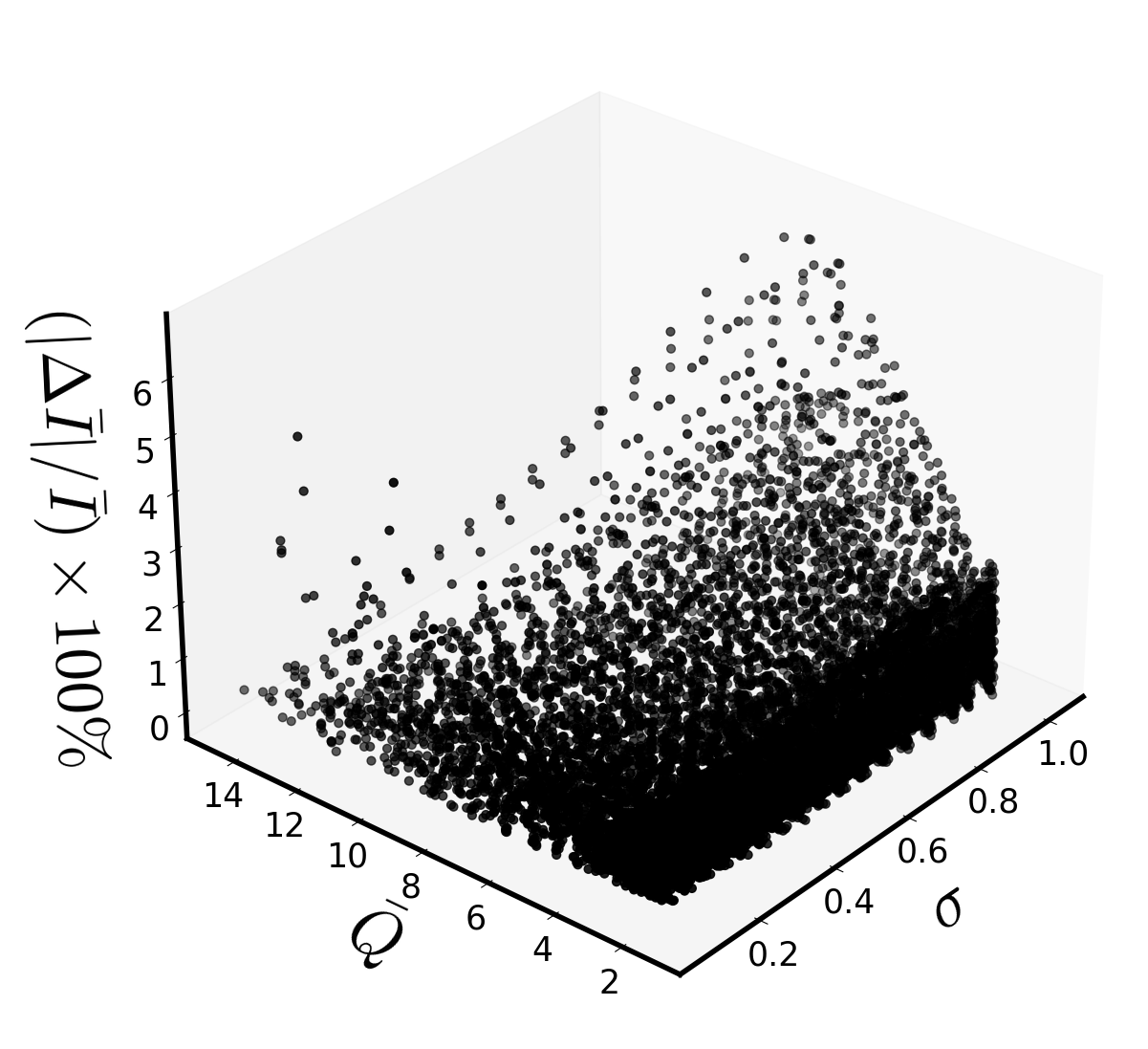}
	\caption{$\bar{I}$ as a function of the dimensionless parameters $\sigma,\ \bar{Q}$ and relative error distribution. The surface corresponds to the regression fitting formula (\ref{eq:Ibar_sigma_Qbar}). The relative errors are given as ($100\%(|\Delta{\bar{I}}|/{\bar{I}})=100\%|{\bar{I}_{fit}}-{\bar{I}}|/{\bar{I}}$).} %
	\label{fig:Ibar_sigma_Qbar_fig}
\end{figure}

The surface (\ref{eq:Ibar_sigma_Qbar}) and the corresponding relative errors are presented in Fig.\ref{fig:Ibar_sigma_Qbar_fig}.
The regression formula (\ref{eq:Ibar_sigma_Qbar}) corresponds to a well-behaved EoS-insensitive relation which gives good results for all the rotating models in equilibrium considered, reproducing most data with an error $\lesssim 4\%$.

\begin{figure}[!ht]
	\includegraphics[width=0.25\textwidth]{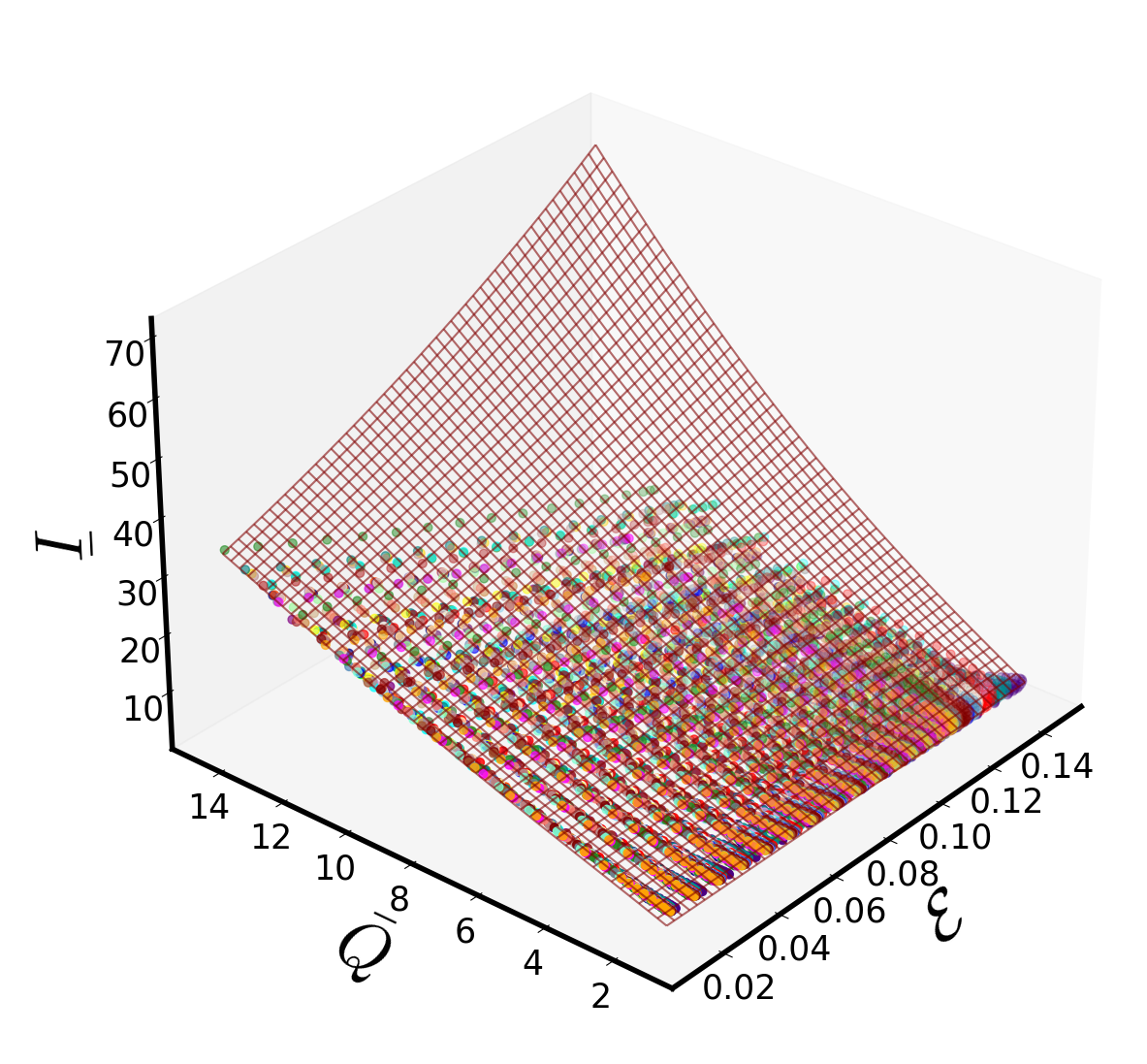}
	\includegraphics[width=0.25\textwidth]{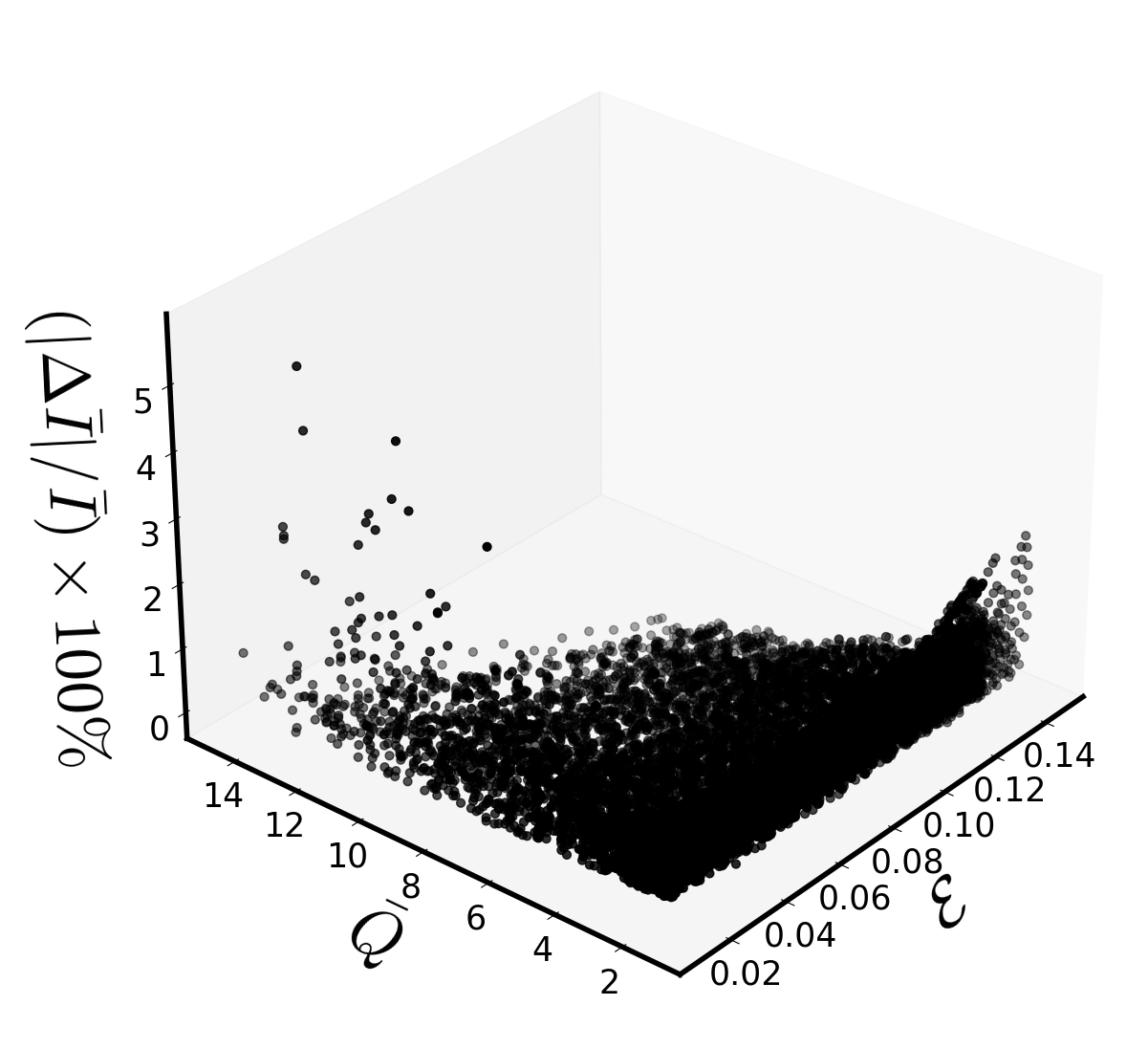}
	\caption{${\bar{I}}$ as a function of the parameters $\ \mathcal{E},\bar{Q}$ and relative errors. The surface corresponds to the regression formula (\ref{eq:Ibar_T_W_Qbar}). The relative errors are given as ($100\%(|\Delta {\bar{I}}|/{\bar{I}})=100\%|{\bar{I}}_{fit}-{\bar{I}}|/{\bar{I}}$).}%
		\label{fig:Ibar_T_W_Qbar_fig}
\end{figure}

In addition to the relations presented so far, it would also be interesting to explore a relation that connects the normalized moment of inertia $\bar{I}$ with the reduced quadrupole moment $\bar{Q}$ and the dimensionless fraction of kinetic to gravitational energy $\mathcal{E}=T/|W|$. The $\bar{I}=\bar{I} (\mathcal{E},\bar{Q})$ parameterization that best describes the data has the functional form
\begin{equation}
\label{eq:Ibar_T_W_Qbar}
\bar{I}(\mathcal{E},\bar{Q})=\sum_{n=0}^{4}\sum_{m=0}^{4-n}\hat{f}_{nm} \ \mathcal{E}^n \ {\bar{Q}}^m.
\end{equation}
From the surface-fit evaluation, the fitting optimizers are given in the table (\ref{tab:Ibar_tw_qbar_optimizers}).
\begin{table}[!h]
	\caption{\label{tab:Ibar_tw_qbar_optimizers} $\hat{f}_{nm}$ regression optimizers for the $\bar{I}(\mathcal{E},\bar{Q})$ parameterization (\ref{eq:Ibar_T_W_Qbar}).}
	\begin{ruledtabular}
		\begin{tabular}{cccc}
			$\hat{f}_{00}$ & $\hat{f}_{01}$ & $\hat{f}_{02}\cdot10^{-1}$ & $\hat{f}_{03}\cdot10^{-2}$   \\
			2.0428593 & 2.1177336 & -1.0563791 & 1.2625985   \\
			\hline\hline	
			$\hat{f}_{04}\cdot10^{-4}$ & $\hat{f}_{10}\cdot10^{1}$ & $\hat{f}_{11}$ & $\hat{f}_{12}\cdot10^{-2}$  \\
			-4.026470 & -1.1344300 & 9.3474163 & -5.0090406  \\
			\hline\hline	
			$\hat{f}_{13}\cdot10^{-2}$ & $\hat{f}_{20}\cdot10^{1}$ & $\hat{f}_{21}\cdot10^{1}$ & $\hat{f}_{22}$ \\
			1.771390401 & 4.2944226 & -1.579430 & 3.0508174 \\
			\hline\hline
			$\hat{f}_{30}\cdot10^{2}$ & $\hat{f}_{31}\cdot10^{1}$ & $\hat{f}_{40}\cdot10^{3}$ &  \\
			-5.312067774 & 7.305432 & 1.8509948 & 
		\end{tabular}
		
	\end{ruledtabular}
\end{table}
The surface (\ref{eq:Ibar_T_W_Qbar}) and the corresponding relative errors are presented in Fig.\ref{fig:Ibar_T_W_Qbar_fig}.

We should note that the relative errors between the fit (\ref{eq:Ibar_T_W_Qbar}) and the observed $\bar{I}$ are $\lesssim 5.634\%$. The largest relative deviations $\geq 2\%$ correspond to only $50$ models. 
Therefore the regression formula (\ref{eq:Ibar_T_W_Qbar}) is an accurate EoS-insensitive relation that gives good results for all the rotating models considered, reproducing most data with an error $\lesssim 2\%$.

\begin{figure}[!ht]
	\includegraphics[width=0.25\textwidth]{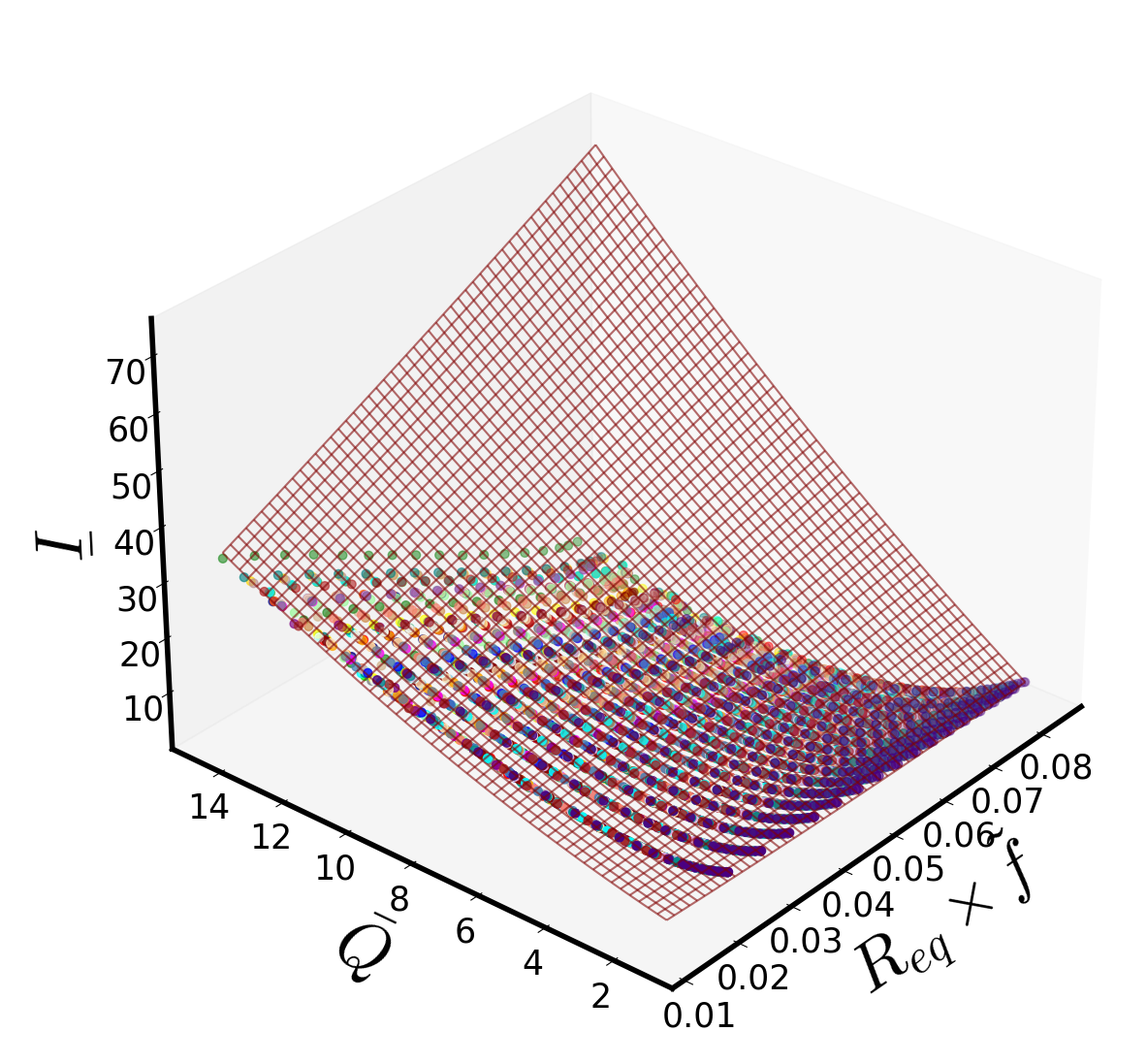}
	\includegraphics[width=0.25\textwidth]{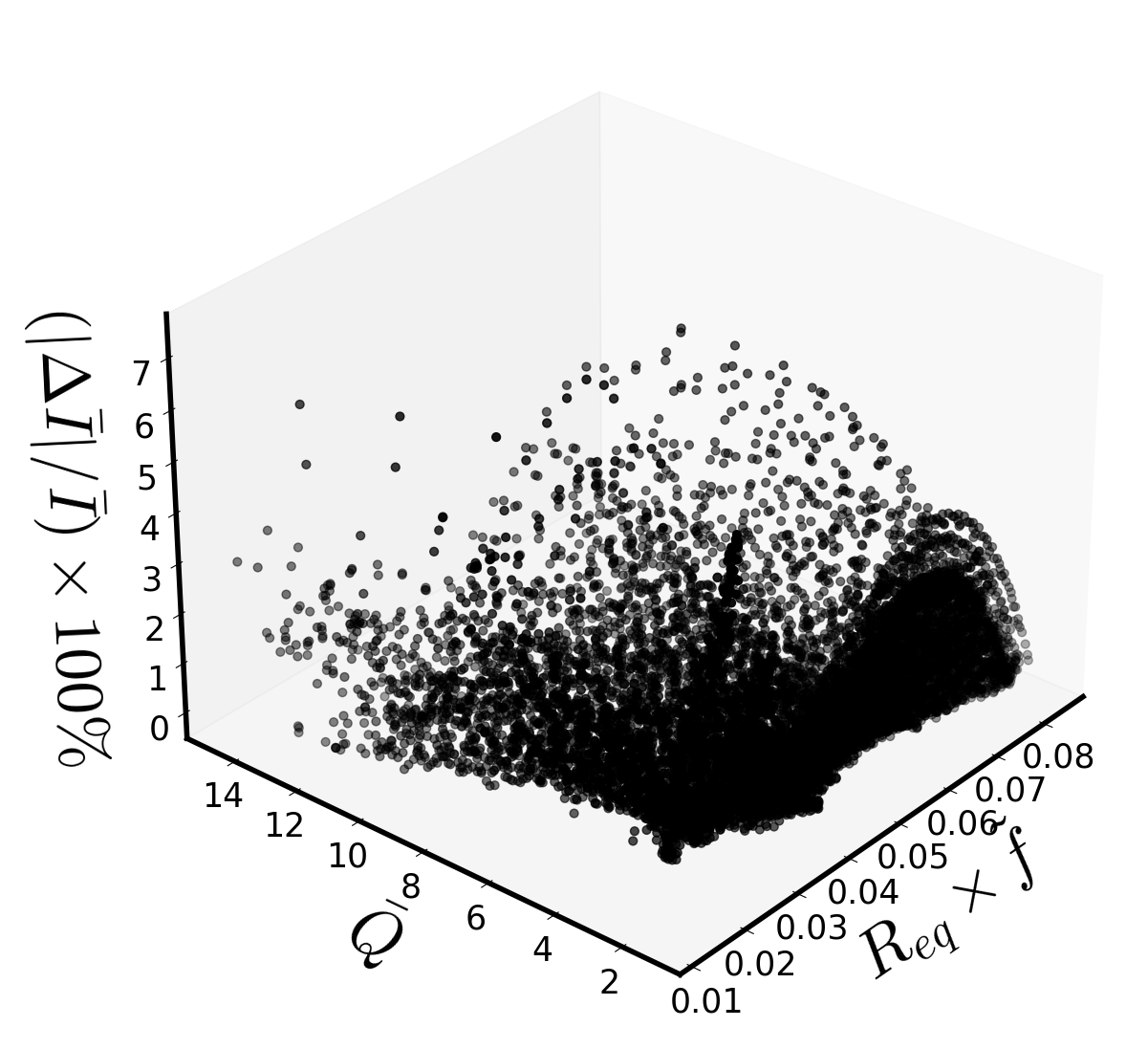}
	\caption{${\bar{I}}$ as a function of the parameters $R_{eq}\times \tilde{f},\bar{Q}$ and relative errors. The surface corresponds to the fit (\ref{eq:Ibar_T_W_Qbar}). The relative errors given as ($100\%(|\Delta {\bar{I}}|/{\bar{I}})=100\%|{\bar{I}}_{fit}-{\bar{I}}|/{\bar{I}}$).} %
		\label{fig:Ibar_Rf_Qbar_fig}
\end{figure}

One last relation concerning  $\bar{I}$ that we are investigating is the one that connects it to the parameters $R_{eq}\times \tilde{f}$ and $\bar{Q}$. 
We remind that the quantity $R_{eq}\times \tilde{f}$ corresponds to one of the possible rotation parameterizations for the star. The $\bar{I}(R_{eq}\times \tilde{f},\bar{Q})$ surface that optimally describes the data has the functional form 
\begin{equation}
\label{eq:Ibar_Rf_Qbar}
\bar{I}(R_{eq}\times \tilde{f},\bar{Q})= \sum_{n=0}^{2}\sum_{m=0}^{2-n}\hat{g}_{nm} \  (R_{eq}\times \tilde{f})^n \bar{Q}^m.
\end{equation}
From the surface-fit evaluation, the polynomial function's (\ref{eq:Ibar_Rf_Qbar}) (best fit) optimizers $\hat{g}_{nm}$ are presented in table (\ref{tab:I_Rf_Ibar_opt_tab}).
\begin{table}[!h]
	\caption{\label{tab:I_Rf_Ibar_opt_tab} $\hat{g}_{nm}$ regression optimizers for the $\bar{I}(R_{eq}\times \tilde{f},\bar{Q})$ parameterization (\ref{eq:Ibar_Rf_Qbar}).}
	\begin{ruledtabular}
		\begin{tabular}{cccc}
			$\hat{g}_{00}$ & $\hat{g}_{01}\cdot10^{-1}$ & $\hat{g}_{02}\cdot10^{-2}$ & $\hat{g}_{10}\cdot10^{2}$   \\
			5.3344693 &  4.7861124 & 7.6686702 & -1.0442463    \\
			\hline\hline	
			$\hat{g}_{11}\cdot10^{1}$ & $\hat{a}_{20}\cdot10^{2}$ &  &   \\
			 3.9934412 & 4.3116877 &  &  \\
		\end{tabular}
		
	\end{ruledtabular}
\end{table}
The surface fit, and the corresponding relative errors are presented in Fig.\ref{fig:Ibar_Rf_Qbar_fig}.
It is evident that the EoS-insensitive formula (\ref{eq:Ibar_Rf_Qbar}) reproduces the vast majority of the data values with an error $\lesssim 5\%$, which makes it a quite accurate relation.


\subsection{Additional universal relations for the reduced spin octupole moment $\bar{S_3}$}
\label{sec:Universal Octupole Moment appendix}


Here, we explore another slightly different parameterization of the spin octupole moment, in terms of the moment of inertia instead of the quadrupole moment.

Specifically we look into a relation of the form $\bar{S_3} = \bar{S_3}(\chi,\bar{I})$. The complete analysis is performed for the whole sample of rapidly rotating stellar models included in our EoS catalog. The surface that best describes the data has a functional form
\begin{equation}
\label{eq:S3bar_chi_Ibar}
\bar{S}_3(\chi,\bar{I})=\sum_{n=0}^{4}\sum_{m=0}^{4-n}\hat{\tilde{a}}_{nm} \  \chi^n\bar{I}^m.
\end{equation}
Again, we choose not to go too high in the polynomial order of the fitting function and set the maximum to $\kappa = 4$. The fitting optimizers $\hat{\tilde{a}}_{nm}$ for the surface $\bar{S_3}=\bar{S_3}(\chi,\bar{I})$ are given in the table (\ref{tab:S3_chi_Ibar_opt_tab}).
\begin{table}[!h]
	\caption{\label{tab:S3_chi_Ibar_opt_tab} $\hat{\tilde{a}}_{nm}$ regression optimizers for the $\bar{S_3}(\chi,\bar{I})$ parameterization (\ref{eq:S3bar_chi_Ibar}).}
	\begin{ruledtabular}
		\begin{tabular}{cccc}
			$\hat{\tilde{a}}_{00}$ & $\hat{\tilde{a}}_{01}$ & $\hat{\tilde{a}}_{02}\cdot10^{-2}$ & $\hat{\tilde{a}}_{03}\cdot10^{-3}$   \\
			-4.9383743 & 1.1325862 & 2.1599976 & -1.1755146    \\
			\hline\hline	
			$\hat{\tilde{a}}_{04}\cdot10^{-5}$ & $\hat{\tilde{a}}_{10}\cdot10^{1}$ & $\hat{\tilde{a}}_{11}\cdot10^{-1}$ & $\hat{\tilde{a}}_{12}\cdot10^{-2}$  \\
			1.4665557 & 1.1619748 & -5.7118845 & 1.7394069 \\
			\hline\hline	
			$\hat{\tilde{a}}_{13}\cdot10^{-5}$ & $\hat{\tilde{a}}_{20}\cdot10^{1}$ & $\hat{\tilde{a}}_{21}$ & $\hat{\tilde{a}}_{22}\cdot10^{-3}$ \\
			1.6073925 & -2.8086884 & -1.0381803 & -6.2358287 \\
			\hline\hline
			$\hat{\tilde{a}}_{30}\cdot10^{1}$ & $\hat{\tilde{a}}_{31}\cdot10^{-1}$ & $\hat{\tilde{a}}_{40}\cdot10^{1}$ &  \\
			4.6577272 & 7.0553777 & -2.6913268 & 
		\end{tabular}
		
	\end{ruledtabular}
\end{table}

\begin{figure}[!h]
	\includegraphics[width=0.26\textwidth]{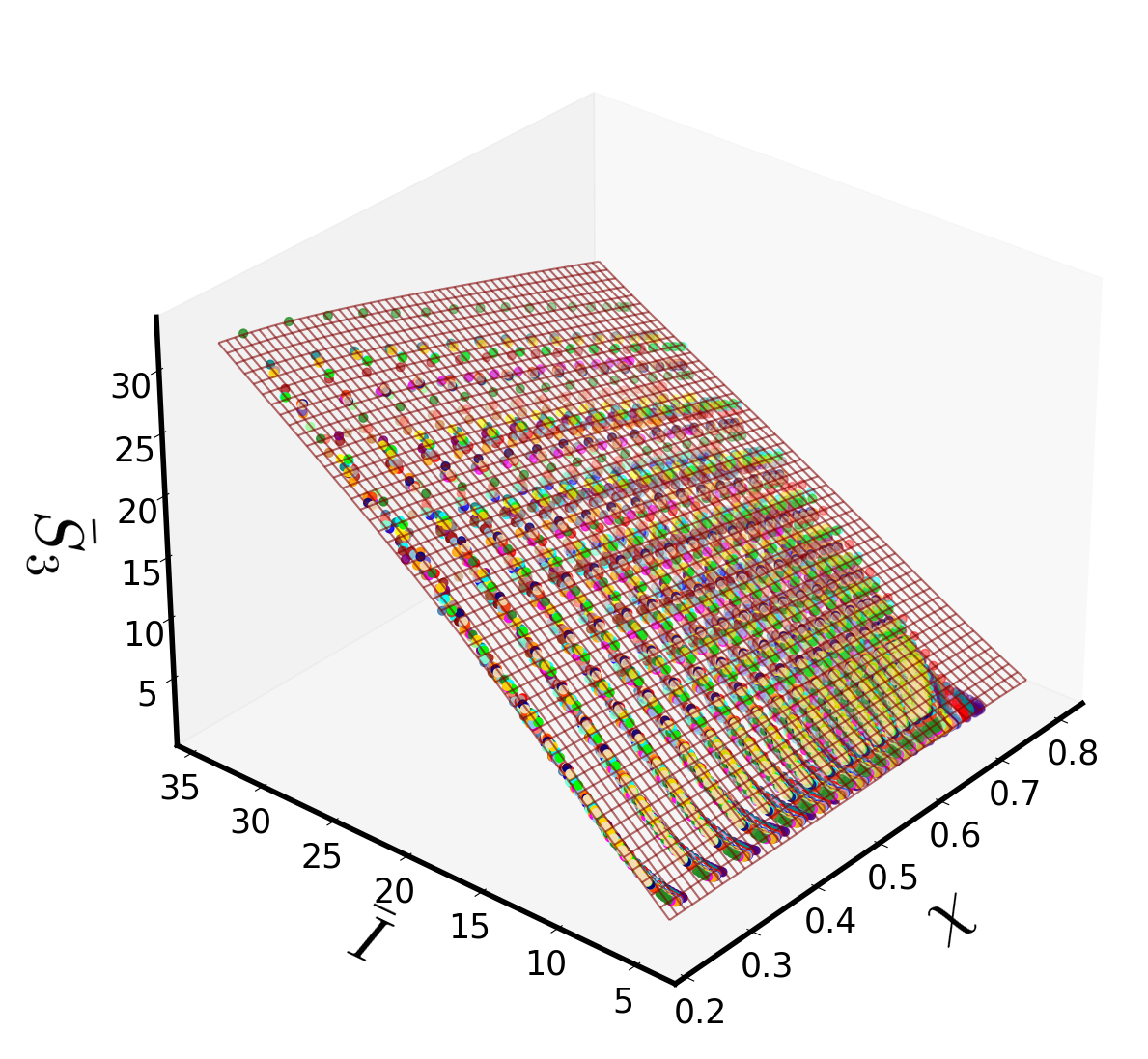}
	\includegraphics[width=0.26\textwidth]{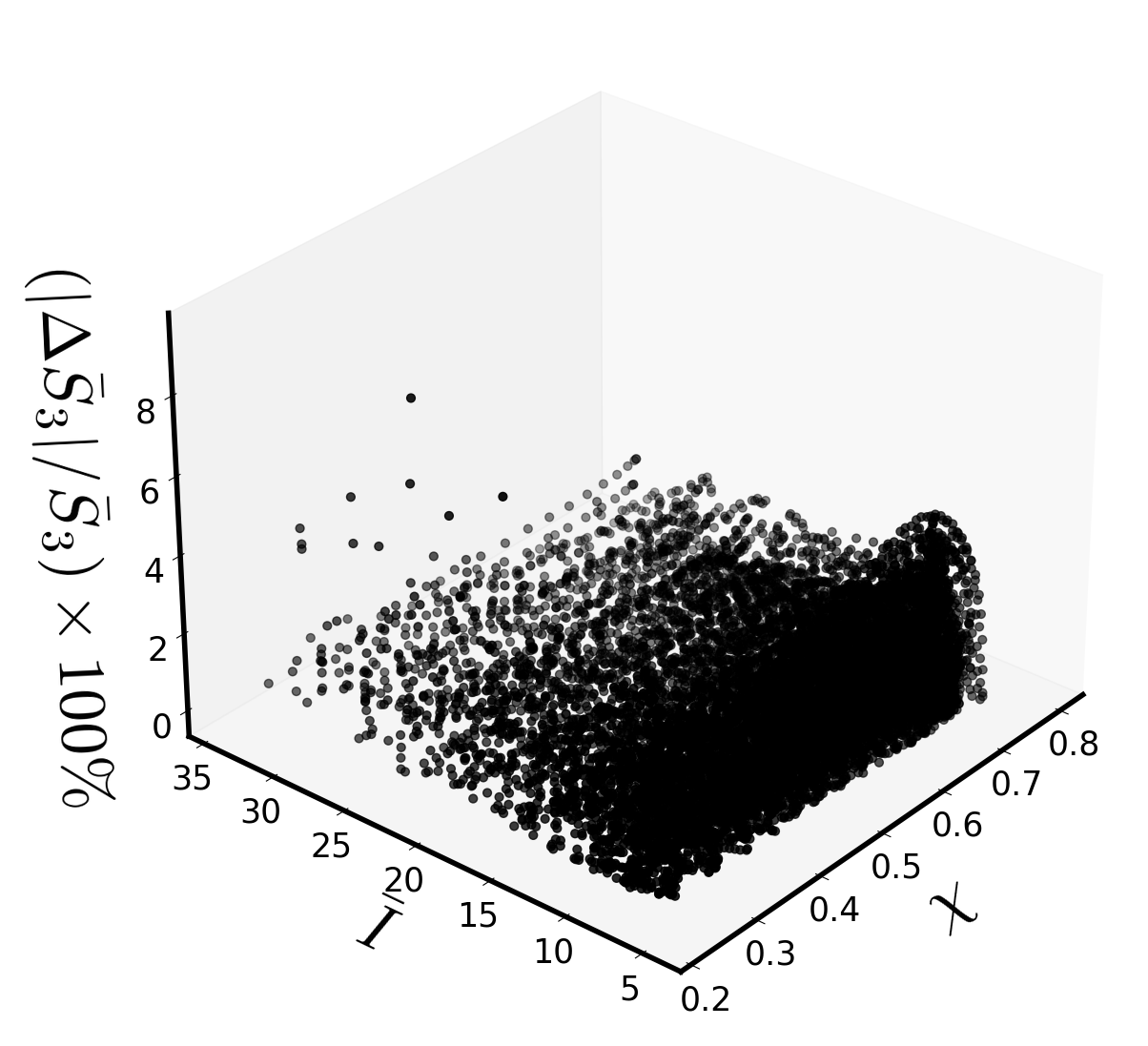}
	\caption{${\bar{S_3}}$ as a function of $\ \chi,\bar{I}$ and relative error distribution. The surface corresponds to the formula (\ref{eq:S3bar_chi_Ibar}). The relative errors are given given as ($100\%(|\Delta {\bar{S_3}}|/{\bar{S_3}})=100\%|{\bar{S}}_{3,fit}-{\bar{S_3}}|/{\bar{S_3}}$).}
	\label{fig:S3_chi_Ibar_opt_fig}
\end{figure}

The surface (\ref{eq:S3bar_chi_Ibar}) that best reproduces the data values and the corresponding relative errors are presented in Fig.\ref{fig:S3_chi_Ibar_opt_fig}.

\begin{figure}[!h]
	\includegraphics[width=0.28\textwidth]{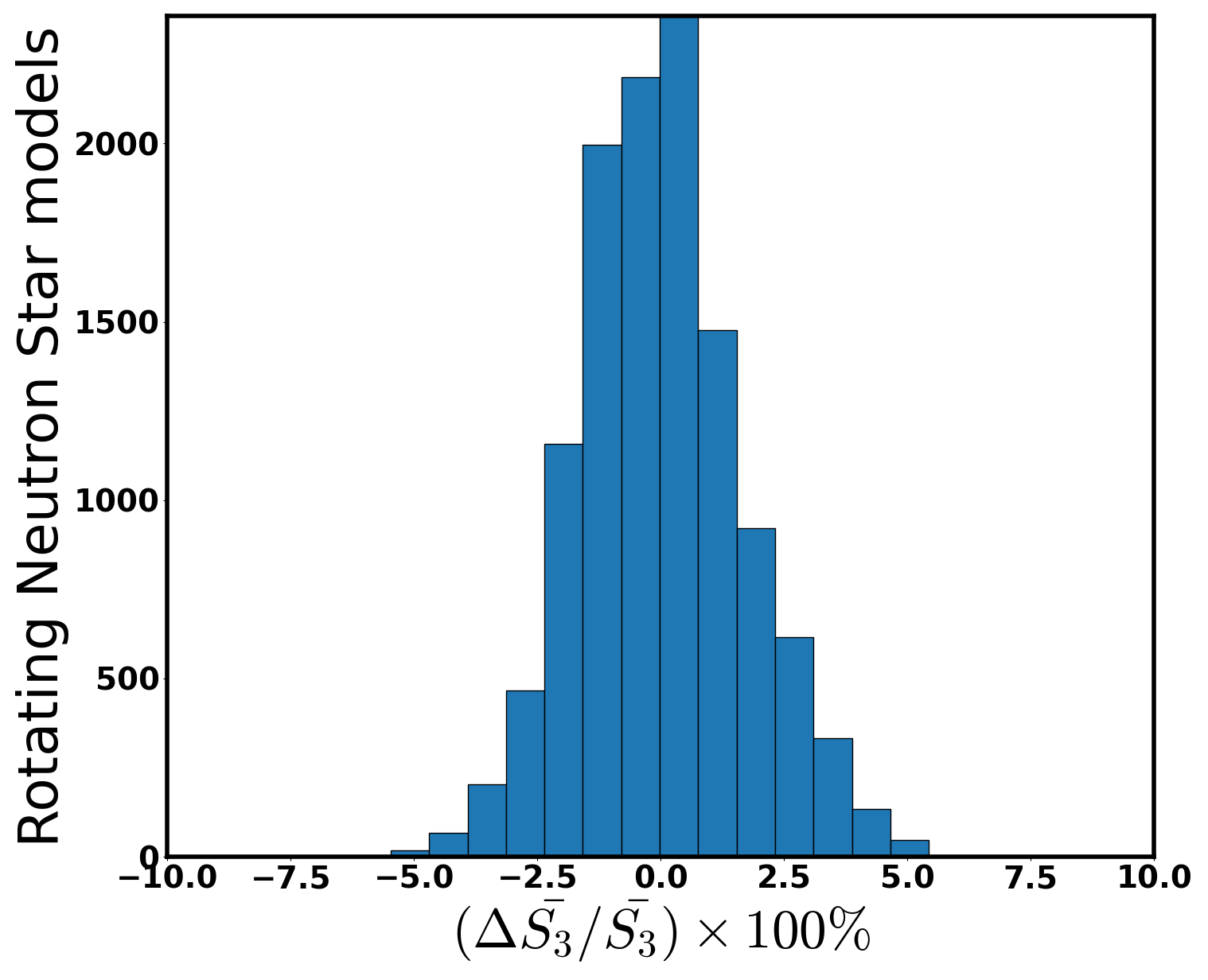}
	\caption{Distribution of rotating NS models vs relative errors for the regression formula (\ref{eq:S3bar_chi_Ibar}).}
	\label{fig:hist_S3_bar_chi_barI_fig}
\end{figure}

The relative deviations between the fit (\ref{eq:S3bar_chi_Ibar}) and the observed $\bar{S}_3$ are $\lesssim 9.328\%$ for all EoSs and NS models considered, with only 30 models out of the total 11983 having relative deviations $\geq 5\%$. These particular models have $\mathcal{C}\in [0.108, 0.290]$ and $\chi\in[0.230, 0.667]$. 
In fig.\ref{fig:hist_S3_bar_chi_barI_fig}, we present the rotating models' distribution concerning the relative errors $100\% \times(\Delta {\bar{S}_3}/{\bar{S}_3})$ derived.
From Fig (\ref{fig:hist_S3_bar_chi_barI_fig}), it is evident that the regression formula (\ref{eq:S3bar_chi_Ibar}) corresponds to a good EoS-insensitive relation that gives good results for the vast majority of the models, reproducing most of the data with an error $\lesssim 5\%$.

\section{\label{app:eos_tables}Equation of state tables}
\newpage

\begin{widetext}

\begin{table}[h]
	\caption{\label{tab:hadronic} Hadronic cold EoS models.}
	\begin{ruledtabular}
		\begin{tabular}{ccccccc}
			EoS & Model & Matter & \texttt{$M_{\max }\left[M_{\odot}\right]$} &  \texttt{$R_{M_{\max }}[\mathrm{km}]$} &  \texttt{$R_{1.4 M_{\odot}}[\mathrm{km}]$}& References\\ 
			
			\hline
			\text{RG(SLY2)}& \text{EI-CEF-Scyrme}& \text{$n,p,e,\mu$} & 2.06& 10.06 & 11.79&\cite{chabanat_skyrme_1998,danielewicz_symmetry_2009,gulminelli_unified_2015} \\
			\hline
			\text{RG(SKb)}& \text{EI-CEF-scyrme}& \text{$n,p,e,\mu$} & 2.20& 10.58 & 12.21& \cite{danielewicz_symmetry_2009,gulminelli_unified_2015,kohler_skyrme_1976}\\
			\hline
			
			\text{RG(SkMp)}& \text{EI-CEF-scyrme}& \text{$n,p,e,\mu$} & 2.11& 10.60 & 12.50& \cite{bennour_charge_1989,danielewicz_symmetry_2009,gulminelli_unified_2015}  \\
			\hline
			\text{RG(SLY9)}& \text{EI-CEF-scyrme}& \text{$n,p,e,\mu$} & 2.16& 10.65 & 12.47& \cite{chabanat_skyrme_1998,danielewicz_symmetry_2009,gulminelli_unified_2015}  \\
			\hline
			\text{RG(SkI3)}& \text{EI-CEF-scyrme}& \text{$n,p,e,\mu$} & 2.25& 11.34 & 13.55& \cite{danielewicz_symmetry_2009,gulminelli_unified_2015,reinhard_nuclear_1995}  \\
			\hline
			\text{RG(KDE0v)}& \text{EI-CEF-scyrme}& \text{$n,p,e,\mu$} & 1.97& 9.62 & 11.42& \cite{agrawal_determination_2005,danielewicz_symmetry_2009,gulminelli_unified_2015}  \\
			\hline
			\text{RG(SK255)}& \text{EI-CEF-scyrme}& \text{$n,p,e,\mu$} & 2.15& 10.84 & 13.15& \cite{agrawal_determination_2005,danielewicz_symmetry_2009,gulminelli_unified_2015}  \\
			\hline
			\text{RG(Rs)}& \text{EI-CEF-scyrme}& \text{$n,p,e,\mu$} & 2.12& 10.76 & 12.93& \cite{danielewicz_symmetry_2009,friedrich_skyrme-force_1986,gulminelli_unified_2015}  \\
			\hline
			\text{RG(SkI5)}& \text{EI-CEF-scyrme}& \text{$n,p,e,\mu$} & 2.25& 11.47 & 14.08& \cite{danielewicz_symmetry_2009,gulminelli_unified_2015,reinhard_nuclear_1995}  \\
			\hline
			\text{RG(SKa)}& \text{EI-CEF-scyrme}& \text{$n,p,e,\mu$} & 2.22& 10.82 & 12.92& \cite{danielewicz_symmetry_2009,gulminelli_unified_2015,kohler_skyrme_1976}  \\
			\hline
			\text{RG(SkOp) 	}& \text{EI-CEF-scyrme}& \text{$n,p,e,\mu$} & 1.98& 10.16 & 12.13& \cite{danielewicz_symmetry_2009,gulminelli_unified_2015,reinhard_nuclear_1995}  \\
			\hline
			\text{RG(SLY230a)}& \text{EI-CEF-scyrme}& \text{$n,p,e,\mu$} & 2.11& 10.18 & 11.83& \cite{chabanat_skyrme_1998,danielewicz_symmetry_2009,gulminelli_unified_2015}  \\
			\hline
			\text{RG(SKI2)}& \text{EI-CEF-scyrme}& \text{$n,p,e,\mu$} & 2.17& 11.25 & 13.48& \cite{danielewicz_symmetry_2009,gulminelli_unified_2015,reinhard_nuclear_1995}  \\
			\hline
			\text{RG(SkI4)}& \text{EI-CEF-scyrme}& \text{$n,p,e,\mu$} & 2.18& 10.66 & 12.38& \cite{danielewicz_symmetry_2009,gulminelli_unified_2015,reinhard_nuclear_1995}  \\
			\hline
			\text{RG(SkI6)}& \text{EI-CEF-scyrme}& \text{$n,p,e,\mu$} & 2.20& 10.71 & 12.49& \cite{danielewicz_symmetry_2009,gulminelli_unified_2015,reinhard_nuclear_1995}  \\
			\hline
			\text{RG(KDE0v1)}& \text{EI-CEF-scyrme}& \text{$n,p,e,\mu$} & 1.98& 9.71 & 11.63& \cite{agrawal_determination_2005,danielewicz_symmetry_2009,gulminelli_unified_2015}  \\
			\hline
			\text{RG(SK272)}& \text{EI-CEF-scyrme}& \text{$n,p,e,\mu$} & 2.24& 11.20 & 13.32& \cite{agrawal_nuclear_2003,danielewicz_symmetry_2009,gulminelli_unified_2015}  \\
			\hline
			\text{RG(SLY4)}& \text{EI-CEF-scyrme}& \text{$n,p,e,\mu$} & 2.06& 10.02 & 11.70& \cite{chabanat_skyrme_1998,danielewicz_symmetry_2009,gulminelli_unified_2015}  \\
			\hline 
			\text{GDTB(DDH$\delta$)}& \text{RMF} & \text{$n,p,e$} & 2.16& 11.19 & 12.58 &\cite{douchin_unified_2001,gaitanos_lorentz_2004,grill_equation_2014}  \\
			\hline
			\text {DS(CMF)-2}& \text{SU(3)-RMF}& \text{$n,p,e$} & 2.13& 11.96 & 13.70 &\cite{bennour_charge_1989,gulminelli_unified_2015,dexheimer_gw190814_2021,dexheimer_novel_2010,dexheimer_proto-neutron_2008,dexheimer_reconciling_2015,dexheimer_tabulated_2017} \\
			\hline
			\text {DS(CMF)-4}& \text{SU(3)-RMF}& \text{$n,p,e$} & 2.05& 11.60 & 13.26 &\cite{bennour_charge_1989,gulminelli_unified_2015,dexheimer_gw190814_2021,dexheimer_novel_2010,dexheimer_proto-neutron_2008,dexheimer_reconciling_2015,dexheimer_tabulated_2017} \\
			\hline
			\text {DS(CMF)-6}& \text{SU(3)-RMF}& \text{$n,p,e$} & 2.11& 11.58 & 13.30 &\cite{bennour_charge_1989,gulminelli_unified_2015,dexheimer_gw190814_2021,dexheimer_novel_2010,dexheimer_proto-neutron_2008,dexheimer_reconciling_2015,dexheimer_tabulated_2017}\\
			\hline
			\text {DS(CMF)-8}& \text{SU(3)-RMF}& \text{$n,p,e,\Delta^{-}$} & 2.09& 11.59 & 13.30 &\cite{bennour_charge_1989,gulminelli_unified_2015,dexheimer_gw190814_2021,dexheimer_novel_2010,dexheimer_proto-neutron_2008,dexheimer_reconciling_2015,dexheimer_tabulated_2017} \\
			\hline 
			\text{BL(chiral)\_2018}& \text{chPT-BBG-BHF}& \text{$n,p,e,\mu$} & 2.08& 10.26 & 12.31 &\cite{bombaci_equation_2018,douchin_unified_2001}  \\
		\end{tabular}
	\end{ruledtabular}
\end{table}
\begin{table}[h]
	\caption{\label{tab:hyperonic} Hyperonic cold EoS models.}
	\begin{ruledtabular}
		\begin{tabular}{ccccccc}
			EoS & Model & Matter & \texttt{$M_{\max }\left[M_{\odot}\right]$} &  \texttt{$R_{M_{\max }}[\mathrm{km}]$} &  \texttt{$R_{1.4 M_{\odot}}[\mathrm{km}]$}& References\\ 
			
			\hline
			\text{OPGR(DDH$\delta$ Y4)}& \text{RMF}& \text{$n,p,e,H=[\Lambda,\Xi^{-}]$} & 2.05& 11.26 & 12.58& \cite{douchin_unified_2001,gaitanos_lorentz_2004,grill_equation_2014,oertel_hyperons_2015}  \\
			\hline 
			\text{OPGR(GM1Y5)}& \text{RMF}& \text{$n,p,e,H=[\Lambda,\Xi^{-},\Xi^0]$} & 2.12& 12.31 & 13.78 &\cite{glendenning_reconciliation_1991,douchin_unified_2001,oertel_hyperons_2015}  \\
			\hline 
			\text{OPGR(GM1Y6)}& \text{RMF}& \text{$n,p,e,H=[\Lambda,\Xi^{-},\Xi^0]$} & 2.29& 12.13 & 13.78& \cite{douchin_unified_2001,glendenning_reconciliation_1991,oertel_hyperons_2015}  \\
			\hline 
			\text {DNS}& \text{SU(3)-CMF}& \text{$n,p,e,\mu,H=[\Lambda,\Sigma^{-}]$} & 2.10& 12.00 &13.58 & \cite{dexheimer_proto-neutron_2008,dexheimer_reconciling_2015,dexheimer_tabulated_2017,schurhoff_neutron_2010}  \\
			\hline
			\text {DS(CMF)-1}& \text{SU(3)-CMF}& \text{$n,p,e,H=[\Lambda,\Sigma^{-}]$} & 2.07& 11.88 & 13.57&\cite{bennour_charge_1989,gulminelli_unified_2015,dexheimer_gw190814_2021,dexheimer_novel_2010,dexheimer_proto-neutron_2008,dexheimer_reconciling_2015,dexheimer_tabulated_2017} \\
			\hline
			\text {DS(CMF)-3}& \text{SU(3)-CMF}& \text{$n,p,e,H=[\Lambda,\Sigma^{-}]$} & 2.00& 11.56 & 13.15&\cite{bennour_charge_1989,gulminelli_unified_2015,dexheimer_gw190814_2021,dexheimer_novel_2010,dexheimer_proto-neutron_2008,dexheimer_reconciling_2015,dexheimer_tabulated_2017} \\
			\hline
			\text {DS(CMF)-5}& \text{SU(3)-CMF}& \text{$n,p,e,H=[\Lambda,\Sigma^{-}]$} & 2.07& 11.43 & 13.20&\cite{bennour_charge_1989,gulminelli_unified_2015,dexheimer_gw190814_2021,dexheimer_novel_2010,dexheimer_proto-neutron_2008,dexheimer_reconciling_2015,dexheimer_tabulated_2017} \\
			\hline
			 \text{DS(CMF)-7}& \text{SU(3)-CMF}& \text{$n,p,e,H=[\Lambda,\Sigma^{-},\Delta^{-}]$} & 2.07& 11.43 & 13.20& \cite{bennour_charge_1989,gulminelli_unified_2015,dexheimer_gw190814_2021,dexheimer_novel_2010,dexheimer_proto-neutron_2008,dexheimer_reconciling_2015,dexheimer_tabulated_2017} 
		\end{tabular}
	\end{ruledtabular}
\end{table}
\begin{table}[h!]
	\caption{\label{tab:hybrid} Hybrid: Quark-Hadron cold EoS models.}
	\begin{ruledtabular}
		\begin{tabular}{ccccccc}
			EoS & Model & Matter & \texttt{$M_{\max }\left[M_{\odot}\right]$} &  \texttt{$R_{M_{\max }}[\mathrm{km}]$} &  \texttt{$R_{1.4 M_{\odot}}[\mathrm{km}]$}& References\\ 
			
			\hline
			\text {OOS(DD2-FRG) (2) flavors}& \text{NP-FRG}& \text{$n,p,e,q$} & 2.05& 12.55 & 13.20&\cite{hempel_statistical_2010,otto_hybrid_2020,typel_composition_2010}  \\
			\hline 
			\text {OOS(DD2-FRG) vec int-(2) flavors}& \text{NP-FRG}& \text{$n,p,e,q$} & 2.14& 12.70 & 13.20&\cite{hempel_statistical_2010,otto_hybrid_2020,otto_nonperturbative_2020,typel_composition_2010}  \\
			\hline 
			\text {BHK(QHC18)}& \text{NJL-MF}& \text{$n,p,e,q$} & 2.05& 10.41 & 11.49&\cite{akmal_equation_1998,baym_hadrons_2018,togashi_nuclear_2017,yu_self-consistent_2020}  \\
			\hline 
			\text {BFH(QHC19-B)}& \text{NJL-MF}& \text{$n,p,e,q$} & 2.07& 10.60 & 11.60&\cite{baym_hadrons_2018,baym_new_2019,togashi_nuclear_2017,yu_self-consistent_2020}  \\
			\hline 
			\text {BFH(QHC19-C)}& \text{NJL-MF}& \text{$n,p,e,q$} & 2.18& 10.80 & 11.60&\cite{baym_hadrons_2018,baym_new_2019,togashi_nuclear_2017,yu_self-consistent_2020} \\
			\hline 
			\text {BFH(QHC19-D)}& \text{NJL-MF}& \text{$n,p,e,q$} & 2.28& 10.90 & 11.60&\cite{baym_hadrons_2018,baym_new_2019,togashi_nuclear_2017,yu_self-consistent_2020} \\
		\end{tabular}
	\end{ruledtabular}
\end{table}

\begin{figure}[!h]
	\includegraphics[width=0.8\textwidth]{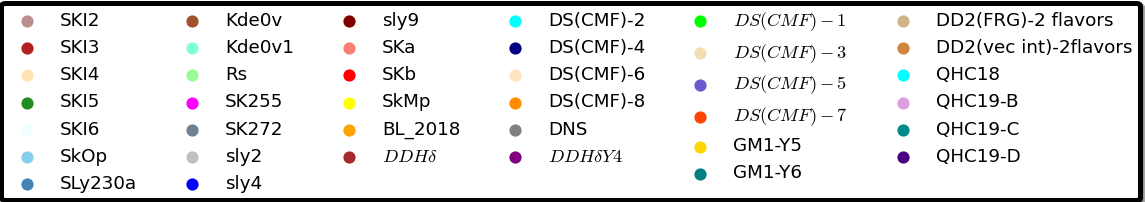}
	\caption{EoS-Color map used for the various figures.}%
	\label{fig:color_band}
\end{figure}

\end{widetext}

\newpage

\bibliography{universal_biblio}

\end{document}